\documentclass[fleqn,usenatbib]{mnras}
\usepackage{mathptmx}
\usepackage[T1]{fontenc}
\usepackage{ae,aecompl}
\usepackage{amsmath}
\usepackage{amssymb}
\usepackage{amsfonts,color,times,url}  
\usepackage[normalem]{ulem}

\usepackage{graphicx}	
\usepackage{amsmath}	
\usepackage{amssymb}	
\usepackage{tikz}
\usepackage{hyperref}

\def\Mpch{~h^{-1} {\rm Mpc}}
\def\Gpch{~h^{-1} {\rm Gpc}}
\def\MpchVolume{~(h^{-1} {\rm Mpc})^3}
\def\GpchVolume{~(h^{-1} {\rm Gpc})^3}

\def\shearUnit{\ensuremath{~{h}\rm{M}_{\odot} \rm{pc}^{-2}}}

\newcommand*\diff{\mathop{}\!\mathrm{d}}

\newcommand{\WVF}{\textsc{wvf}}
\newcommand{\SVF}{\textsc{svf}}
\newcommand{\zobov}{\textsc{zobov}}
\newcommand{\SVFtwoD}{\textsc{svf\_2d}}
\newcommand{\tunnels}{tunnels}
\newcommand{\troughs}{troughs}

\newcommand{\Euclid}{\textsc{euclid}}
\newcommand{\LSST}{\textsc{lsst}}

\newcommand{\CMASS}{\textsc{cmass}}
\newcommand{\HOD}{\textsc{HOD}}

\newcommand{\CMASSdef}{{\sc cmass-default}}
\newcommand{\CMASSn}{\textsc{cmass-fixed-$n_{g}$}}
\newcommand{\CMASSnxi}{\textsc{cmass-fixed-$n_{g}$-$\xi_{g}$}}

\newcommand{\GR}{\textsc{GR}}
\newcommand{\FR}{\ensuremath{f(R)}}
\newcommand{\Ffive}{\textsc{F5}}
\newcommand{\Fsix}{\textsc{F6}}

\newcommand {\xig} {\ensuremath{ \xi_{g} }}

\newcommand {\Reff} {\ensuremath{ R_{\rm eff} }}
\newcommand {\Rpeff} {\ensuremath{ R_{p\;{\rm eff}} }}
\newcommand {\meanSigma} {\ensuremath{ \langle \Sigma \rangle }}

\newcommand{\eq}[1]{Eq. \eqref{#1}}

\newcommand{\refsec}[1]{Sec. \ref{#1}}

\newcommand{\refappendix}[1]{Appendix \ref{#1}}

\newcommand{\reffig}[1]{Fig. \ref{#1}}

\newcommand{\figDir}{fig_pdf/}
\newcommand{\figwidth}{0.45\linewidth}
\newcommand{\verticaloffset}{\\[-.79cm]} 
\newcommand{\whitespace} { \hspace{0.8cm} \tikz \fill [white] (0,0) rectangle (0.43\linewidth,.8cm); }
\newcommand{\whitespaceFull} { \hspace{0.8cm} \tikz \fill [white] (0,0) rectangle (0.92\linewidth,.8cm); }
\newcommand{\reduceVerticalOffset} { \vspace{-.3cm} }

\newcommand{\figwidthTwo }{0.44\linewidth}
\newcommand{\figwidthTwoR}{0.428\linewidth}
\newcommand{\whitespaceTwo} { \hspace{.8cm} \tikz \fill [white] (0,0) rectangle (0.44\linewidth,0.8cm); }
\newcommand{\verticaloffsetTwo}{\verticaloffset}

\newcommand{\figwidthTwoTwo }{0.38\linewidth}
\newcommand{\whitespaceTwoTwo} { \hspace{.8cm} \tikz \fill [white] (0,0) rectangle (0.38\linewidth,0.1cm); }
\newcommand{\verticaloffsetTwoTwo}{\\[-.12cm]}

\definecolor{mycolor}{rgb}{.0,.3,1.}
\newcommand{\changed}[1]{{#1}}

\newcommand{\changedNew}[1]{{#1}}

\title[Void comparison in modified gravity]{The Santiago-Harvard-Edinburgh-Durham void comparison I: SHEDding light on chameleon gravity tests}

\author[]
{\parbox{\textwidth}{
	Marius Cautun$^{1}$\thanks{E-mail : m.c.cautun@durham.ac.uk}, Enrique Paillas$^{2,3}$, Yan-Chuan Cai$^{4}$, Sownak Bose$^{1,5}$, Joaquin Armijo$^{2,3}$, Baojiu Li$^{1}$ and Nelson Padilla$^{2,3}$
	}
\vspace{.2cm}\\
$^{1}$ Institute of Computational Cosmology, Department of Physics, Durham University, South Road, Durham, DH1 3LE, UK \\
$^{2}$ Instituto de Astrof\'isica, Ponticia Universidad Cat\'olica de Chile, Av. Vicu\~na Mackenna 4860, Santiago, Chile \\
$^{3}$ Centro de Astro-Ingenier\'ia, Ponticia Universidad Cat\'olica de Chile, Av. Vicu\~na Mackenna 4860, Santiago, Chile \\
$^{4}$ Institute for Astronomy, University of Edinburgh, Royal Observatory, Edinburgh EH9 3HJ, UK \\
$^{5}$ Harvard-Smithsonian Center for Astrophysics, 60 Garden Street, Cambridge, Massachusetts 02138, USA
}

\begin{document}
\label{firstpage}
\pagerange{\pageref{firstpage}--\pageref{lastpage}}
\maketitle

\begin{abstract}
We present a systematic comparison of several existing and new void finding algorithms, focusing on their potential power to test a particular class of modified gravity models -- chameleon $f(R)$ gravity. These models deviate from standard General Relativity (GR) more strongly in low-density regions and thus voids are a promising venue to test them. We use Halo Occupation Distribution (HOD) prescriptions to populate haloes with galaxies, and tune the HOD parameters such that the galaxy two-point correlation functions are the same in both $f(R)$ and GR models. We identify both 3D voids as well as 2D underdensities in the plane-of-the-sky to find the same void abundance and void galaxy number density profiles across all models, which suggests that they do not contain much information beyond galaxy clustering. However, the underlying void dark matter density profiles are significantly different, with $f(R)$ voids being more underdense than GR ones, which leads to $f(R)$ voids having a larger tangential shear signal than their GR analogues. We investigate the potential of each void finder to test $f(R)$ models with near-future lensing surveys such as \Euclid{} and \LSST{}. The 2D voids have the largest power to probe $f(R)$ gravity, with a \LSST{} analysis of tunnel (which is a new type of 2D underdensity introduced here) lensing distinguishing at 80 and 11$\sigma$ (statistical error) $f(R)$ models with parameters, $|f_{R0}|=10^{-5}$ and $10^{-6}$, from GR.
\end{abstract}



\section{Introduction}

After billions of years of evolution since the Big Bang, the Universe has developed \changed{rich large-scale structures, the so-called cosmic web. This forms a complex and space-filling pattern,} and is composed of lumpy knots at the intersection of long filamentary bridges, which in turn form at the intersection of tenuous sheets \citep[][see \citealt{Cautun2013} for a visualisation]{Bond1996}. These structures are populated by dark matter haloes, some of which are illuminated by galaxies such as our own Milky Way. 

Such rich structures are a consequence of the interplay of various ingredients, most of which have their roots in fundamental physics. The tiny initial density fluctuations are believed to be seeded by quantum effects in a hypothetical period of inflationary expansion. The matter species are predicted by the standard model of particle physics, along with its extensions which provide theoretical candidates for the dark matter. The effect of gravity is felt on almost all scales, from determining the expansion rate of the Universe as a whole to governing the formation and evolution of the cosmic web itself and the galaxies and stars inside them. Finally, there is the mysterious cosmic acceleration -- the observation that the expansion rate of the Universe has been accelerating instead of decelerating as predicted originally -- discovered initially almost two decades ago \citep{Riess1998, Perlmutter1999}, which could be due to a small positive cosmological constant, a new exotic matter species (dark energy) or a breakdown of Einsteinian General Relativity (GR) on cosmic scales (modified gravity). This interplay suggests that the accurate determination of the properties of the cosmic web could help to improve our understanding of the laws of fundamental physics that have shaped the Universe. However, it also poses great challenges to find optimal ways of extracting accurate information from a highly entangled observational reality. 

One of the key properties of the cosmic web, which has been a subject of growing interest in recent years, is related to the vast space in between {sheets and filaments}, nearly empty of galaxies and dark matter. These are commonly referred to as `cosmic voids' \citep{vandeWeygaert2009}.  By definition, these are low density regions that contain little matter and few galaxies, however they still encode detailed information about the underlying cosmological model \citep[e.g.][]{Li2011,Bos2012,Cai2014,Massara2015,Pisani2015,Zivick2015,Achitouv2016,Banerjee2016,Demchenko2016}.
Voids make for an attractive cosmological probe since their properties are hardly affected by galaxy formation physics, which is still a major unknown, and are well reproduced by dark matter only simulations (after suitably populating haloes with galaxies) \citep{Paillas2017}.
Their abundance, for example, can be qualitatively understood using semi-analytical models \citep[see, e.g.,][]{Sheth2003, Paranjape2012, Jennings2013}. However, as we will discuss below, there is an uncertainty in the precise definition of voids, which inevitably makes void abundance a less well defined quantity. In recent years, other observables associated with voids have attracted attention, e.g. void redshift-space distortions, which is the anisotropy in the observed correlation of voids and galaxies in redshift space induced by the peculiar velocities of galaxies \citep{Hamaus2015,Hamaus2016,Cai2016}. An additional observable is void lensing, an anti-lensing effect caused by the underdense voids deflecting light away from their centres, which has been detected in both the Sloan Digital Sky Survey ({\sc sdss}) and the Dark Energy Survey ({\sc des}) \citep{Clampitt2014, Gruen2015, Sanchez2016}. 

The surge of interest in cosmic voids is also partly thanks to a community that works on modified gravity theories \citep[see, e.g.,][for the latest reviews]{Joyce2014, Koyama2016}.  This growing field explores the possibility that the cosmic acceleration is due to new gravitational physics on cosmological scales, and attempt to use precision cosmological data to verify the validity of GR in a new regime, beyond traditional tests of the theory \citep{Will2014}. In environments similar to the Solar System (e.g. deep or large gradient of the gravitational potential, high matter density), small-scale tests of GR have placed stringent constraints on the viability of any new gravity theory (or extension to GR). As a result, many of the newly proposed theories have a so-called screening mechanism, by which standard GR is recovered, and existing small-scale tests of GR satisfied. As a result of this, the difference between these models and GR are usually suppressed in dense regions where most cosmological data is currently available. It is because of this that voids offer an alternative and promising venue to test them. In the literature, there have been many studies trying to understand the properties of voids in non-standard cosmological models. While simple semi-analytical models \citep[see, e.g.,][for examples]{Clampitt2013,Lam2014} can offer useful insight into the essential physics, most studies so far have relied on numerical simulations to get more precise predictions for a larger range of void properties \citep[e.g.,][]{Li2011, Li2012, Bos2012, Cai2014, Zivick2015, Barreira2015, Pisani2015, Barreira2017a, Falck2017, Paillas2017}.

The name `cosmic voids' is used to describe generic low-density regions in the Universe, and it is difficult to make a precise definition without having a clear idea about how to determine the spatial boundary of such regions. For example, looking at a 3D distribution of galaxies, it is immediately apparent that regions devoid of galaxies usually take on irregular shapes and there is no straightforward way to define their spatial domains. In the literature, two frequently used methods to identify voids are based on `spherical underdensity' \citep[e.g.][]{Padilla2005} and `space tessellation' \citep[e.g.][]{Platen2007,Neyrinck2008}. In the former approach, one centres voids at positions of lowest density and grows spheres until some density threshold, $\rho_{\rm v}$ is attained (e.g. the average density inside the sphere is less than $\rho_{\rm v}$). In the second approach, one covers the space with tessellation cells (such as Voronoi and Delaunay) in such a way that larger cells correspond to low-density regions. The cells are then joined to form voids according to certain prescriptions. This broad distinction, however, does not cover all possible algorithms of void finding, and actually allows for many variations in the exact implementation. For example, the threshold $\rho_{\rm v}$ in spherical underdensity void finders can take different values, the method for joining tessellation cells to form voids can vary, the tracers (e.g., which galaxy population and what tracer number density) used to identify voids differ in different works, and so on. Such ambiguities in void finding could be seen as a disadvantage: if different groups cannot agree on how voids are defined, and the results from them display discrepancies among each other, how can any of these results be useful?

However, it has been suggested \citep[see, e.g.,][]{Cai2014} that a positive point of view could be taken: the cosmic web is a highly complicated structure and one should not expect that a single statistic can characterise it completely. Rather, each void finding algorithms represents a different ways to quantify the emptiness -- there is no {\it a priori} way to know which algorithm is the `best', since this will depend on both the observables being used and the questions being asked. The latter consideration adds a new dimension to void studies and is a main motivation of this work: given a certain cosmological model such as a specific modified gravity (which usually has distinct physical properties that imprint different observational features compared to other models), what is the void finder that best captures these differences and hence presents the best opportunity to tell it apart from other models? To address this question, in this paper we will compare the predictions of some key void properties using a number of void finders.

This is the first of a series of papers, with which we endeavour to make a detailed exploration of the performances of {\it various} void finders in predicting {\it various} void properties for {\it various} theoretical models. The emphasis here is that the different void finders should be compared (1) in specific and clear contexts, i.e., the comparison is done for certain theoretical models so that the conclusion would apply to those models and not necessarily to all models, (2) at equal footing, i.e., wherever possible, the properties of voids identified by different void finders are calculated using the same pipelines to exclude spurious differences that can bias the comparison, and (3) with a reasonable degree of reality, i.e, we focus on voids identified from galaxy fields in typical current and near-future galaxy surveys, and assume that the different models being tested all predict the same galaxy clustering -- the observed one. In practice, this is achieved by tuning the parameters governing how galaxies populate dark matter haloes; we do this because there is only one observed Universe, and if the galaxy clustering predicted by a model doesn't agree with observations, then that model should have already been ruled out. Some of the void finding algorithms being compared here are new, and we hope that this further adds to the usefulness of this work to the community. We want to stress that it is not our intention to distinguish {\it good} versus {\it poor} void finders, but rather to find the {\it right} ones for specific questions at hand.

The only other comparison study of cosmic voids, to the best of our knowledge, is the Aspen Void Finder Comparison Project by \citet[][AVFCP]{Colberg2008}, which compared a total of 13 void finders; it differs from this work in several ways. First, AVFCP also compared voids identified from the dark matter field, while we only use galaxies as tracers since these are directly observable in galaxy surveys. Second, AVFCP identify voids from a selected region of size $60~h^{-1}$Mpc from the Millennium simulation, while we use a whole simulation box of size $1024~h^{-1}$Mpc to get a large number of voids spanning a range of sizes, and study different void observables. Finally, as we have mentioned above, this is a void code comparison project in the context of testing gravity, and thus we run our void finders on several sets of simulations which differ (only) by their underlying gravity theories; in particular, we have tuned our mock galaxy catalogues so that they have the same correlation functions, to avoid the contamination of having different tracers for the different models. This work will focus on the comparison of void finders in testing a particularly popular class of modified gravity models: the so-called chameleon model. 

This paper is organised as follows: In \S~\ref{sect:theory} we briefly review the $f(R)$ model, which is a representative example of chameleon models, and the simulations and mock galaxy catalogues used in the comparison.  In \S~\ref{sect:voids} we introduce the \changed{six} void finding algorithms to be studied and present a visual comparison of the various void catalogues. The main results of this paper are presented in \S~\ref{sect:results}, where we compare the void abundance, void galaxy number density profile, matter density profile, and lensing tangential shear profiles; we also estimate the signal-to-noise for each void finder to determine how well they are able to discriminate this particular class of models from GR. Finally, we discuss and conclude in \S~\ref{sect:discussion}.

Throughout this paper, we use the unit convention $c=1$, with $c$ the speed of light, except where $c$ is explicitly included. We use an overbar to denote the background value, and a subscript $_0$ to denote the present-day value of a quantity, unless otherwise stated.

\section{Theory}
\label{sect:theory}

We start by presenting a brief description of the $f(R)$ gravity model studied in this paper (\S\ref{subsect:MG_theory}), the simulations used and the catalogues of tracers of the large-scale structure which are needed for finding voids (\S\ref{subsect:simulation}). Many of the details -- such as the model and simulations -- can be found elsewhere; therefore this section is mainly for completeness and will be kept concise.

\subsection{Theory, simulations and halo/galaxy catalogues}
\label{subsect:MG_theory}

\subsubsection{$f(R)$ gravity}

$f(R)$ gravity is an alternative possibility to dark energy in explaining the accelerated Hubble expansion. It is constructed by replacing the usual Ricci scalar $R$ in the Einstein-Hilbert action with an algebraic function of $R$:
\begin{equation}
S = \int{\rm d}^4 x \sqrt{-g}~\frac{1}{16\pi G}\left[R+f(R)\right],
\label{equ:EH_action}
\end{equation}
in which $G$ is the gravitational constant, $g$ is the determinant of the metric tensor $g_{\mu\nu}$, and, for simplicity, we have ignored the \changed{matter Lagrangian term which is the same as in GR}.

The action in Eq.~(\ref{equ:EH_action}) can be extremised with respect to $g_{\mu\nu}$, which leads to a modified version of the Einstein equations:
\begin{equation}
G_{\mu\nu}+f_{R}R_{\mu\nu}-g_{\mu\nu}\left[\frac{1}{2}f-\nabla^2 f_{R}\right]-\nabla_{\mu}\nabla_{\nu}f(R) = 8\pi G T^m_{\mu\nu}.
\label{equ:E_equaiton}
\end{equation}
Here, $G_{\mu\nu}\equiv R_{\mu\nu}-\frac{1}{2}g_{\mu\nu}R$ is the Einstein tensor, $\nabla_\mu$ denotes the covariant derivative compatible with $g_{\mu\nu}$, $f_R\equiv{\rm d}f/{\rm d}R$, $\square\equiv\nabla^\alpha\nabla_\alpha$ and $T^m_{\mu\nu}$ is the stress-energy tensor for matter. Eq.~(\ref{equ:E_equaiton}) is a 4th-order equation because $R$ contains 2nd-order derivatives of the metric field. For this reason, it is useful to rewrite it as the usual Einstein equation in GR with a scalar field (called the scalaron), $f_R$. The equation of motion of the scalaron can be obtained by taking the trace of Eq.~(\ref{equ:E_equaiton}):
\begin{equation}
\square f_R = \frac{1}{3}\left(R-f_R R+2f+8\pi G \rho_m\right),
\label{equ:fR}
\end{equation}
where $\rho_m$ is the density for non-relativistic matter.

As we will restrict ourselves to subhorizon scales in a spatially flat universe, it is possible to use the quasi-static approximation which assumes that the time derivatives of $f_R$ are negligible compared with their spatial derivatives \citep[see, e.g.,][for numerical tests of this approximation]{Bose2015}. Under this approximation, Eq.~(\ref{equ:fR}) becomes:
\begin{equation}
\vec\nabla^2 f_R=-\frac{1}{3}a^2\left[R(f_R)-\bar R+8\pi G(\rho_m-\bar \rho_m)\right],
\label{equ:fR_no_T}
\end{equation}
where $\vec\nabla$ is the spatial gradient, $a$ is the scale factor, with the usual convention $a=1$ today. 

On the other hand, the Newtonian potential $\Phi$ in this model is governed by a modified Poisson equation, which now becomes:
\begin{equation}
\vec\nabla^2 \Phi=\frac{16\pi G}{3}a^2(\rho_m-\bar \rho_m)+\frac{1}{6}\left[R(f_R)-\bar R\right].
\label{equ:X}
\end{equation}

In some sense, the scalar field $f_R$ plays the role of the potential of a fifth force which is the difference between the modified gravitational force as predicted by Eq.~(\ref{equ:X}) and the standard Newtonian force produced by the same matter density field $\rho_m$. There exist two opposite limits of the behaviour of this fifth force (though solutions anywhere in between could exist in regions and/or epochs of a real universe):
\newline\indent$\bullet$ When $|f_R|\ll|\Phi|$, we approximately recover the GR solution $R=-8\pi G\rho_m$, which simplifies Eq.~(\ref{equ:X}) to the usual Poisson equation: 
\begin{equation}
\vec{\nabla}^2\Phi = 4\pi G\delta\rho_m,
\end{equation}
where $\delta\rho_m\equiv\rho_m-\bar{\rho}_m$ is the matter density perturbation. The effect of the fifth force is strongly suppressed in this regime, which is the consequence of what now is known as the chameleon mechanism \citep{Khoury2004}.

$\bullet$ In contrast, in the case of $|f_R|\geq|\Phi|$, we have $|\delta R|\ll\delta\rho_m$ where $\delta R\equiv R-\bar{R}$, leading to a modified Poisson equation: 
\begin{equation}
\vec{\nabla}^2\Phi\approx16\pi G\delta\rho_m/3,
\end{equation} 
which describes a $1/3$ enhancement compared to the strength of Newtonian gravity. This enhancement is independent of the functional form of $f(R)$, though the specific choice does affect the transition between this regime and the previous one. In the literature, these are often called the {\it screened} and {\it unscreened} regimes respectively.

The chameleon mechanism earns its name because of the apparently environmental dependence of the fifth force; namely, that it is screened in regions where the Newtonian potential is deep enough. Intuitively, the scalar field $f_R$ is the mediator of the fifth force, and has a mass due to its self-interaction (as described by its potential), so that the fifth force is of Yukawa type. The mass is environment-dependent and becomes quite heavy in deep Newtonian potentials, causing the Yukawa force to decay exponentially with distance, \changed{which would make it difficult to detect  experimentally the fifth force on Earth.}

The Solar System is an example of a scale where screening {\it could} occur, making $f(R)$ gravity viable (i.e., not yet ruled out by fifth force experiments). However, determining whether the screening has {\it indeed} taken place is much more tricky, because this also depends on the environment of the Solar System itself, such as the Galaxy and its underlying dark matter: even if the Solar System itself is incapable of producing a deep enough Newtonian potential to trigger screening, the Galaxy and its dark matter halo might just be able to do that. 

In this paper, we shall not pursue this line of research. Instead, we'll focus on cosmological scales and, in particular, regions of low matter density -- the so-called {\it cosmic voids}. The idea is that even if a given $f(R)$ model passes all local tests, it can still deviate substantially from GR in voids, where the chameleon screening does not happen. This opens a window of testing gravity independently. 

\subsubsection{The choice of $f(R)$ model}
\label{subsect:fR_theory}

The mere requirement of chameleon screening taking place in the Solar System for the model to be viable does not place a strong restriction on the functional form of $f(R)$ which, as mentioned above, determines how the fifth force transitions from screened to unscreened regimes. Thus, in the literature many possibilities have been studied to different levels of detail.

However, there are good reasons for focusing on one particular example in a detailed study. First, at this point, neither the motivation for the functional form of $f(R)$ nor the general understanding of the origin of the cosmic acceleration has matured to allow a connection to a more fundamental theory, and so any choice of $f(R)$ is phenomenological. Indeed, \changed{$f(R)$} models are known to face difficulties if inspected from a theoretical point of view \citep[see, e.g.,][]{Erickcek:2013oma}.

Second, although the exact details differ, many $f(R)$ models share some common properties -- chameleon screening in dense regions being one -- resulting in qualitatively similar behaviour on cosmological scales. As a result, instead of asking `{what do observations tell us about the form of $f(R)$?}', we can ask the question `{to what extent are deviations from GR in the way that is prescribed by $f(R)$ gravity allowed by observations?}'. This question can be answered by choosing one particular model and testing it against as many observations as possible, as accurately as possible. 

For these considerations, in this paper we choose to work with the model first proposed by \citet[][hereafter HS]{Hu2007}, which has three advantages. First, it is the most well-studied choice of $f(R)$ gravity in the literature, so that this paper is building upon many existing tests of the model. Second, its particular $f(R)$ form allows an efficient algorithm to be used to greatly speed up numerical simulations of the model \citep{Bose:2016wms}, making its study easier, more efficient, and more accurate. Finally, not only is this model a representative example of $f(R)$ gravity, but it also -- as we shall see shortly -- has the flexibility to cover very different behaviours by varying its parameters. This last property means that studies of this particular model can inform us of general implications beyond the $f(R)$ class of models. 

The HS model is specified as following:
\begin{equation}\label{eq:hs}
f(R) = -M^2\frac{c_1\left(-R/M^2\right)^n}{c_2\left(-R/M^2\right)^n+1},
\end{equation}
where $c_1, c_2$ \changed{and $n$} are dimensionless \changed{model} parameters, and $M^2\equiv8\pi G\bar{\rho}_{m0}/3=H_0^2\Omega_m$, with $H$ the expansion rate and $\Omega_m$ the current fractional density of non-relativistic matter.

When $|\bar{R}|\gg M^2$, $\bar{f}=f(\bar{R})\rightarrow-\frac{c_1}{c_2}M^2$, and $f_R$ and its derivatives become small, so that Eq.~(\ref{equ:fR}) can be approximated as: 
\begin{equation}\label{equ:R_bg}
-\bar{R} \approx 8\pi G\bar{\rho}_m-2\bar{f} \approx 3M^2\left(a^{-3}+\frac{2c_1}{3c_2}\right),
\end{equation}
which mimics the background cosmology of the standard $\Lambda$CDM model, with the mapping:
\begin{equation}
\frac{c_1}{c_2} = 6\frac{\Omega_\Lambda}{\Omega_m},
\end{equation}
where $\Omega_\Lambda\equiv1-\Omega_m$ (we consider a spatially flat universe throughout this paper). To reproduce the observed accelerated expansion rate with $\Omega_\Lambda\approx0.7$ and $\Omega_m\approx0.3$, we have $|\bar{R}|\approx40M^2\gg M^2$ today, which is also true at all earlier times because $|\bar{R}|$ increases with redshift, and so the approximation in Eq.~(\ref{equ:R_bg}) is always valid.

Under the same approximation, and by taking the derivative of Eq.~(\ref{eq:hs}) with respect to $R$, we find a simplified relation between $f_R$ and $R$:
\begin{equation}\label{eq:temp}
f_R \approx -n\frac{c_1}{c_2^2}\left(\frac{M^2}{-R}\right)^{n+1} < 0,
\end{equation}
which can be readily inverted to obtain the quantity $R(f_R)$ that appears in the scalar field and modified Poisson equations. As a result, we only need to specify $n$ and $c_1/c_2^2$ -- along with cosmological parameters including $\Omega_m, \Omega_\Lambda, H_0$ -- to fully fix the model. However, it is often more convenient to use $f_{R0}$, the present-day value of the scalar field $f_R$, instead of $c_1/c_2^2$.
The two parameters are related by:
\begin{equation}
\frac{c_1}{c_2^2} = -\frac{1}{n}\left[3\left(1+4\frac{\Omega_\Lambda}{\Omega_m}\right)\right]^{n+1}f_{R0}.
\end{equation}

\changed{In this paper}, we will focus on the particular choice, $n=1$, and consider two values of $f_{R0}$ -- $|f_{R0}|=10^{-6}, 10^{-5}$, which are referred to as F6 and F5 in the literature. These choices of $f_{R0}$ cover a broad range of model behaviours: F6 is the one which deviates least strongly from GR, while F5 deviates the most. Recent studies show that F5 could already be in tension with various cosmological observations \citep[see, e.g.,][for a review of current constraints on $f(R)$ gravity]{Lombriser2014}, but for completeness we will still consider F5 in what follows.


\subsection{Simulations and halo/galaxy catalogues}
\label{subsect:simulation}



The simulations used in this paper ({\sc elephant} simulations, or Extended LEnsing PHysics using ANalaytic ray Tracing) were performed using the {\sc ecosmog} code \citep{ecosmog}, which is an augmented version of the popular simulation code {\sc ramses} \citep{ramses}, with added new routines to solve the scalar field and Einstein equations in modified gravity models. It inherits {\sc ramses}'s efficient {\sc mpi} parallelisation and adaptive-mesh-refinement (AMR) feature. The code starts with a uniform domain grid which covers a cubic simulation volume and the grid has $N^{1/3}_{\rm dc}$ cells a side. The cells are refined (split into eight daughter cells) if the particle number in them exceeds some criterion ($N_{\rm ref}$); in this way, the code hierarchically refines the domain grid to achieve high resolutions for solving the fifth force when the particle number density is high.
The rest of the code algorithm is the same as in simulations for standard gravity, and interested readers are referred to \citep[e.g.,][for more details]{ramses}.

\begin{table}
\caption{The parameters and technical specifications of the $N$-body simulations used in this work. 
Note that $N_{\rm ref}$ is an array because we take different values at different refinement levels, and that $\sigma_8$ is for the $\Lambda$CDM model and only used to generate the initial conditions -- its value for $f(R)$ gravity is different but is irrelevant here.}
\begin{tabular}{@{}lll}
\hline\hline
parameter & physical meaning & value \\
\hline
$\Omega_m$  & present fractional matter density & $0.281$ \\
$\Omega_{\Lambda}$ & $1-\Omega_m$ & $0.719$ \\
$h$ & $H_0/(100$~km~s$^{-1}$Mpc$^{-1})$ & $0.697$ \\
$n_s$ & primordial power spectral index & $0.971$ \\
$\sigma_{8}$ & r.m.s. linear density fluctuation & $0.820$ \\
\hline
$n$ & HS $f(R)$ parameter & $1.0$ \\
$|f_{R0}|$ & HS $f(R)$ parameter & $10^{-6}, 10^{-5}$ \\
\hline
$L_{\rm box}$ & simulation box size & 1024~$h^{-1}$Mpc\\
$N_{\rm p}$ & simulation particle number & $1024^3$\\
$m_{\rm p}$ & simulation particle mass & $7.78\times 10^{10}h^{-1}M_{\odot}$\\
$N_{\rm dc}$ & domain grid cell number & $1024^3$\\
$N_{\rm ref}$ & refinement criterion & 8, 8, 8, 8, 8, 8, 8, 8...\\
\hline\hline
\end{tabular}
\label{table:simulations}
\end{table}

The cosmological and numerical parameters are listed in Table~\ref{table:simulations}, and the former are chosen to correspond to the best-fitting \citep{WMAP9} $\Lambda$CDM parameters. The simulations were initialised at $z_{\rm ini}=49$, with the initial conditions generated using the publicly available {\sc mpgrafic} code \citep{mpgrafic}. As a control, for each realisation of $f(R)$ simulations we also ran a $\Lambda$CDM simulation with the same parameters and specifications. Since the F5 and F6 models only deviate from $\Lambda$CDM at late times, at $z_{\rm ini}$ the effect of the fifth force is negligible, and so we let the simulations of all models start from exactly the same initial conditions. For $\Lambda$CDM, we had 5 independent realisations of boxes whose initial conditions differ only in their random phases, and we will use these to estimate errors in this paper. For $f(R)$ gravity, only the first two realisations were run up to $z=0$ and used for the analysis here.

\subsubsection{Halo and galaxy catalogues}
\label{subsec:HOD}

Dark matter haloes and the self-bound substructures associated with them are identified using the publicly-available {\sc rockstar} halo finder \citep{Behroozi2013}\footnote{\href{https://bitbucket.org/gfcstanford/rockstar}{https://bitbucket.org/gfcstanford/rockstar}}.  {\sc rockstar} uses the six-dimensional phase-space information from the dark matter particles to identify haloes. Note that, in principle, the presence of the fifth force in $f(R)$ gravity would require a modification to the unbinding procedure employed by {\sc rockstar} (or indeed any halo finding algorithm); however, \cite{Li2010} found the effect of this modification to be quite small and, as such, we use identical versions of {\sc rockstar} for the GR and $f(R)$ simulations. In this paper, we make use of only independent (`main') haloes, and not their substructures. 

In order to map the halo catalogues to a corresponding galaxy distribution, we populate haloes with galaxies using the Halo Occupation Distribution (HOD) method \citep{Peacock2000,Scoccimarro2001,Benson2000,Berlind2002,Kravtsov2004}, in which it is assumed that the probability for a halo to host a certain number of galaxies can be computed through a simple functional dependence on the mass of the host halo. We use the form of the HOD suggested by \cite{Zheng2007}, in which the mean number of central galaxies, $\left< N_{{\rm cen}} (M) \right>$, and the mean number of satellite galaxies, $\left< N_{{\rm sat}} (M) \right>$, hosted in a halo of mass $M$, are given by:
\begin{eqnarray} \label{eq:HOD_eqns}
\left< N_{{\rm cen}} (M) \right> &=& \frac{1}{2} \left[ 1 + {\rm erf} \left( \frac{\log M - \log M_{{\rm min}}}{\sigma_{\log M}} \right) \right], \nonumber \\ 
\left< N_{{\rm sat}} (M) \right> &=& \left<N_{{\rm cen}}\right> \left( \frac{M-M_0}{M_1} \right)^{\alpha},
\end{eqnarray}
in which $M_{{\rm min}}$, $M_0$, $M_1$, $\sigma_{\log M}$ and $\alpha$ are free parameters of the HOD model. Once their values have been specified, the mean number of galaxies in a halo of mass $M$ is then given by $\left<N(M)\right> = \left<N_{{\rm cen}} (M) \right> + \left< N_{{\rm sat}} (M) \right>$. From Eq.~\eqref{eq:HOD_eqns}, it can be seen that $M_{{\rm min}}$ and $M_0$, respectively, denote the threshold halo mass required to host at least one central or one satellite galaxy. When placing HOD galaxies in haloes, central galaxies are assumed to reside at the potential minima of their host halo. Satellites, on the other hand, are distributed between $\left[0,r_{200}\right]$\footnote{The radius within which the enclosed density is 200 times the critical density of the universe at that redshift.} of the host halo centre, according to a NFW profile with a concentration equal to the concentration of the host halo as computed by {\sc rockstar}. Furthermore, central galaxies are assigned the centre of mass velocity of the host halo, $V_{{\rm CM}}$; the velocity of a satellite galaxy is $V_{{\rm CM}}$ plus a perturbation along the $x, y$ and $z$ axes sampled from a Gaussian distribution with a dispersion equal to the r.m.s. velocity dispersion of the host halo.

\begin{table*}
\centering
\caption{Table of HOD parameters for the {\sc cmass-fixed-$n_g$-$\xi_g$} catalogue for GR, F6 and F5. The HOD parameters for F5 and F6 are obtained by requiring $n_g = 3.8 \times 10^{-4} h^{3} {\rm Mpc}^{-3}$ and $n_g = 3.2 \times 10^{-4} h^{3} {\rm Mpc}^{-3}$ at $z = 0$ and $z=0.5$, respectively, in addition to requiring that the projected clustering of galaxies in these models is nearly identical to that in GR. The parameters for the HOD catalogue corresponding to the GR run are the same as those obtained by \citet{Manera2013}. For F5 and F6 the HOD parameters outside (inside) the parentheses are for HOD catalogues generated using the simulation Box 1 (Box 2).}
\begin{tabular}{@{}llllll}
\hline \hline
 Model & $\log(M_{{\rm min}})$ & $\log(M_0)$ & $\log(M_1)$ & $\sigma_{\log M}$ & $\alpha$ \\ \\ 
 \hline
 & & \qquad\qquad\qquad\qquad\qquad\qquad $z=0.5$ \\ 
 \hline
 GR & 13.090 & 13.077 & 14.000 & 0.596 & 1.0127 \\
 F6 & 13.093 (13.073) & 13.077 (13.060) & 14.000 (13.983) & 0.540 (0.516) & 1.0127 (1.0127) \\
 F5 & 13.107 (13.130) & 13.077 (13.177) & 14.143 (14.040) & 0.439 (0.510) & 0.7444 (1.0127) \\
 \hline
 & & \qquad\qquad\qquad\qquad\qquad\qquad $z=0.0$ \\ 
 \hline
 GR & 13.090 & 13.077 & 14.000 & 0.596 & 1.0127 \\
 F6 & 13.103 (13.095) & 13.077 (13.082) & 14.000 (14.005) & 0.504 (0.486) & 1.0127 (1.0127) \\
 F5 & 13.106 (13.139) & 13.077 (13.126) & 14.098 (14.049) & 0.486 (0.546) & 1.0127 (1.0127) \\
 \hline\hline
	\end{tabular}
	\label{table:HOD1}
	\vspace{-.2cm}
\end{table*}

\begin{figure}
     \centering
     \begin{tabular}{c}
        \includegraphics[width=.93\linewidth,angle=0]{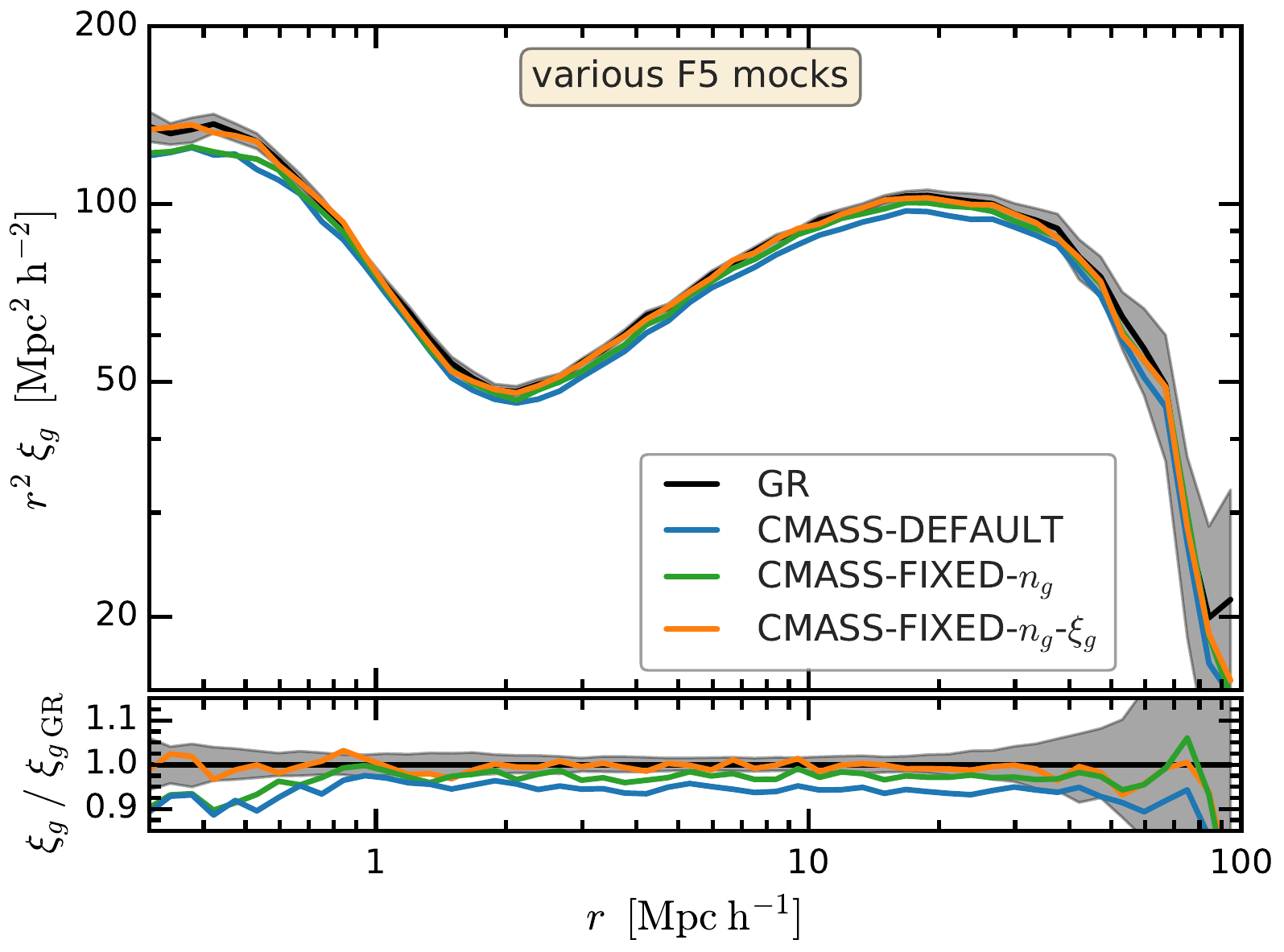} \verticaloffset
        \whitespaceFull \verticaloffset
        \includegraphics[width=.93\linewidth,angle=0]{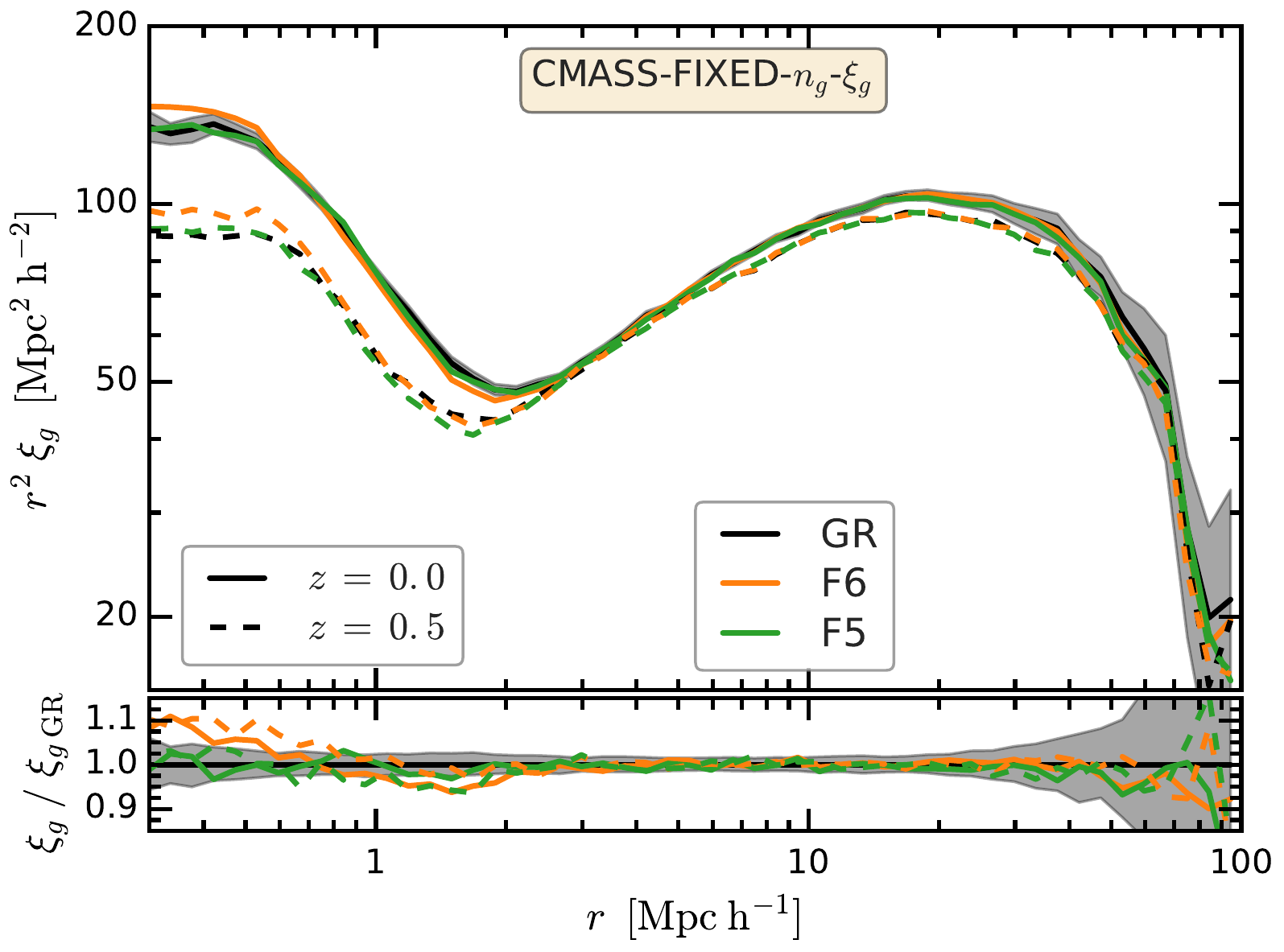}
     \end{tabular}
     \reduceVerticalOffset{}
     \caption{ \textit{Top panel:} the two-point galaxy correlation function, \xig{}, in \Ffive{} for various $z=0$ \HOD{} catalogues. The solid black curves show the \CMASS{} \HOD{} for \GR{}. The other curves show the following \Ffive{} \HOD{} catalogues: the same galaxy number density $n_{g}$ and \xig{} as \GR{} (\CMASSnxi{}), only the same $n_{g}$ as \GR{} (\CMASSn{}) and default \GR{} \HOD{} parameters (\CMASSdef{}). The secondary panel shows the ratio of the \Ffive{} \xig{} to the \GR{} one. \textit{Bottom panel:} \xig{} for the \CMASSnxi{} \GR{} and \FR{} catalogues at $z=0$ and $z=0.5$. The secondary panel shows the ratio of the \FR{} correlation function to the \GR{} one. The grey shaded region show the $1\sigma$ uncertainty for the \GR{} two-point correlation function. }
     \label{fig:2pt_corr_func}
\end{figure}
\begin{figure}
     \centering
     \begin{tabular}{c}
        \includegraphics[width=.93\linewidth,angle=0]{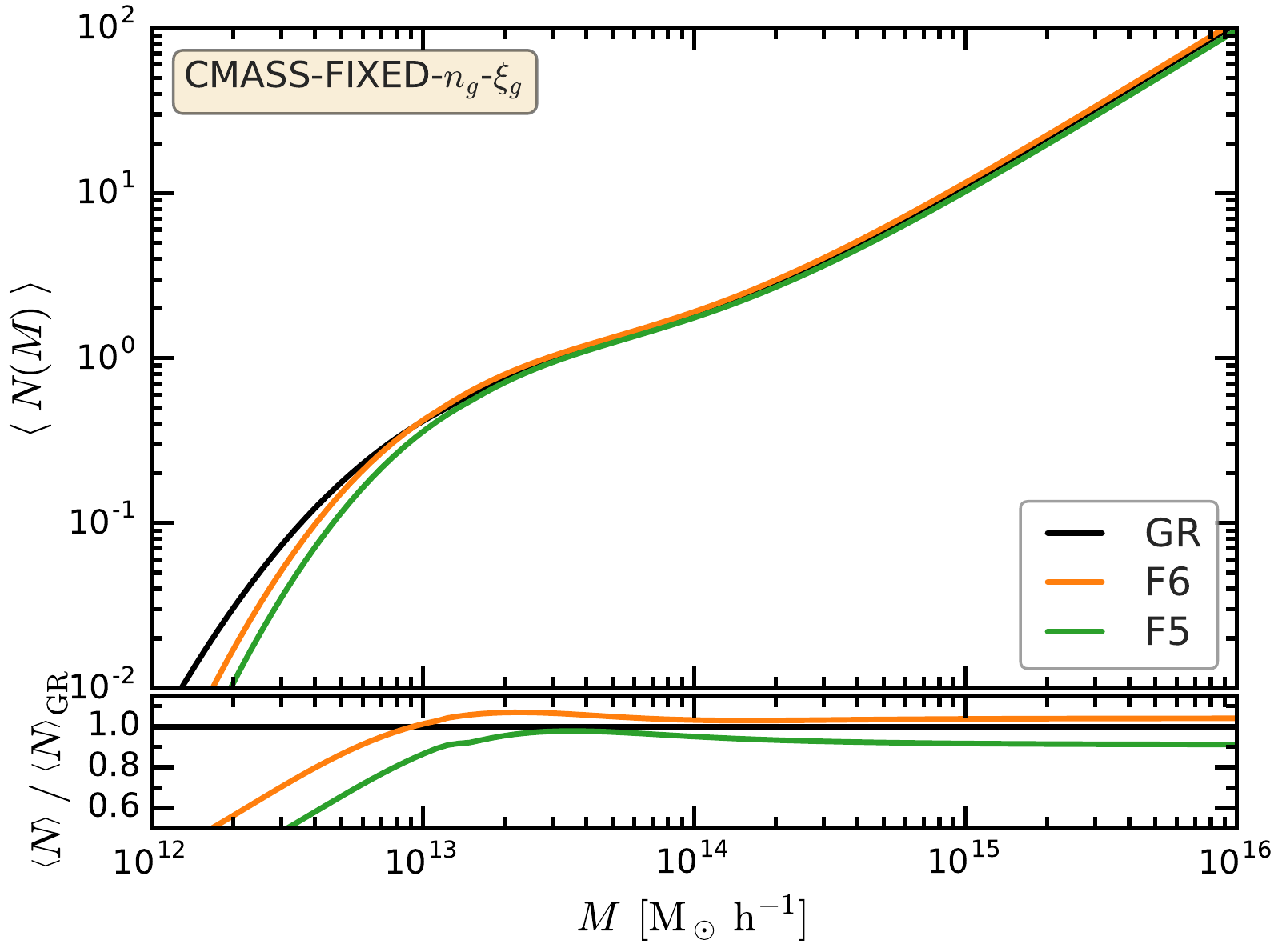} 
     \end{tabular}
     \reduceVerticalOffset{}
     \caption{ The mean number of galaxies, $\left< N \right>$, as a function of halo mass, $M$, for our \CMASSnxi{} catalogues. The different curves show the \GR{}, \Fsix{} and \Ffive{} \HOD{} models. The secondary panel shows the ratio of the \FR{} \HOD{} to the \GR{} one. }
     \label{fig:HOD}
     \reduceVerticalOffset{}
\end{figure}

The parameters of the HOD are calibrated by requiring that the galaxy catalogue produced matches up with the galaxy distribution obtained from redshift surveys according to various metrics: most commonly, the number density of galaxies, $n_g$, and their projected two-point clustering. For our catalogues, the target (comoving) number density of galaxies were chosen to be appropriate for the {\sc boss cmass dr9} galaxy sample \citep{Anderson2012}: $n_g = 3.8 \times 10^{-4} h^{3} {\rm Mpc}^{-3}$ and $n_g = 3.2 \times 10^{-4} h^{3} {\rm Mpc}^{-3}$ at $z = 0$ and $0.5$, respectively. Simply specifying the target number density is not enough to constrain the HOD, so we additionally require that the galaxy distribution exhibit the same two-point clustering across all gravity models at each redshift. 
Imposing this criterion in addition to the number density target substantially reduces the degree of degeneracy between different permutations of HOD parameter values. In total, we have constructed three types of HOD catalogues:
\begin{enumerate}
\item \CMASSdef{}: in which the same set of HOD parameters are used for {\it all} gravity models; 
\item \CMASSn{}: in which the parameters of the HOD are tuned (separately for each gravity model) to reproduce the same number density of galaxies {\it only};
\item \CMASSnxi{}: in which the parameters of the HOD are tuned (separately for each gravity model) to reproduce the same number density of galaxies {\it and} projected clustering.
\end{enumerate}

For catalogue (i), we use the same set of HOD parameters as in \cite{Manera2013}, which were tuned to create a mock galaxy catalogue representative of the {\sc boss cmass dr9} observational sample at $z=0.5$. The catalogues in (ii) were tuned separately at $z=0.5$ and $z=0$ to match the target number densities for that redshift. This was done by varying $M_{{\rm min}}$ while keeping the mass ratios, $M_0/M_{{\rm min}}$ and $M_1/M_{{\rm min}}$, constant.
The catalogues in (iii) were obtained by varying all five HOD parameters for each $f(R)$ model, redshift and realisation so that it has the same galaxy number density and two-point correlation as the GR mock catalogue for the same redshift and realisation. \changed{The tunning of the \HOD{} parameters was achieved using a search with the Nelder-Mead simplex algorithm through the 5-dimensional parameter space that minimized the r.m.s. difference between the \FR{} and \GR{} two-point correlation function in the distance range $[2,80]\Mpch$ \citep[for more details see][]{Li2017}. We tried different starting points for the search algorithm and all converged to the same value, suggesting that the choice of \HOD{} parameters is reasonably unique.}
Table~\ref{table:HOD1} summarises the HOD parameter values used to construct the \CMASSnxi{} catalogues for the first (Box-1) and second (Box-2) realisations. Note that for both $z=0$ and $z=0.5$, the GR parameter values are the same as those in \cite{Manera2013} (i.e., the same as catalogue i).

The difference in galaxy clustering induced by the different HODs is shown in Figure \ref{fig:2pt_corr_func}, which compares the galaxy two-point correlation function, $\xi_g$, in GR, with the F5 model for catalogues (i)-(iii). The top panel shows that simply fixing the F5 HOD parameters to the same values as in GR (\CMASSdef{}), or fixing them to achieve the same number density (\CMASSn{}) does not guarantee that the clustering is the same in the two models, and can in fact leave differences as large as 5\% across all separations. The bottom panel of Fig.~\ref{fig:2pt_corr_func} compares $\xi_g$ in GR, F6 and F5 at $z=0$ and $z=0.5$, but for the \CMASSnxi{} catalogue only; the agreement between the models now improves to better than 2\% for separations larger than 2 $h^{-1}$Mpc. Fig.~\ref{fig:HOD} displays the mean galaxy occupancy, $\left< N(M) \right>$, as a function of halo mass, $M$, for the \CMASSnxi{} catalogues in GR, F5 and F6. Finally, we note that catalogues (i) and (ii) are only used for illustration purposes in Fig.~\ref{fig:2pt_corr_func}; the fiducial catalogue used in the rest of the analysis is catalogue (iii).

\subsubsection{Mock observations}
\changed{We study the populations of voids at two redshifts, $z=0.0$ and $0.5$. The voids are identified using \HOD{} galaxy catalogues that are built from the distribution of dark matter haloes in the simulation snapshots corresponding to the two redshift choices. For simplicity, \textcolor{blue}{when} we apply the void finders to the distribution of galaxies in the periodic simulation box, which has a side length of $1024\Mpch$, we work in the distant observer approximation, which is when all the line-of-sights can be taken to be parallel, and we fix the line-of-sight to be along a preferential axis of the simulation box. This means that the 2D underdensities, which are identified in the projected galaxy distribution, correspond to geometrical cylinders along the line-of-sight; in actual observations or in more realistic mock catalogues, the 2D underdensities would correspond to conical frustums (cones with their top sliced off). We work within the same distant observer approximation when calculating projected matter density profiles and the tangential shear signal. Our mock galaxy catalogues also neglect redshift space distortions, which, while not important for 2D voids, can affect the identification of 3D voids. A detailed study of the redshift-space distortions effect on void finding and void properties will be left for future work.
}

\vspace{-.15cm}
\section{Void finders}
\label{sect:voids}

\begin{figure*}
     \vspace{-.05cm}
     \centering
     \begin{tabular}{cc}
        \includegraphics[width=\figwidthTwo,angle=0]{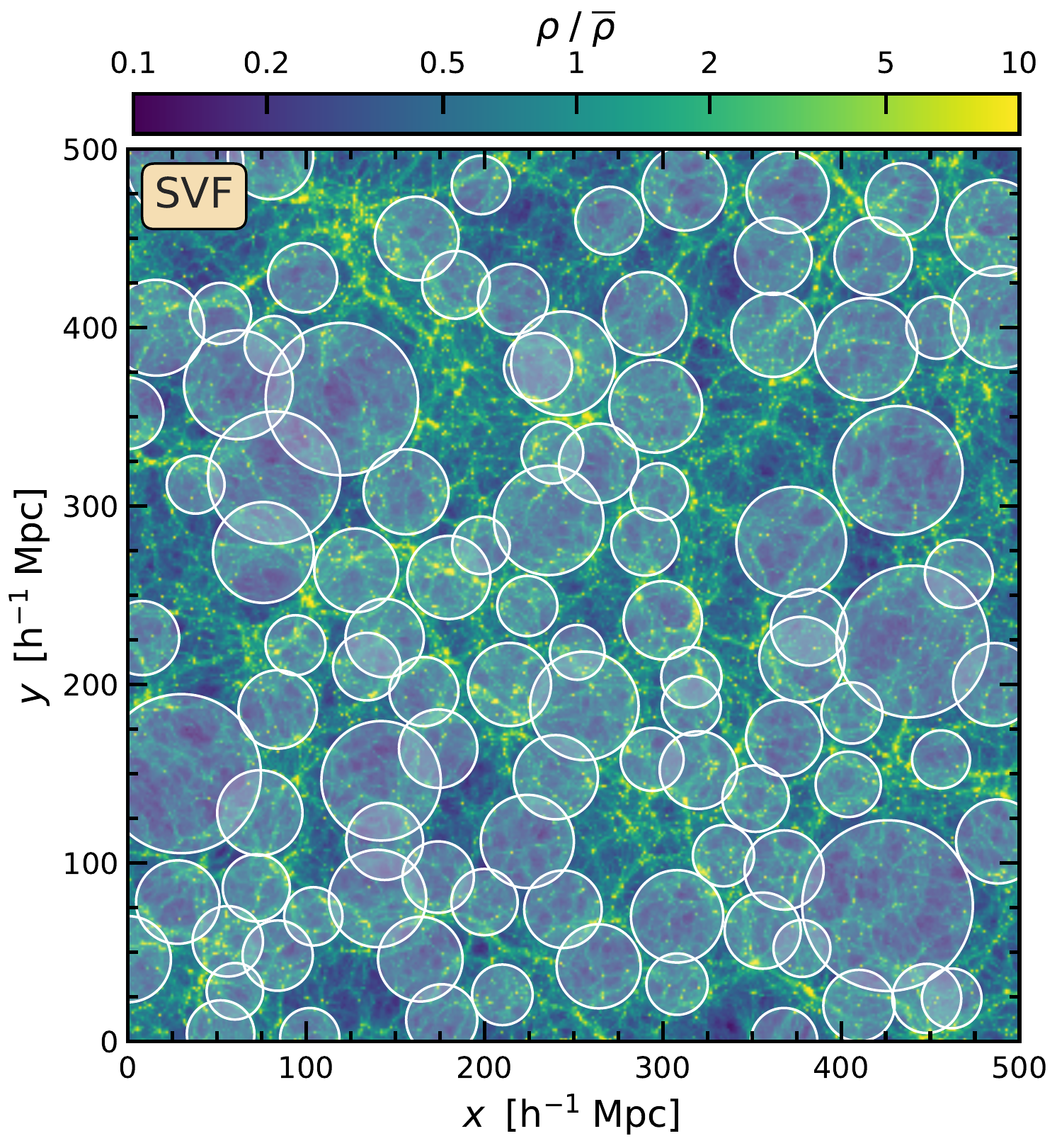} &
        \includegraphics[width=\figwidthTwoR,angle=0]{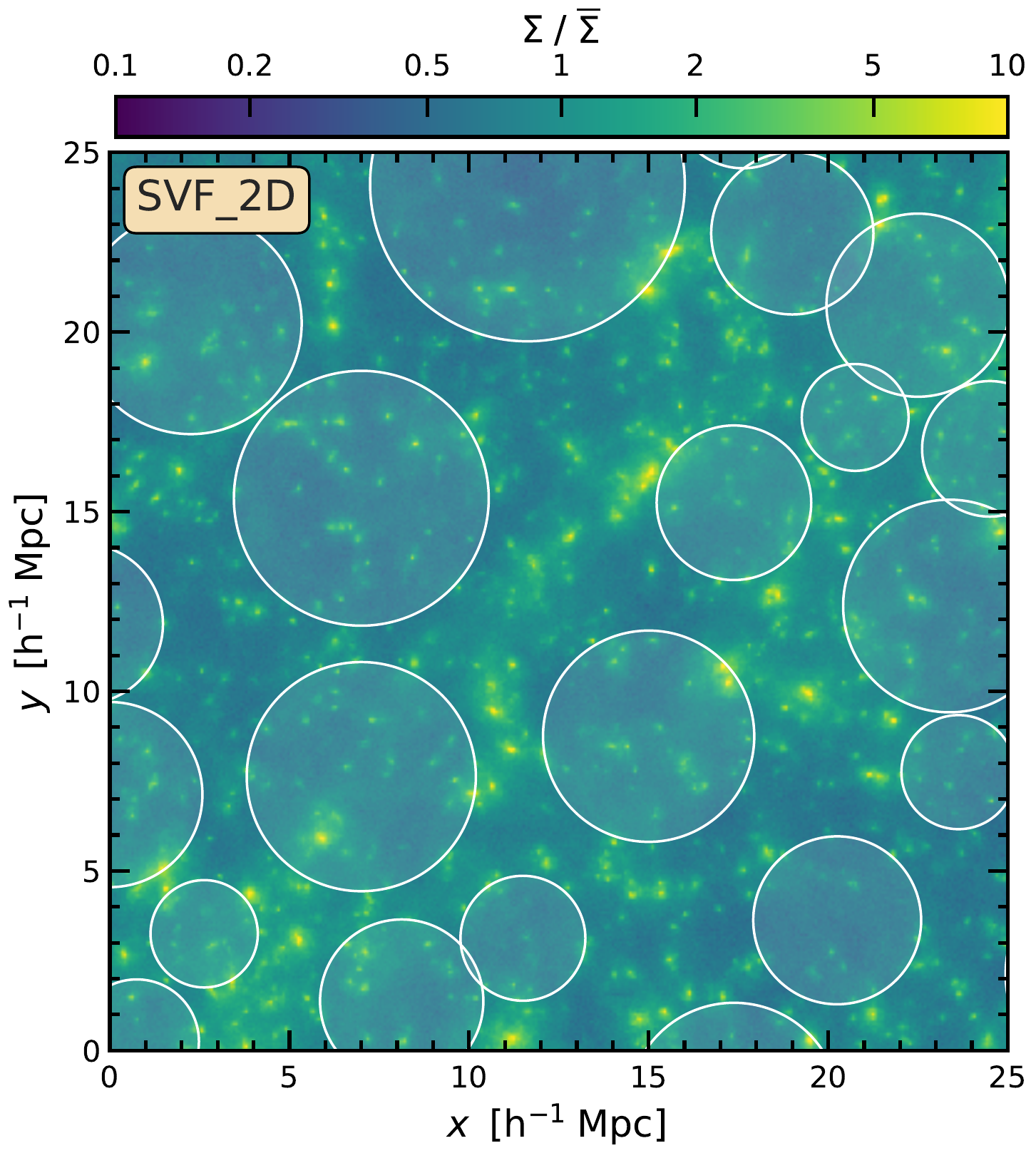} \verticaloffsetTwo
        \whitespaceTwo & \whitespaceTwo \verticaloffsetTwo
        \includegraphics[width=\figwidthTwo,angle=0]{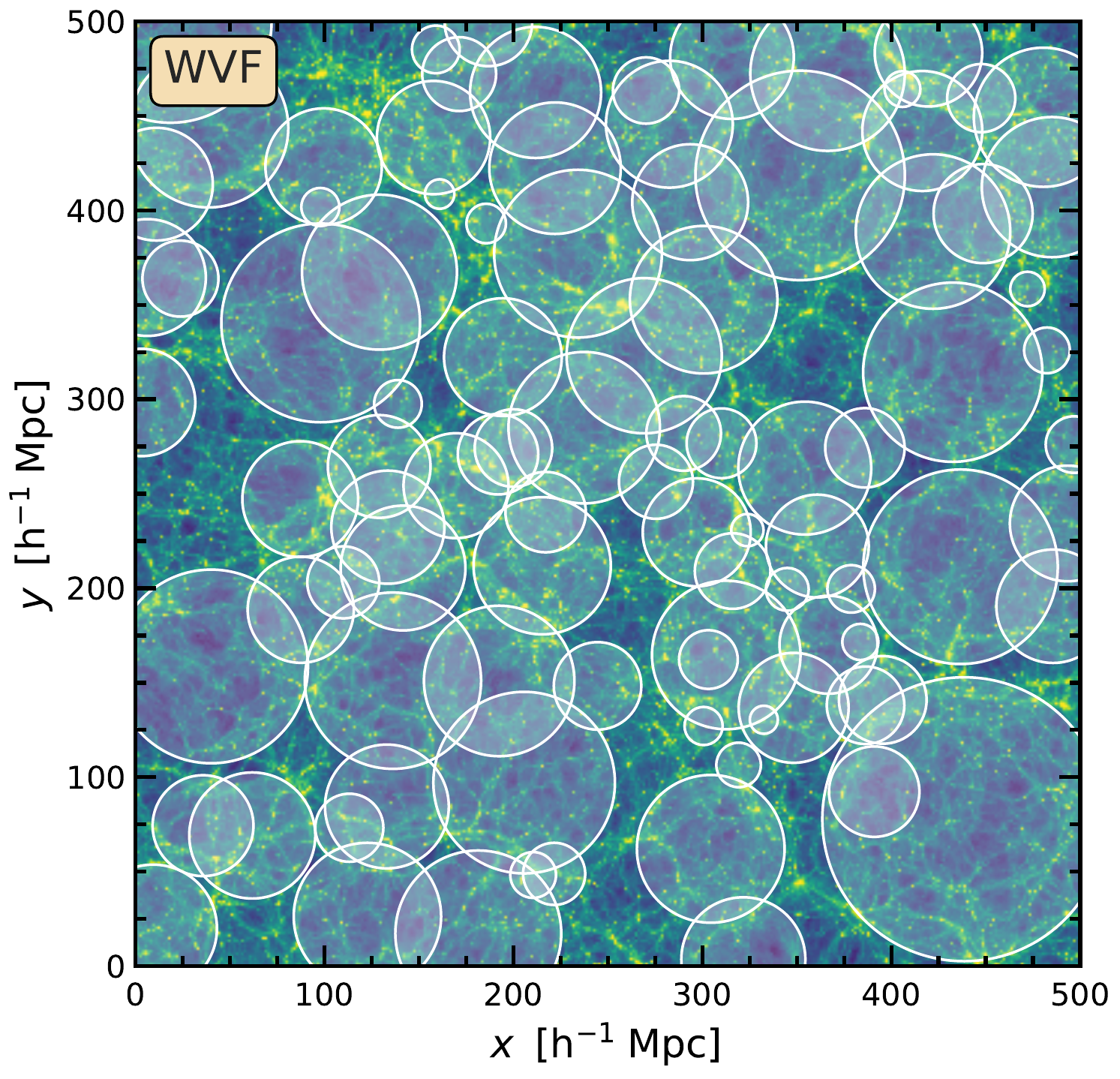} &
        \includegraphics[width=\figwidthTwoR,angle=0]{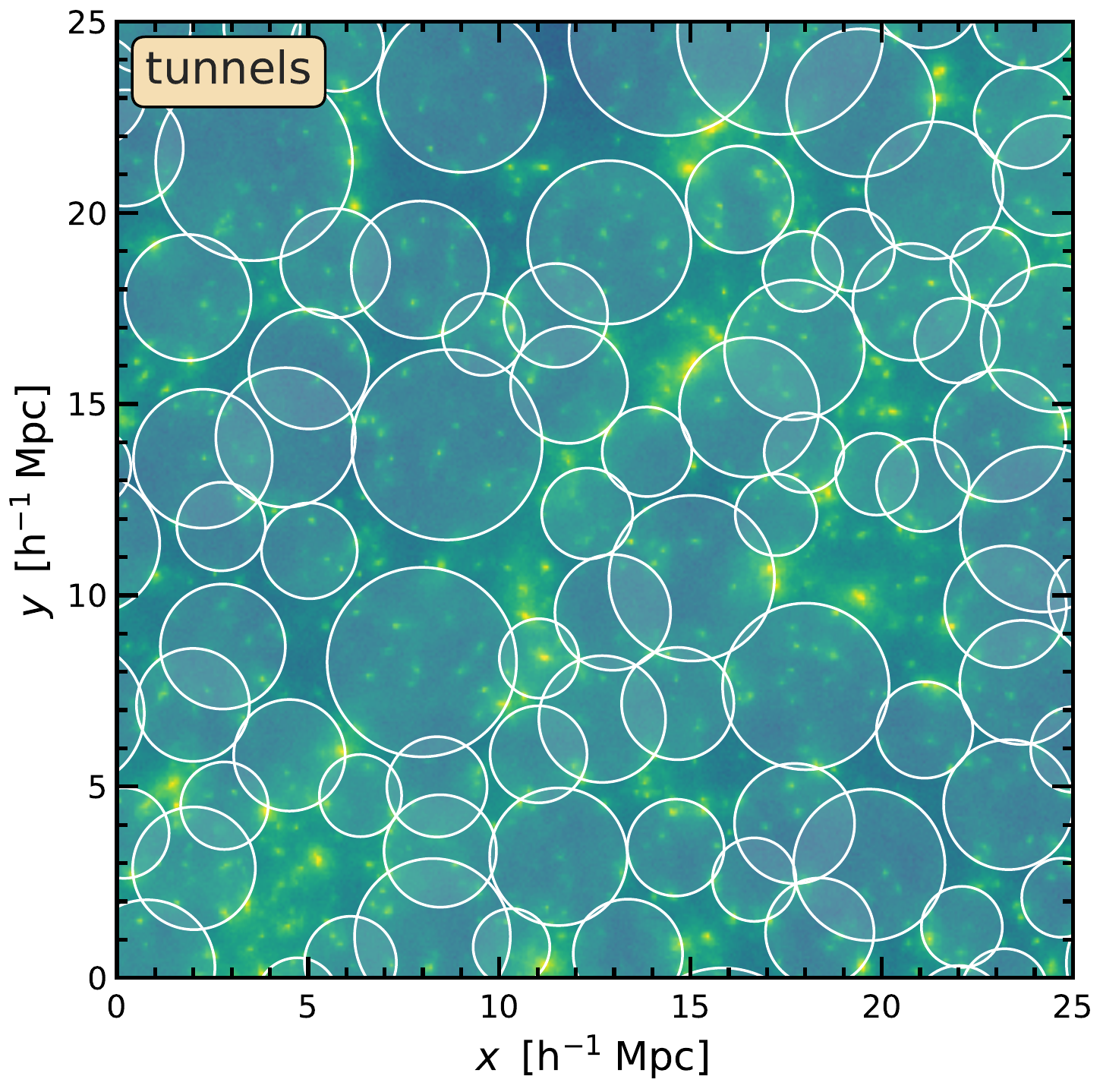} \verticaloffsetTwo
        \whitespaceTwo & \whitespaceTwo \verticaloffsetTwo
        \includegraphics[width=\figwidthTwo,angle=0]{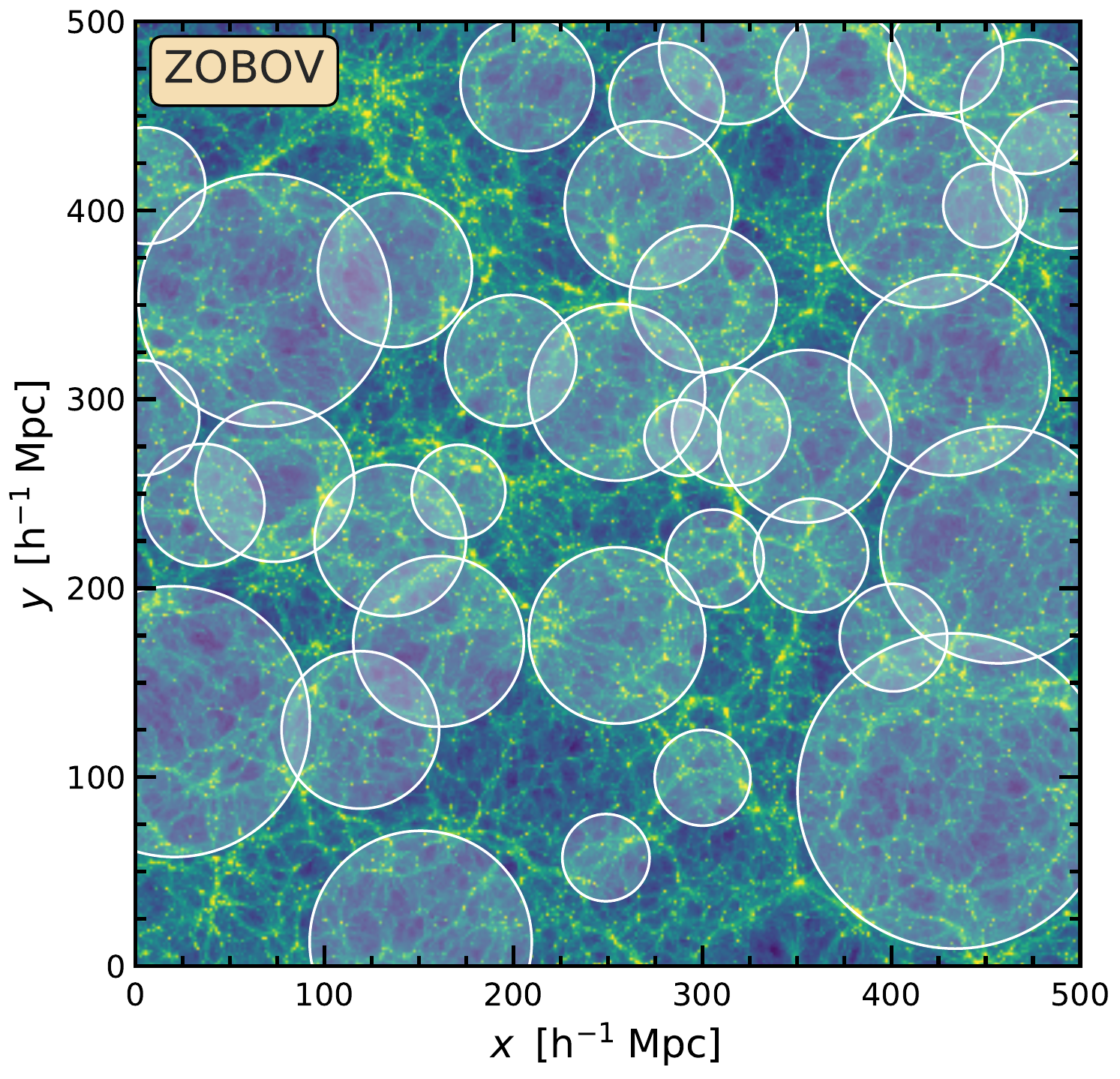} &
        \includegraphics[width=\figwidthTwoR,angle=0]{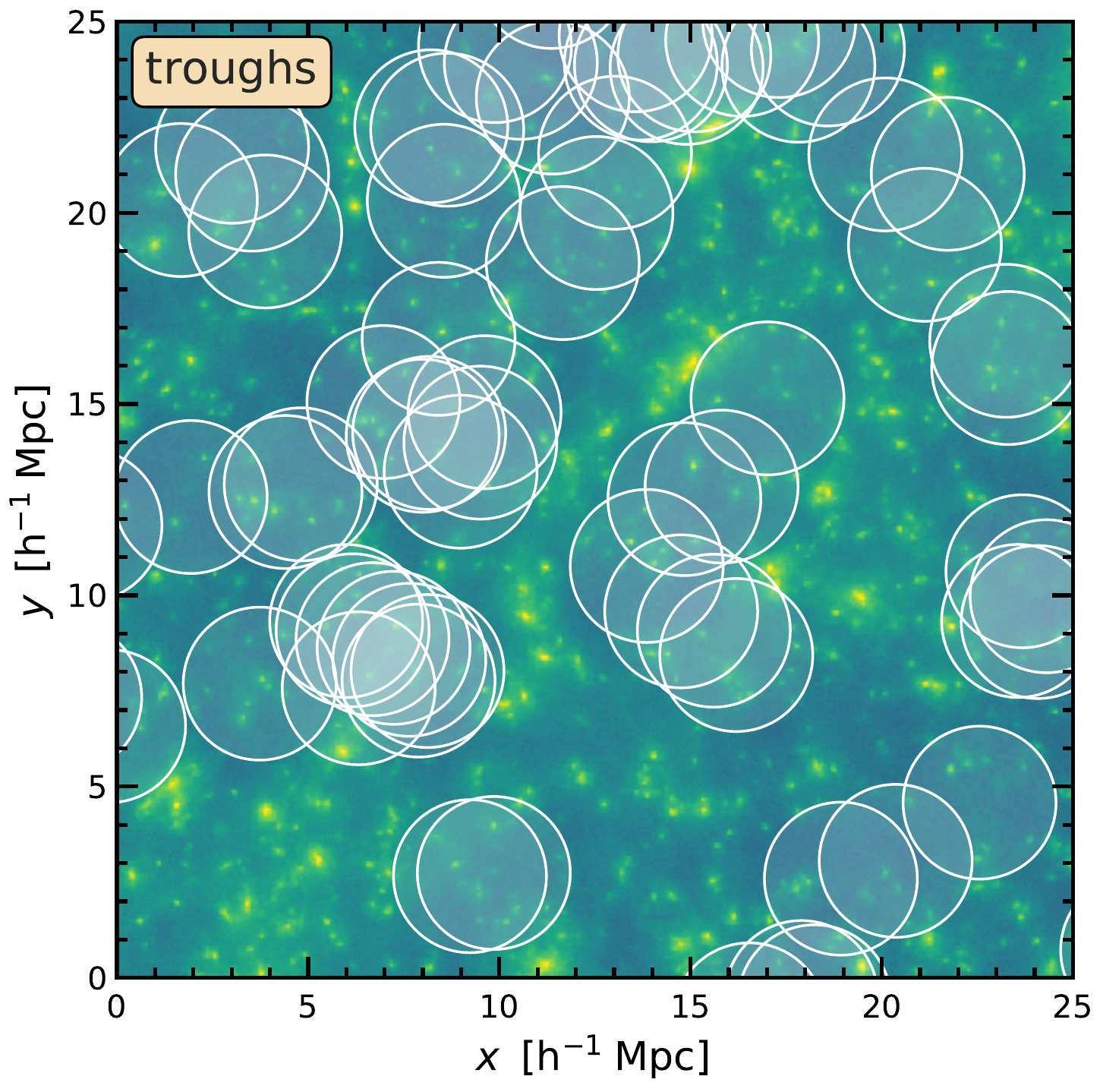}
     \end{tabular}
     \reduceVerticalOffset{}
     \caption{ Illustration of the structures identified by the six void finding methods employed in this paper. Each circle corresponds to an underdensity of radius equal to the one shown in the plots. The \textit{left column} plots the \SVF{} (top-left), \WVF{} (centre-left) and \zobov{} (bottom-left) 3D voids in a 50$\Mpch$ slice, with the background image showing the density in that slice. The \textit{right column} plots the 2D \SVFtwoD{} voids (top-right), tunnels (centre-right) and troughs (bottom-right), with the background image showing the projected density \changed{of the full box (which has a 1024$\Mpch$ side length)} along the line-of-sight. Note the different scales for the left and right columns. }
     \label{fig:images_3D_voids}
\end{figure*}

Our study makes use of several void finders with the goal of determining which ones are best suited for probing chameleon-type modified gravity models. Broadly, we can split the void finders into two categories: the ones that identify 3D underdensities (i.e. voids) and the ones that identify 2D underdensities in the plane-of-the-sky (2D spherical void finders, tunnels and troughs). The underdensities identified by the latter methods are not formally called voids, since voids are 3D objects, but nonetheless, we can think of them as 2D voids that are very elongated along the line of sight. \changed{We identify the 2D underdensities by projecting the full simulation box along one of the principal axes of the simulations. The box has a side length of $1024\Mpch$, which is approximately the comoving distance between redshift $0.3$ and $0.7$ -- the latter is the redshift range we use to make predictions for \Euclid{}- and \LSST{}-like surveys (see \refsec{subsec:euclid_lsst_predictions}).} The structures identified by each method are illustrated in \reffig{fig:images_3D_voids}. The rest of this section provides a short description of each of the six methods used to classify 2D and 3D underdensities. 

\vspace{-.15cm}
\subsection{3D Spherical underdensity void finder (\SVF{})}
The 3D spherical underdensity void finder (from hereon \SVF{}) used in this work is a substantially modified version of the algorithm presented in \citet{Padilla2005}. It searches for spherical regions that satisfy a specified density criterion in a simulation.

To find the void centres, a rectangular grid is constructed over the simulation volume. The number of galaxies in each cell of the grid is counted, and the centres of the cells that are empty of galaxies are considered to be prospective void centres. The integrated density profile about each centre is then calculated. The radius at which the integrated density is equal to 20 per cent of the mean number density of galaxies is considered to be the radius of a void. Only the largest sphere satisfying this condition about any centre is kept in the void catalogue. This density threshold is commonly adopted in the literature of void studies, motivated by a calculation presented in \cite{Blumenthal1992}, who used linear theory to show that the voids we observe at the present time correspond to mass underdensities of about 20 per cent of the cosmic mean. After this step, one might end up with a large number of \changed{voids} with similar centres, sharing a fraction of their volume. If \changed{the distance between the centres of any two voids is less than 80 per cent of the sum of their radii, we keep only the largest of the two.}  This naturally favours a single, large void over many smaller voids of comparable sizes and similar centres. A conservative degree of overlap between adjacent voids is still required, though, in order to trace the low density regions of the cosmic web to a good extent.

There are two free parameters in the 3D spherical underdensity void finder, the density criterion and the overlapping threshold for excluding voids. As mentioned above, the former is chosen as 20 per cent of the mean number density of galaxies for physical considerations. For the latter, \citet{Cai2014} have tested several different choices and found that the signal-to-noise is not sensitively dependent on the exact value, and so we choose 80 per cent as our default.

\vspace{-.15cm}
\subsection{3D Watershed void finder (\WVF{})}

The \WVF{} \citep{Platen2007} associates the voids to the watershed basins of the large scale density field without imposing {\it a priori} constraints on the size, shape and the mean underdensity of the objects it identifies. The method starts by constructing a galaxy density field using the Delaunay Tessellation Field Estimator \citep{Schaap2000,Cautun2011}, which uses a Delaunay triangulation with the galaxies at its vertices to extrapolate a volume-filling density field. The resulting density is defined on a $1024^3$ regular grid with a grid cell size of $1\Mpch$. The density is then smoothed with a Gaussian filter of $2\Mpch$ radius to reduce small scale structures inside and at the boundaries of voids, which could potentially give rise to artificial voids. The $2\Mpch$ filter corresponds to the typical width of the filaments and walls forming the void boundaries \citep[e.g.][]{Cautun2013,Cautun2014a}. The smoothed density field is then segmented into watershed basins. This process is equivalent to following the path of a rain drop along a landscape: each volume element, in our case the voxel of a regular grid, is connected to the neighbour with the lowest density, with the same process repeated for each neighbour until a minimum of the density field is reached. Finally, a watershed basin is composed of all the voxels whose path ends at the same density minimum.

The resulting \WVF{} void catalogue is characterized in terms of the void centres, which are chosen as the volume-weighted barycentre of all the voxels associated to each void, and the void sizes. Since voids have irregular shapes, the latter is given in terms of the effective void radius, $\Reff=(\tfrac{3}{4\pi}V)^{1/3}$, which is the radius of a sphere with the same volume as the void volume.

The \WVF{} has a single free parameter -- the Gaussian smoothing size. As mentioned above, our choice of $2~h^{-1}$Mpc corresponds roughly to the size of filaments and helps to reduce artificial voids. The smoothing size is much smaller than the radius of most of the \WVF{} voids considered in this paper, and, we do not expect it to strongly affect the properties of the \WVF{} voids.

\vspace{-.15cm}
\subsection{3D \zobov{} void finder}

We also run a slightly modified version of the \zobov{} algorithm \citep{Neyrinck2008} on our \HOD{} galaxy mocks. \zobov{} uses Voronoi tessellations to estimate the galaxy density field at each galaxy position and then identifies the density minima by comparing each Voronoi cell with its neighbours. Starting from the density minima, neighbouring Voronoi cells of increasing densities are grouped together to form `zones'. The growth of `zones' stops when the density of the next neighbouring Voronoi cell decreases. The original algorithm goes one step further to group neighbouring `zones' together if their boundary is below a density threshold, i.e., as this threshold density increases, more and more `zones' are grouped together to form larger and larger voids. This step can lead to unwanted effects; \citet{Cai2017} showed that when applying `zone' merging some of the largest \zobov{} voids found in their \CMASS{} mocks do not seem to correspond to true matter underdensities. To avoid such spurious effects, we treat every `zone' as a void in this work, i.e., we do not merge zones. By definition, our zones (voids) do not overlap with each other.

The void volume, $V$, is defined as the sum of the Voronoi cell volumes that are part of that given `zone'. The effective radius of the void is then defined as $\Reff=(\tfrac{3}{4\pi}V)^{1/3}$. Voronoi cells belonging to each zone are weighted by their volumes to define the void centre.

Our implementation of the \zobov{} void finder has no free parameters given a tracer catalogue.

\vspace{-.15cm}
\subsection{2D Spherical underdensity void finder (\SVFtwoD{})}

2D spherical voids are obtained using a slightly modified version of the \SVF{} algorithm. To find the void centres, a rectangular grid is constructed over the projected distribution of \HOD{} galaxies along one of the axes of the simulations, and the number of galaxies in each grid cell is counted. The centres of empty grid cells are considered as prospective void centres, and circles are grown from those centres until the integrated number density of galaxies at the circle radius is equal to 40 per cent of the mean density. After this step, if two voids overlap more than 80 per cent of the sum of their radii, only the largest void is kept in the catalogue.

The \SVFtwoD{}, as the 3D one, has three parameters: the density criterion for defining a void, the criterion \changed{for removing overlapping voids, and the length of projected redshift range}. For the former, we have tested density criteria equal to 0.2, 0.3, 0.4, 0.5 and 0.8 of the mean projected galaxy number density respectively, and found that choosing the 0.4 value results in the strongest detection of weak lensing by 2D underdensities \changed{(when considering only sample variance uncertainties)}. \changed{Similar to the 3D case, if the centres of two neighbouring voids are closer than 80 per cent of the sum of their radii, we remove the smaller one. Here, we identify voids by projecting in the distant observer approximation the entire simulation box ($1024\Mpch$ in length) along one of its preferential axes.}

\vspace{-.15cm}
\subsection{2D Tunnels}

The tunnels correspond to elongated line-of-sight regions that intersect one or more voids without passing through overdense regions (Cautun et al. in prep). Using galaxies as tracers of the matter distribution, the tunnels are identified as circles in the plane-of-the-sky that are devoid of galaxies. The typical size of tunnels depends on the number density of tracer galaxies and on the line-of-sight depth used to identify them; a higher tracer density or a larger line-of-sight depth results in smaller tunnels. In the distant observer approximation, which we use in this work, the tunnels consist of line-of-sight cylinders that are empty of galaxies\footnote{For realistic surveys, the tunnels correspond to conical frustums. For all practical purposes, the angular opening of tunnels is very small and they can be thought of as having planar top and bottom circular bases.}.

To identify the tunnels, we start by projecting the \HOD{} galaxy catalogue along one of the simulation axes to obtain a 2D distribution of galaxies. To identify the largest circles empty of galaxies, we build a Delaunay tessellation with the galaxies at its vertices. By definition, the circumcircle of every Delaunay triangle is empty of galaxies, with the closest galaxies being the three that give the triangle vertices and that are found exactly on the circumcircle. The tunnels consist of the circumcircles whose centres are not inside a larger circumcircle. We also discard candidates for which the Delaunay triangle has an area of 0.2 or less than that of its circumcircle; such cases correspond to triangles that either have a side much shorter than the other two or that have one very large angle. The tunnel centre and radius are given by the centre and radius of its corresponding circumcircle. 

Since galaxies are biased tracers of dark matter, tunnels do not necessarily correspond to empty regions of dark matter. Nevertheless, the tunnel radius is correlated with their projected matter density, as shown in \refappendix{appendix:tunnels_troughs}. On average, large tunnels correspond to underdense regions while small ones correspond to overdense ones, with the transition taking place at a tunnel radius of $1\Mpch$, \changed{which corresponds to $0.4$ times the mean projected galaxy separation}. Here we only consider tunnels larger than this transition radius since we are interested in the modified gravity signature of underdense regions.

The tunnel catalogue depends on two free parameters, the transition radius between underdense and overdense tunnels, and the line-of-sight projection length. The former is determined by analysing the enclosed projected matter density inside tunnels. The latter, which affects the size distribution of \SVFtwoD{} voids too, should be selected according to the details of the observational survey we want to match. For this study, we project the entire simulation box \changed{($1024\Mpch$ in length) using the distant observer approximation}.

\vspace{-.15cm}
\subsection{2D Troughs}
\label{subsec:troughs}

The troughs \citep{Gruen2015} are somewhat similar to tunnels in that they also correspond to elongated line-of-sight regions that are underdense. In contrast to tunnels, all troughs have the same radius and are given by randomly positioned circles in the plane-of-the-sky that contain very low galaxy counts.

The troughs are identified using the same projected \HOD{} galaxy catalogues as the tunnels, which represent our simulated plane-of-the-sky in the distant observer approximation. We choose to study troughs of $2\Mpch$ in radius, which is similar to the typical radius of tunnels. Moreover, troughs of this size contain a similar galaxy count at the 5$''$ troughs studied by \citet{Gruen2015} and \citet{Barreira2017a}. The trough identification starts by randomly positioning $10^6$ circles of $2\Mpch$ in radius on our simulated plane-of-the-sky. The troughs are given by the circles which contain 2 or fewer galaxies inside them; these correspond to $23$ and $30$ per cent of population at $z=0$ and $z=0.5$, respectively. This selection is similar to the one used by \citeauthor{Gruen2015} and it is also the one that gives a good compromise between selecting very underdense regions and covering a large fraction of the available simulation. See \refappendix{appendix:tunnels_troughs} for a more detailed discussion.

\subsection{Comparison of the different void finders}

\reffig{fig:images_3D_voids} presents a visual comparison of the underdensities identified by the six void finders. For the 3D voids, the figure shows the voids whose centres lie in $500\times 500\times 50 \MpchVolume$ region of the simulation. 
The background image shows the density, $\rho$, of that region expressed in terms of the mean background density, $\overline{\rho}$. 
While some voids seem to have overdense regions inside them, in most cases it is either a projection effect or it is due to representing the \WVF{} and \zobov{} voids as circles when they \changed{typically have non-spherical shapes}.  A closer inspection of the positions of 3D voids reveals that, in a large number of cases, we can find a match between individual objects identified by different void finders. This is especially striking for the \WVF{} and \zobov{} methods, since both are based on the watershed transform of the galaxy density field. Nonetheless, many of the matched voids have different centres and radii, and thus no two methods result in similar void catalogues (see \citealt{Colberg2008} for a comparison of more void finders).

The 2D underdensities generally have much smaller radii than their 3D counterparts, so the right column of \reffig{fig:images_3D_voids} shows 2D voids in a smaller region of size $25\times25$ ($h^{-1}$Mpc)$^2$ in the $xy$ plane of the whole simulation box. The 2D voids were identified by projecting the full simulation box \changed{($1024\Mpch$ side length)} along the $z$-direction \changed{in the distant observer approximation}, and the background colours show the projected matter density field (in unit of mean projected density) along the $z$-direction for the full simulation box. The three right-hand panels show that the 2D voids, which were selected as underdensities in the projected galaxy distribution, correspond to underdensities in the projected matter density field too. Most matter underdense regions are identified as 2D voids, but the centre and size of those 2D voids show a large variation between the three types of objects we study: \SVFtwoD{} voids, tunnels and troughs. Both \SVFtwoD{} voids and tunnels show relatively little overlap, but on average the tunnels are smaller and cover a larger fraction of the plane-of-the-sky than \SVFtwoD{} voids. The troughs are very clustered in underdense regions and show a large degree of overlap; in fact, for readability purposes, the bottom-right panel of \reffig{fig:images_3D_voids} shows only half of the troughs we identified, with the half not shown being right on top of the ones shown in the figure.

\vspace{-.15cm}
\section{Results}
\label{sect:results}

In this section we compare the distribution of void properties, e.g., void abundance, galaxy and matter density profiles, and lensing signal, between the standard $\Lambda$CDM cosmology and \FR{} models. We perform this analysis using the \CMASSnxi{} \HOD{} catalogues, which means that both the \GR{} and \FR{} mocks have the same number density of tracers {\it and} the same real-space two-point correlation function of these tracer galaxies. We perform this comparison for voids defined using six different methods: three that identify 3D underdensities (\SVF{}, \WVF{} and \zobov{}) and three that identify 2D underdensities using the projected distribution of galaxies (\tunnels{}, \troughs{} and 2D \WVF{}).

We characterise voids in terms of their radius, which we denote with $\Reff$ and $\Rpeff$ (i.e., projected radius) for the 3D and 2D structures, respectively. This nomenclature is motivated by two of the void finders, \WVF{} and \zobov{}, that identify irregularly shaped voids and thus these voids are characterized in terms of an effective radius, which is the radius of a sphere with the same volume as that of the void. For consistency, we use the same $\Reff$ notation also for the radius of the other voids, even though those are by definition spherical/circular objects.

\begin{figure*}
     \centering
     \begin{tabular}{cc}
        \includegraphics[width=\figwidth,angle=0]{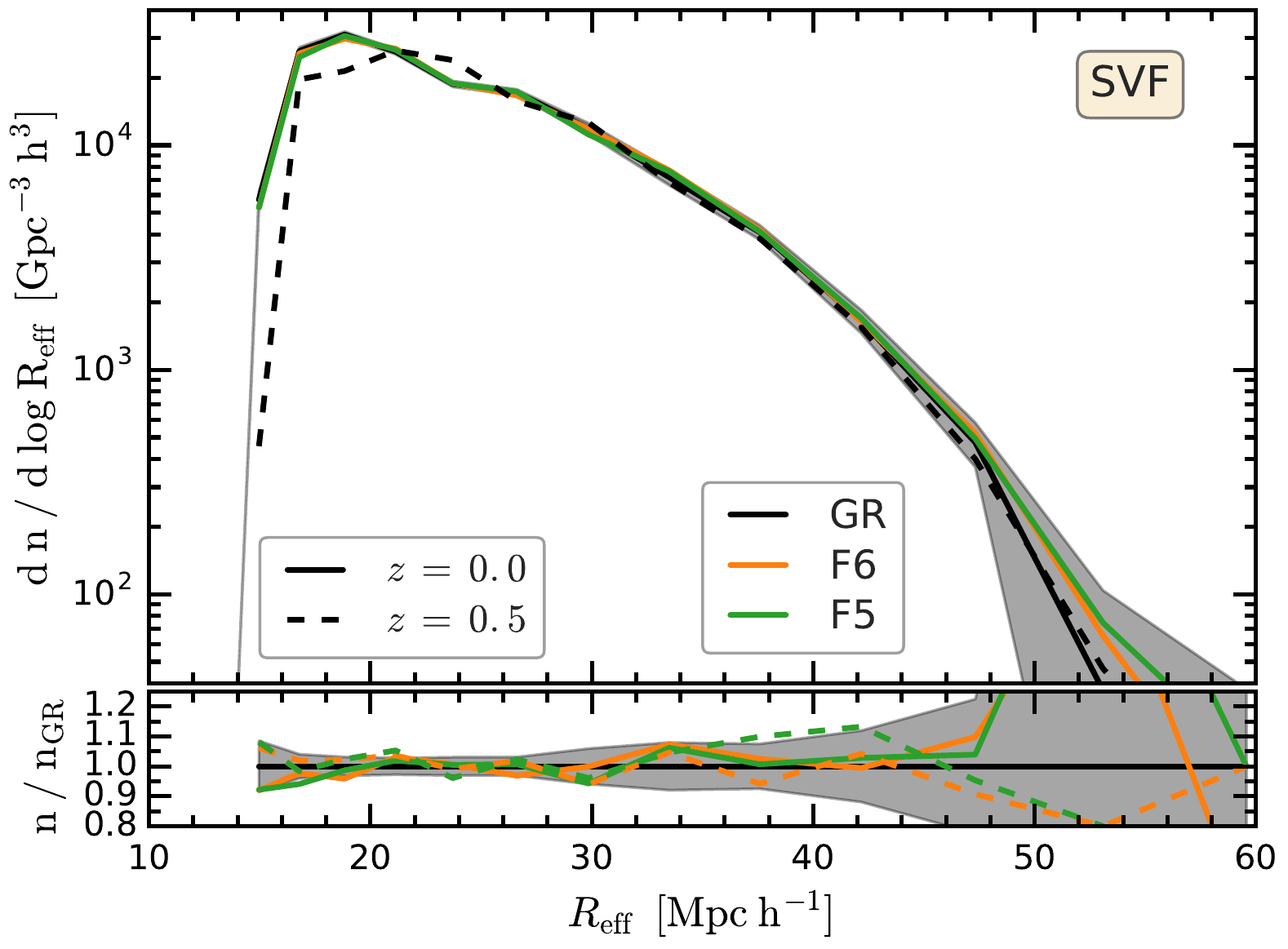}  & 
        \includegraphics[width=\figwidth,angle=0]{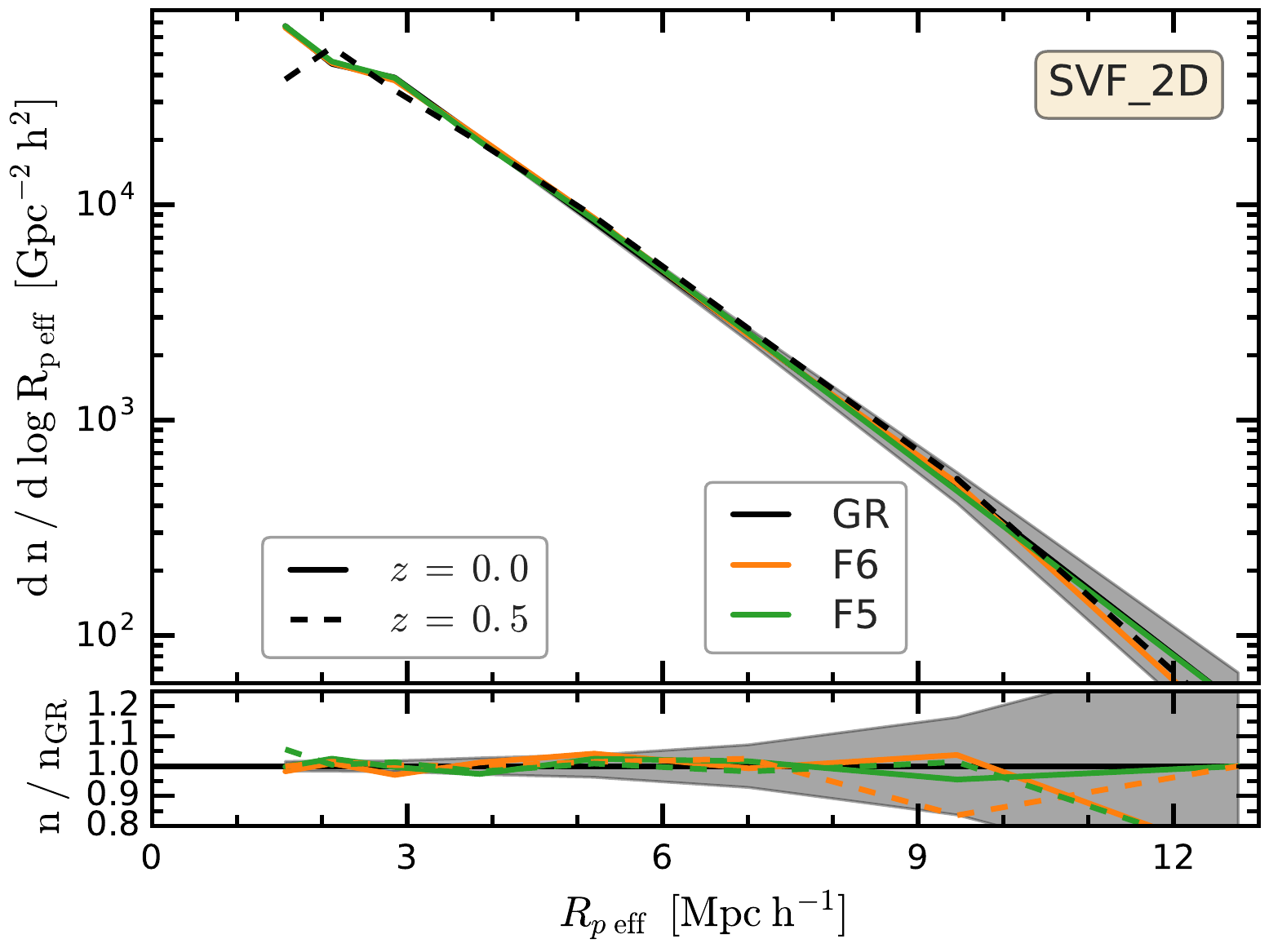}  \\[-.18cm]
        \includegraphics[width=\figwidth,angle=0]{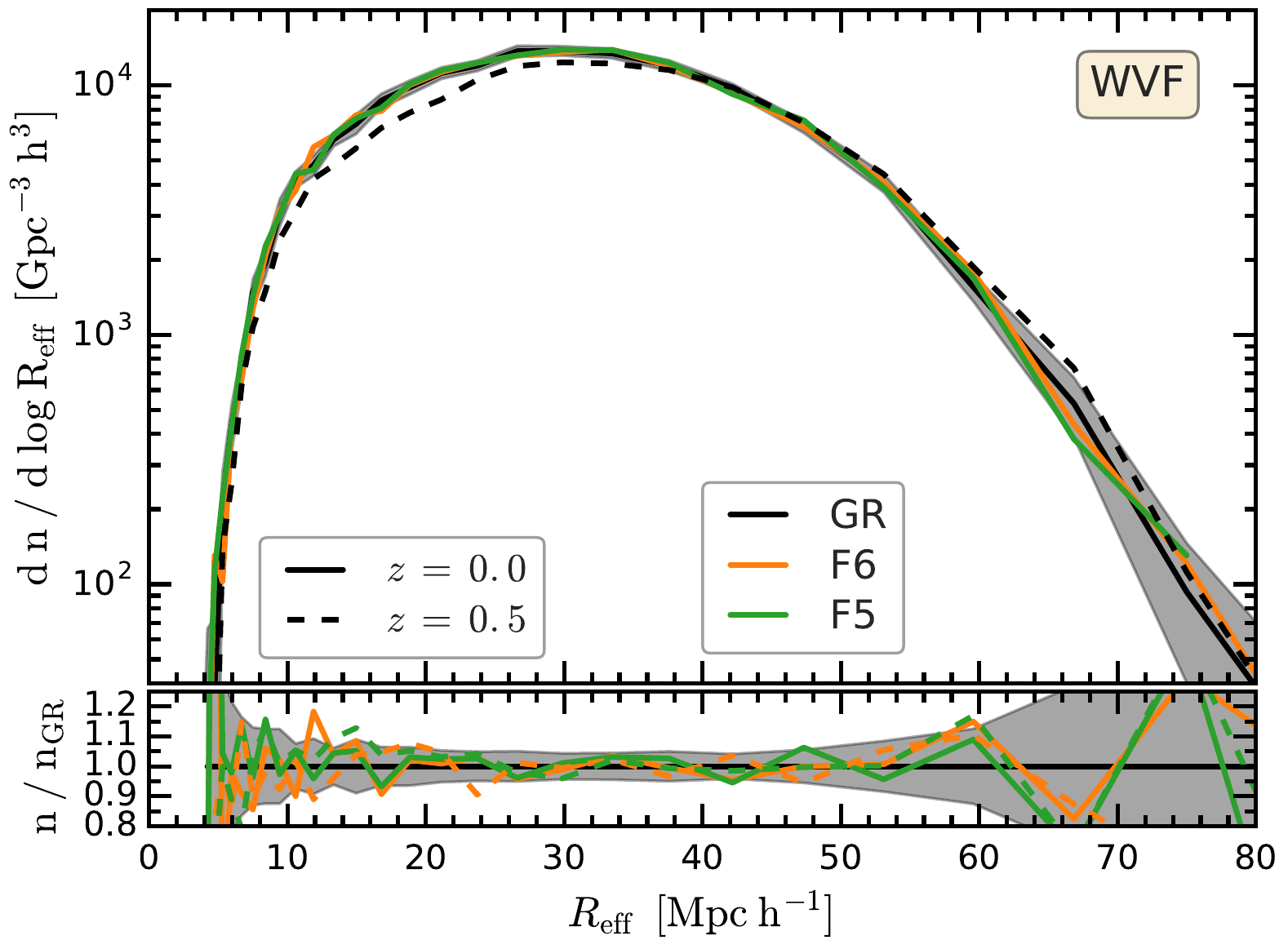}  & 
        \includegraphics[width=\figwidth,angle=0]{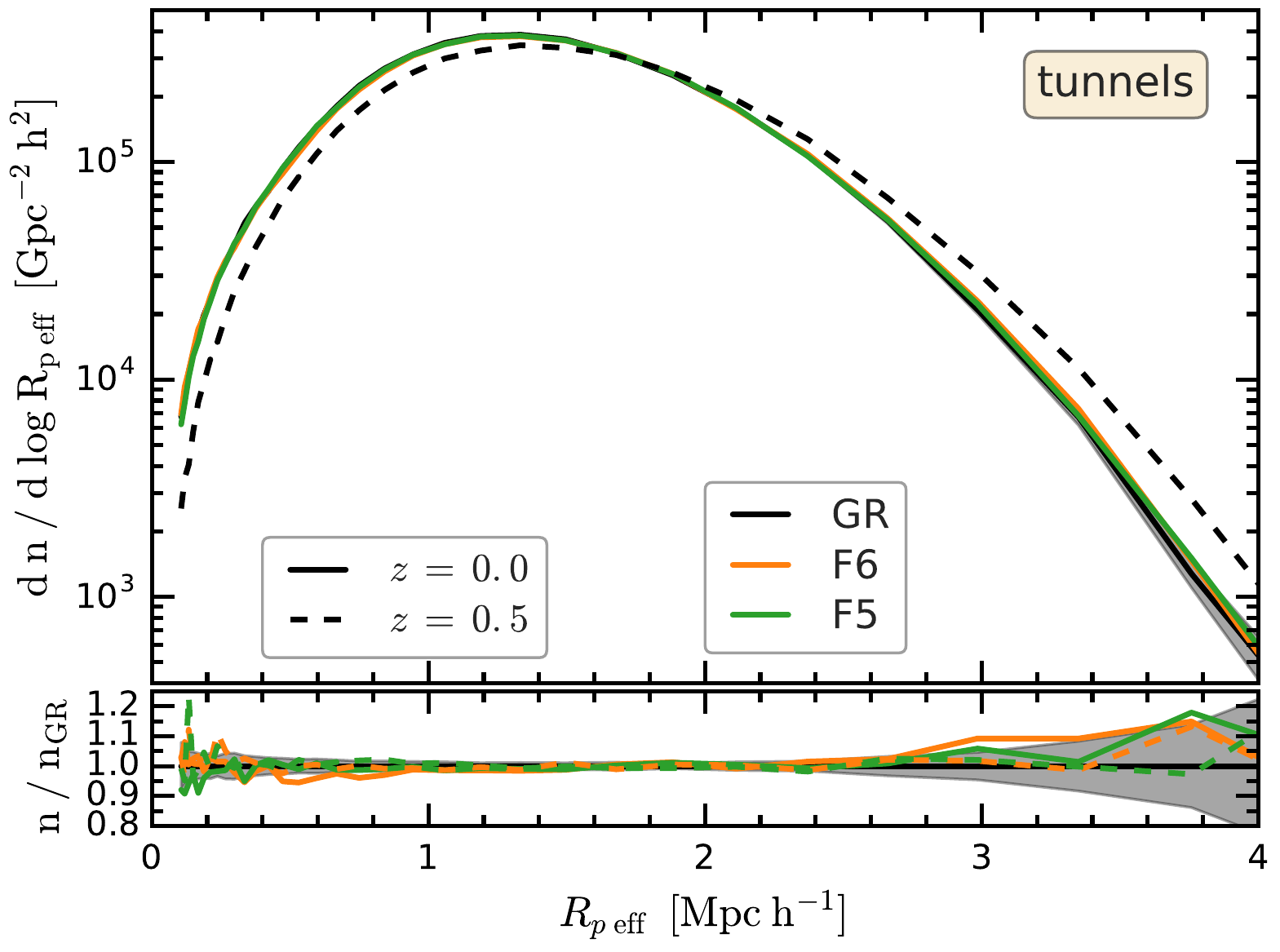} \\[-.18cm]
        \includegraphics[width=\figwidth,angle=0]{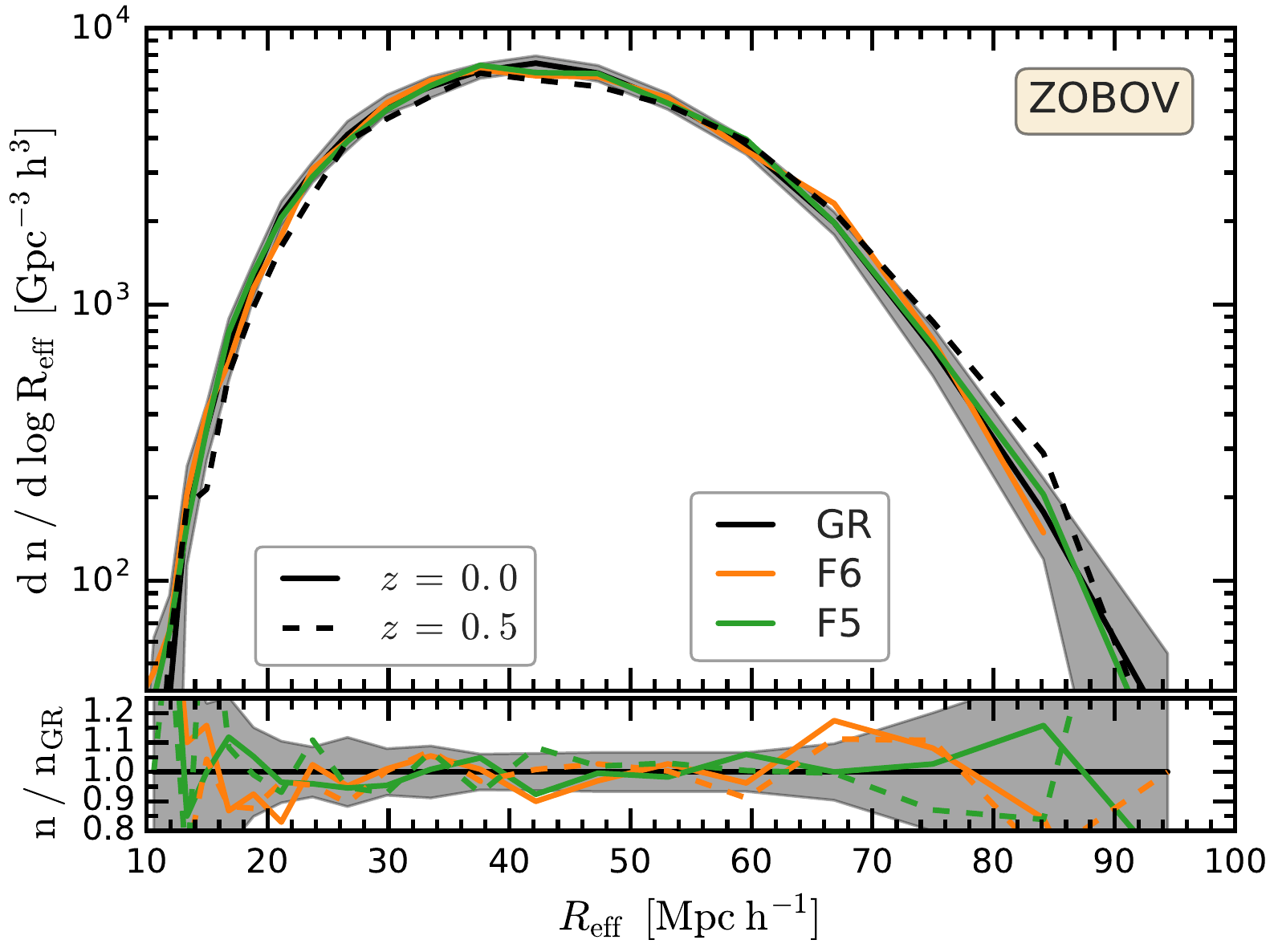} &
        
     \end{tabular}
     \reduceVerticalOffset{}
     \caption{ Comparison of the void abundance, i.e. number density of voids as a function of their radius, in \GR{} and \FR{} models \changed{using \HOD{} galaxy catalogues that were tuned to have the same number density and two-point correlation function across all models}. The figure shows the abundance of 3D voids identified using \SVF{} (top-left panel), \WVF{} (centre-left panel) and \zobov{} (bottom-left panel); and that of 2D \SVFtwoD{} voids (top-right panel) and tunnels (centre-right panel). All the 2D troughs have by definition the same $2\Mpch$ radius. \changed{The 2D voids were obtained by projecting the entire simulation box ($1024\Mpch$ side length) along one of its preferential axes in the distant observer approximation.} For each panel, the various curves show the results for \GR{} and for two \FR{} models at $z=0$ (solid curves) and $z=0.5$ (dashed curves). For clarity, only the \GR{} $z=0.5$ are shown in each of the main panels. The secondary panels shows the ratio of the \FR{} results to the \GR{} one for both $z=0$ and $z=0.5$. The shaded region shows the $1\sigma$ uncertainty interval computed using multiple \GR{} realisations. }
     \label{fig:abundance_fixed_n_xi}
     \reduceVerticalOffset{}
\end{figure*}

\begin{figure*}
     \centering
     \begin{tabular}{cc}
        \includegraphics[width=\figwidth,angle=0]{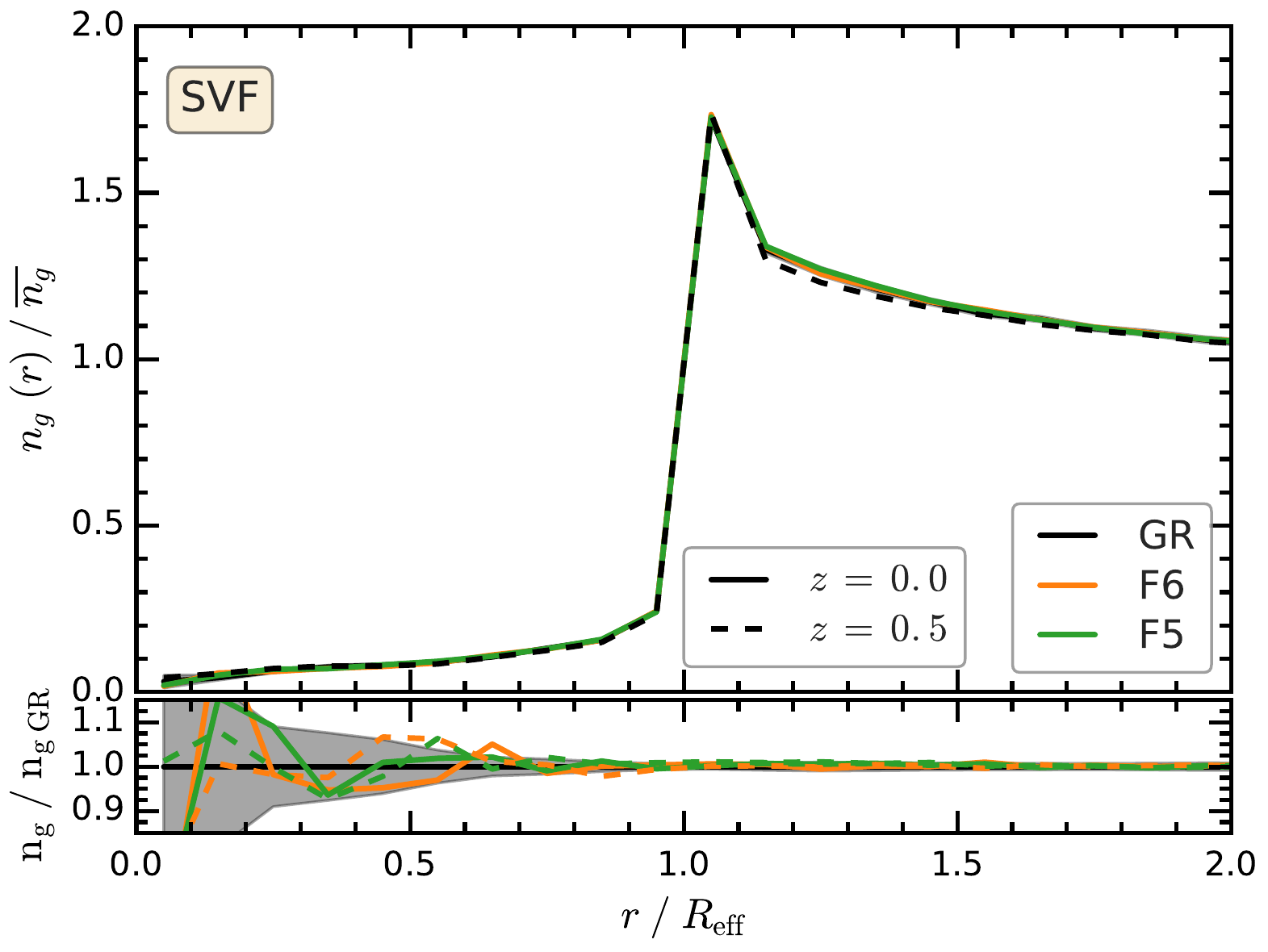} &
        \includegraphics[width=\figwidth,angle=0]{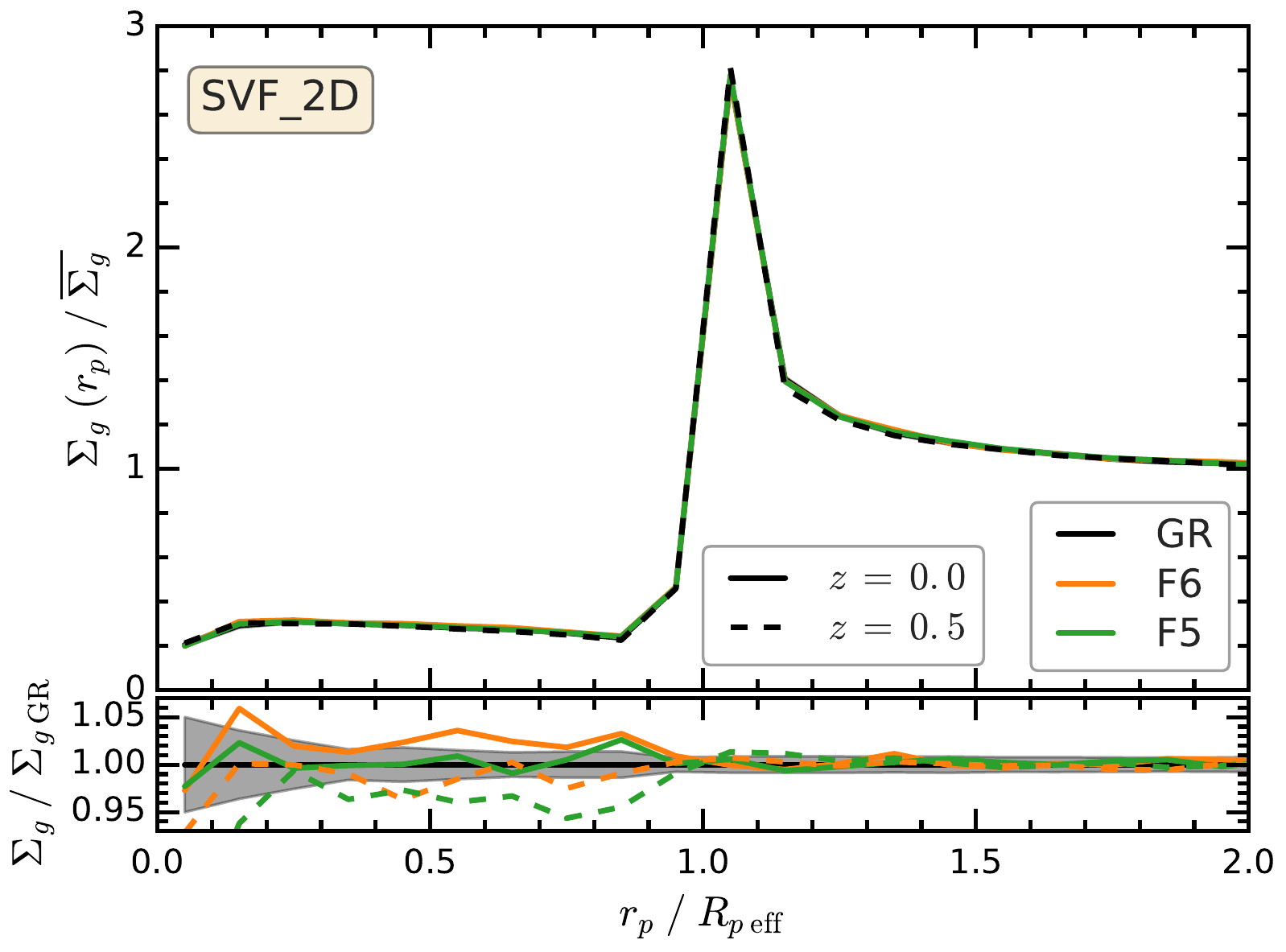} \verticaloffset
        \whitespace & \whitespace \verticaloffset
        \includegraphics[width=\figwidth,angle=0]{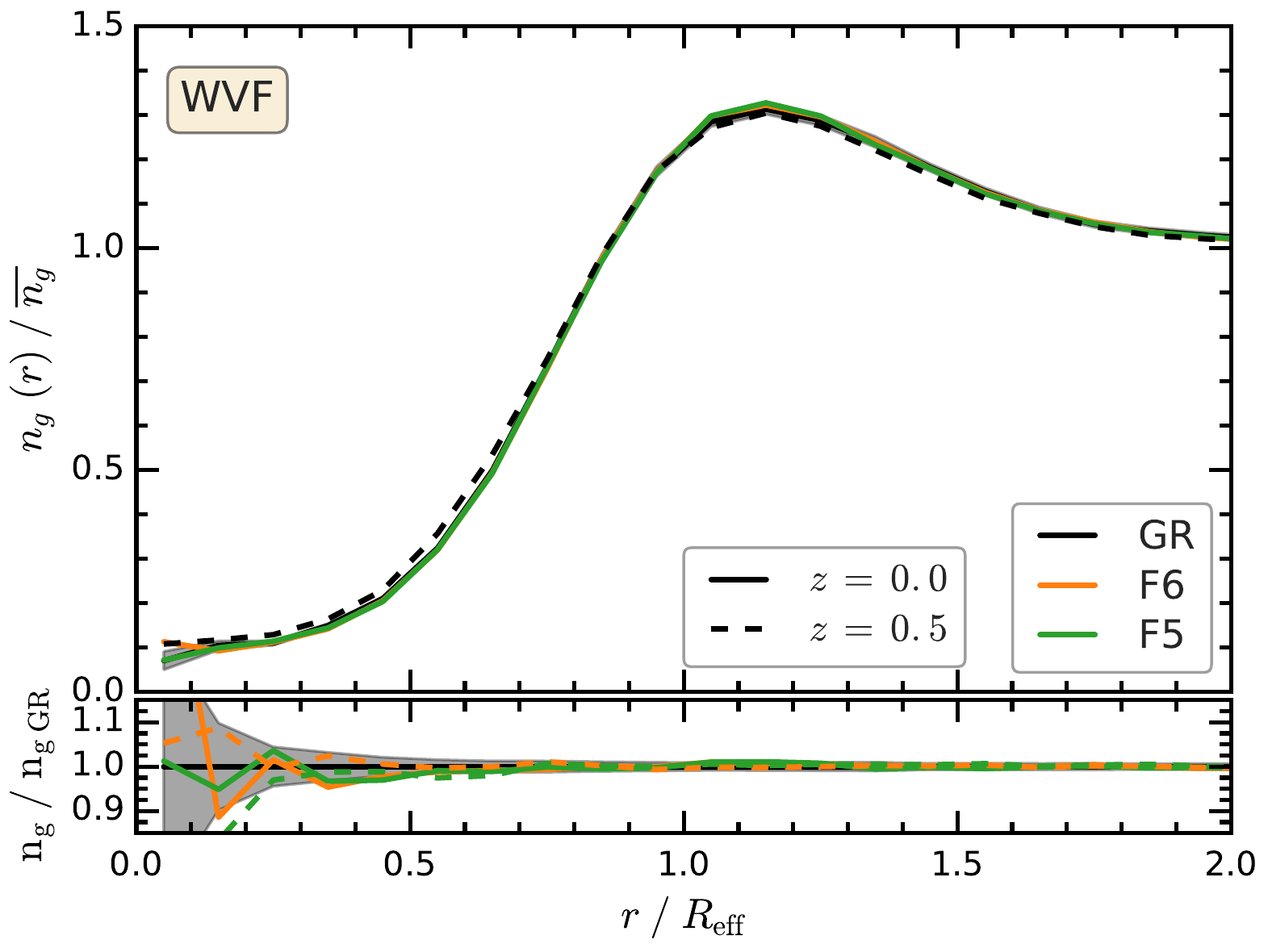} &
        \includegraphics[width=\figwidth,angle=0]{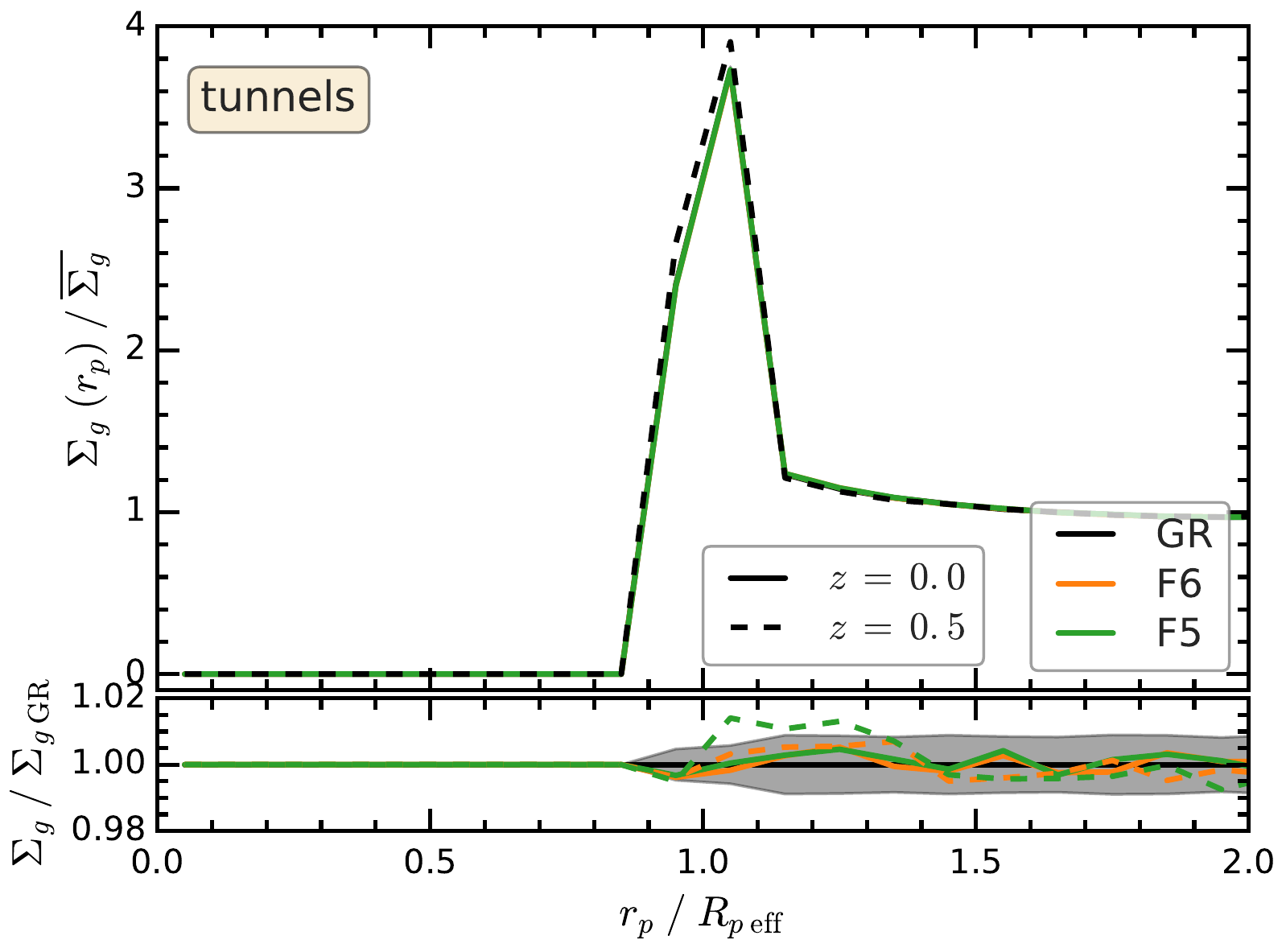} \verticaloffset
        \whitespace & \whitespace \verticaloffset
        \includegraphics[width=\figwidth,angle=0]{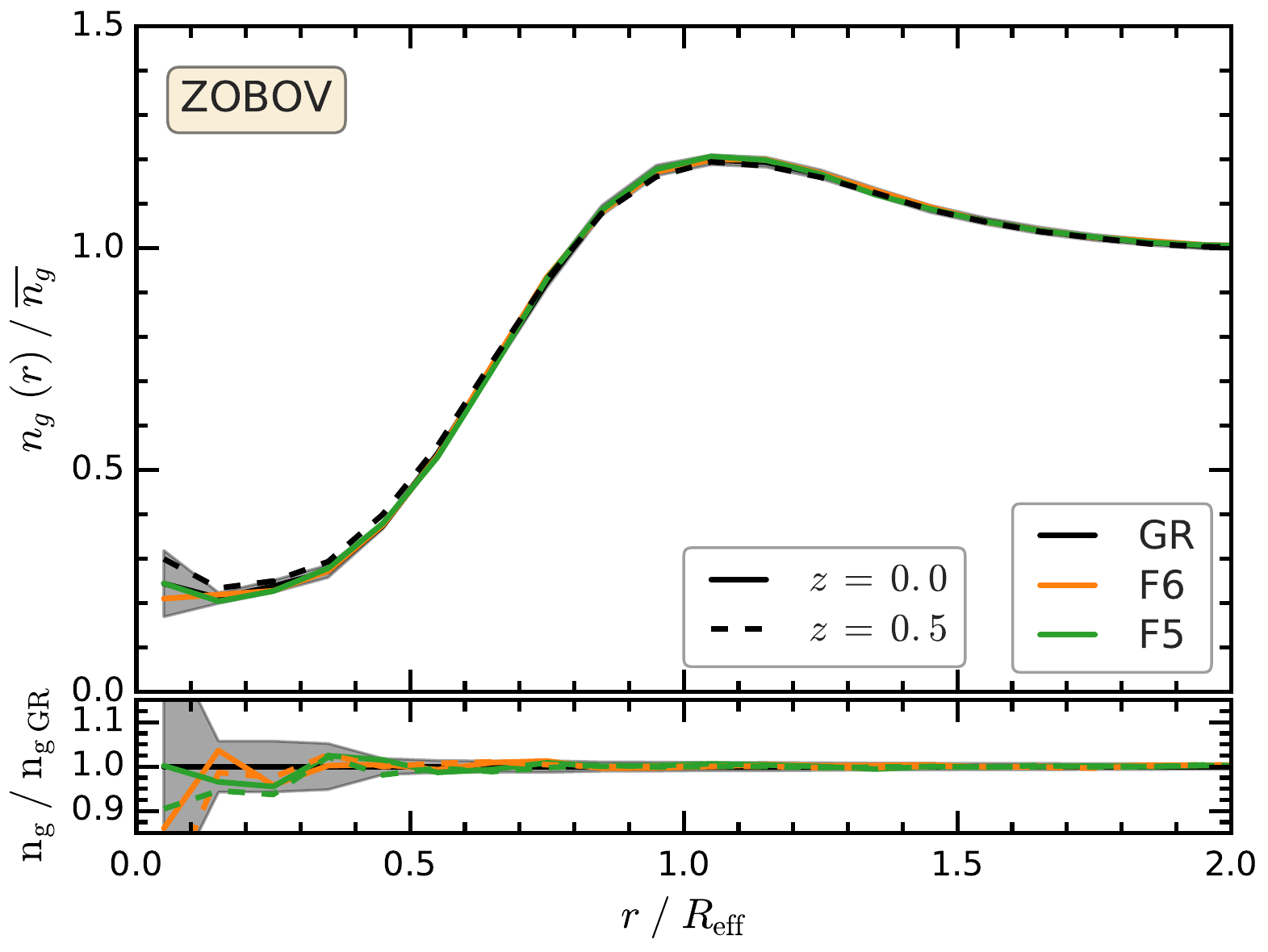} &
        \includegraphics[width=\figwidth,angle=0]{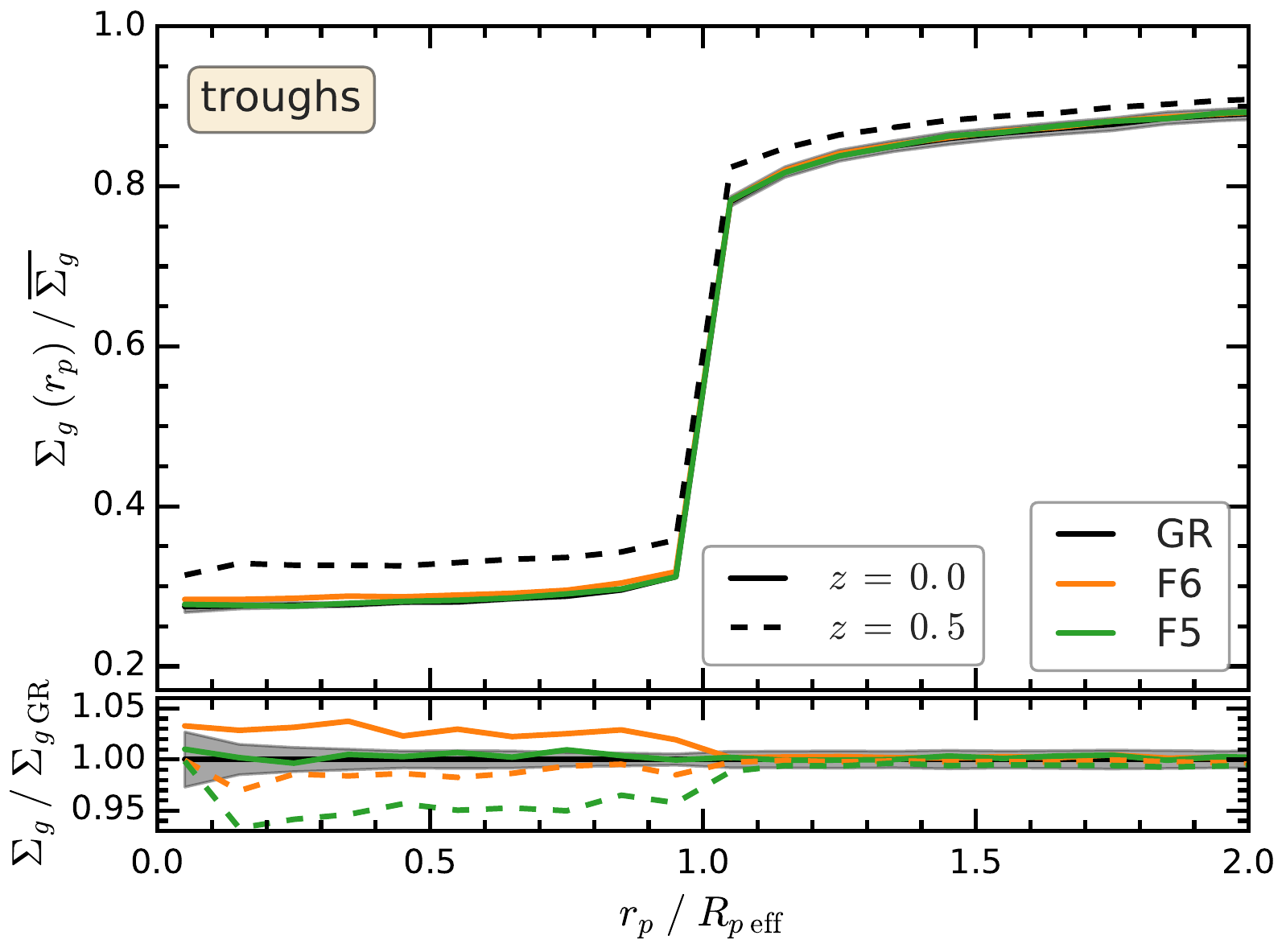}
     \end{tabular}
     \reduceVerticalOffset{}
     \caption{ Comparison of the radial galaxy density profile of voids between \GR{} and \FR{} models. The left column shows the 3D galaxy number density profile, $n_{\rm g}/\overline{n_{\rm g}}$, of 3D voids. The right column shows the projected surface density of galaxies, $\Sigma_{\rm g}/\overline{\Sigma_{\rm g}}$, of 2D line-of-sight underdensities. The symbols and curves are the same as in \reffig{fig:abundance_fixed_n_xi}. }
     \label{fig:gal_den_fR_models}
     \reduceVerticalOffset{}
\end{figure*}

We present stacked void profiles that are averaged over voids of all sizes, unless we specify otherwise. We have checked that similar differences between \GR{} and \FR{} are present when stacking voids in a narrower range of sizes, albeit with a lower signal-to-noise due to reduced sample size and thus poorer statistics. When averaging, each void is given the weight, $w=\Reff^2$, such that larger voids are weighed more compared to the more numerous small voids. This weight is motivated by measurements of the tangential shear profile of voids in observations. The number of source galaxies inside a void scales approximately with $\Reff^2$ and thus large voids have a larger contribution to the stacked tangential shear measurements. We note that when simulating other observables, different weights may be appropriate; for example in the case of the 3D void galaxy density profile, the contribution of each void to the stacked signal scales with $\Reff^3$. Nonetheless, for simplicity, we keep the same $w=\Reff^2$ weight for all our stacked profiles.

We present mean profiles by averaging over two realizations of a cosmological volume of $1024\Mpch$ in length, Box 1 and 2, which were simulated using both \GR{} and \FR{} gravity models. We estimate uncertainties using five realisations of the \GR{} box. For each realisation, we split the volume into $4^3$ non-overlapping regions and perform $100$ bootstrap samples over these regions. The procedure leads to $5\times100$ samples which we use to compute the correlation matrix and estimate errors. For tunnels and troughs we split the mock plane-of-the-sky into \changed{$8^2$} non-overlapping regions, after which we perform exactly the same analysis. Such a procedure accounts for correlations between neighbouring voids that are especially important in the case of troughs since many troughs overlap. In practice, the procedure is implemented by first computing the stacked profile and the total weight of the voids in each of the regions defined above. The total weight of all voids inside a region is assigned as the weight of that region. Then, all the regions selected for a bootstrap sample are combined according to their weights.

\vspace{-.15cm}
\subsection{Void abundances}
\label{sect:abundance}
We start by comparing the distribution of void sizes between \GR{} and \FR{} models. \reffig{fig:abundance_fixed_n_xi} shows these distributions for all the void finders employed here with the exception of troughs which are defined to have a constant $2\Mpch$ radius. The distribution of void sizes varies between finders (note that the horizontal axes have different ranges in the different panels of Fig.~\ref{fig:abundance_fixed_n_xi}), but, when considering the same identification method, there is  no statistically relevant difference between \GR{} and \FR{} gravity. This result agrees with that in  \citet{Cai2014}, where it was found that while voids identified in the matter density field are larger in \FR{} models compared to the \GR{} case, these differences largely disappear when identifying voids using the halo distribution. \citet{Falck2017} obtained a similar result for the DGP modified gravity model: when identifying voids using haloes, there is no difference in void abundance between $\Lambda$CDM and DGP models. In contrast, \citet{Zivick2015} found that \FR{} models boost the number of large voids; the discrepancy could be due to their neglect of halo and galaxy bias since they identified voids using a subsampled distribution of dark matter particles \changed{(e.g. see \citealt{Pollina2016})}.

By analysing our various \HOD{} catalogues (see \refsec{subsec:HOD}), we find that the void abundance is most sensitive to the number density of tracer galaxies, $n_{\rm g}$. Once the \FR{} galaxy catalogues have the same $n_{\rm g}$ as the \GR{} ones, there is little difference in void abundance. Further matching also the two point correlation function of the \FR{} and \GR{} catalogues does not lead to a significant change. \textcolor{black}{We have checked explicitly that (not shown here) the same is true if dark matter haloes with fixed number density are used as tracers to find voids}. The differences in void abundance between $z=0$ and $z=0.5$ (the solid versus dashed curves from \reffig{fig:abundance_fixed_n_xi}) could also be due to differences in galaxy number density, which are $n_g = 3.8 \times 10^{-4}$ and $3.2 \times 10^{-4} h^{3} {\rm Mpc}^{-3}$ at $z = 0$ and $0.5$, respectively. A lower tracer number density results in systematically larger voids and fewer small voids due to them merging to form larger ones. 

\begin{figure*}
     \centering
     \begin{tabular}{cc}
        \includegraphics[width=\figwidth,angle=0]{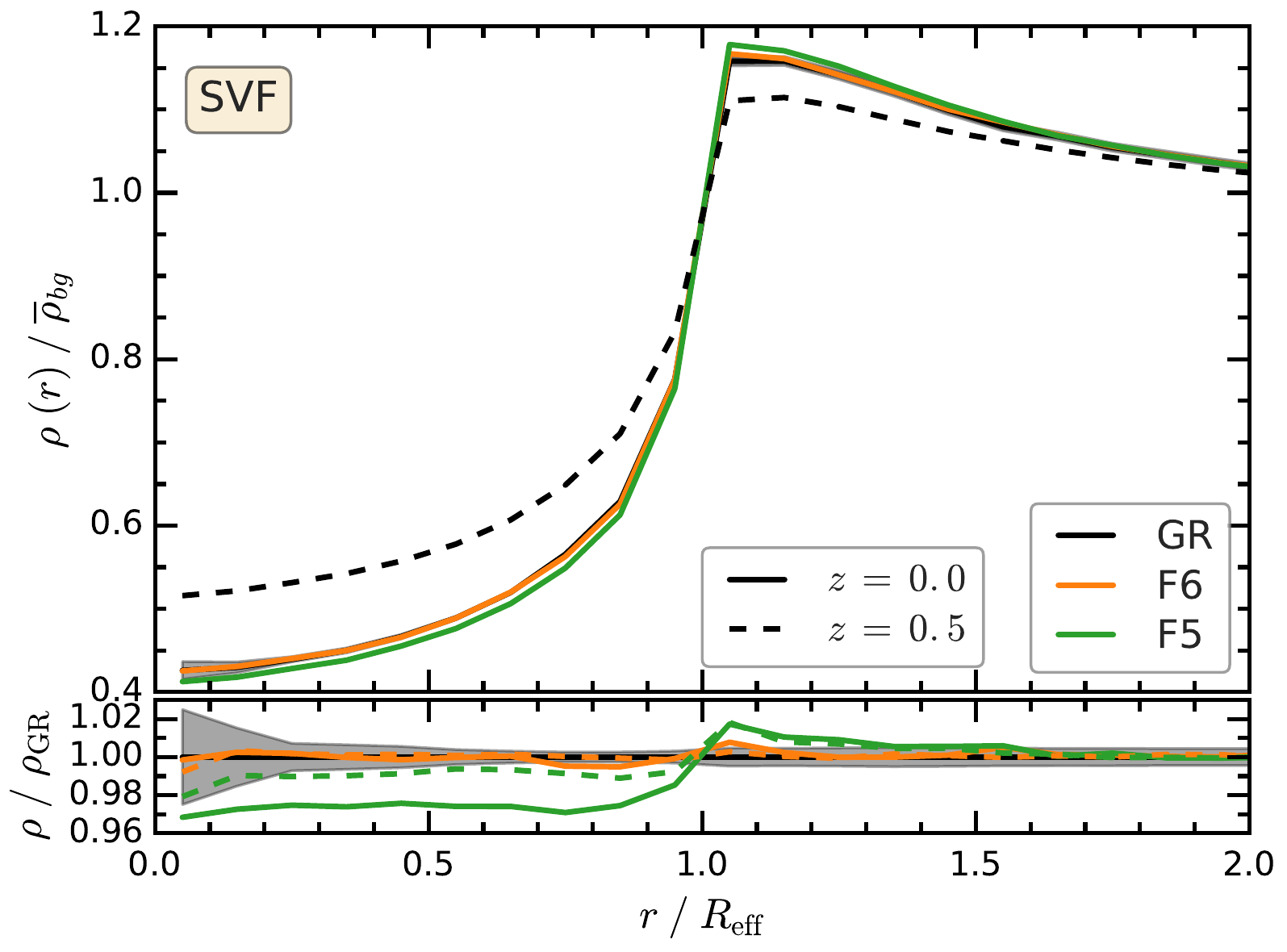} &
        \includegraphics[width=\figwidth,angle=0]{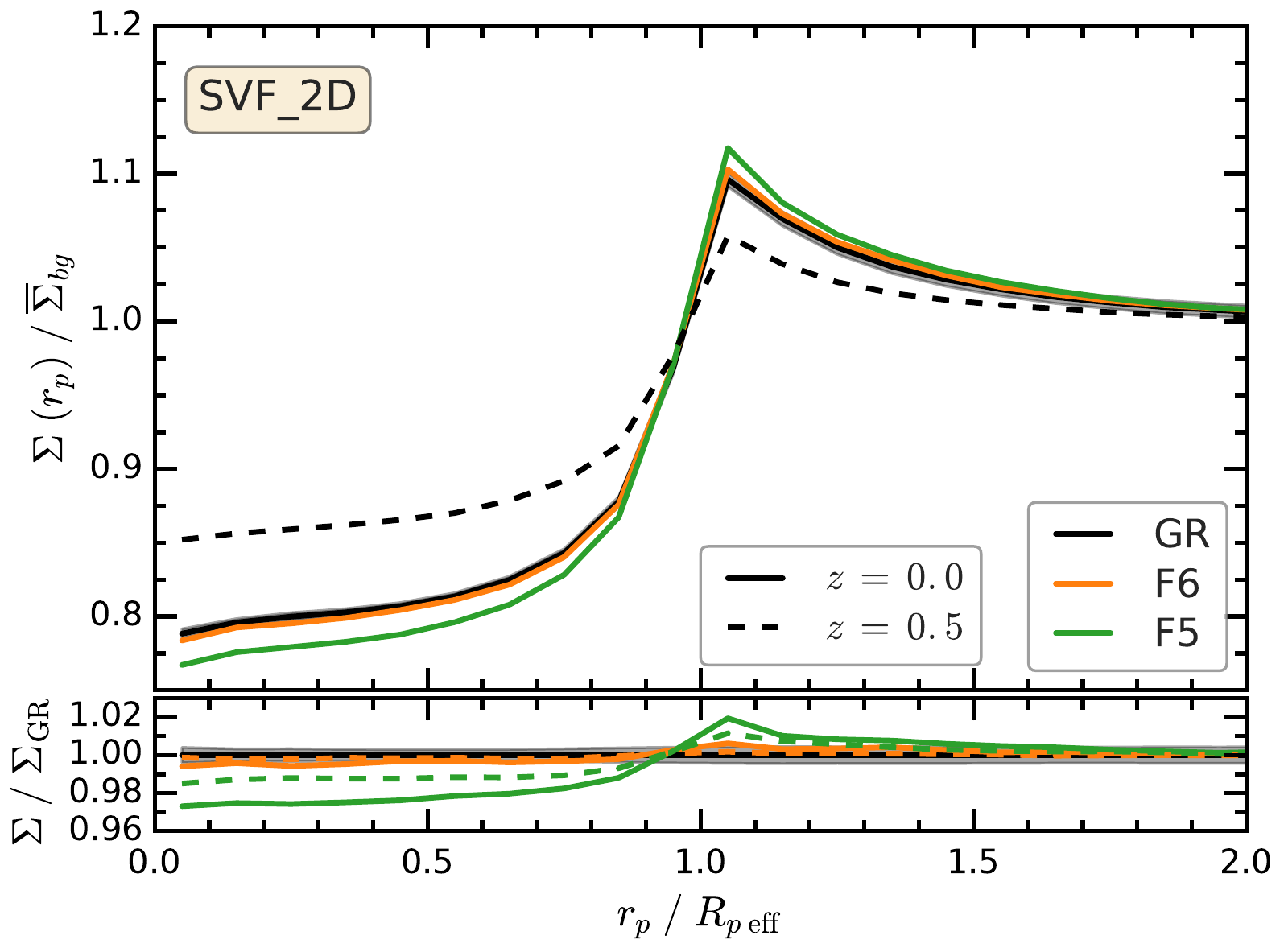} \verticaloffset
        \whitespace & \whitespace \verticaloffset
        \includegraphics[width=\figwidth,angle=0]{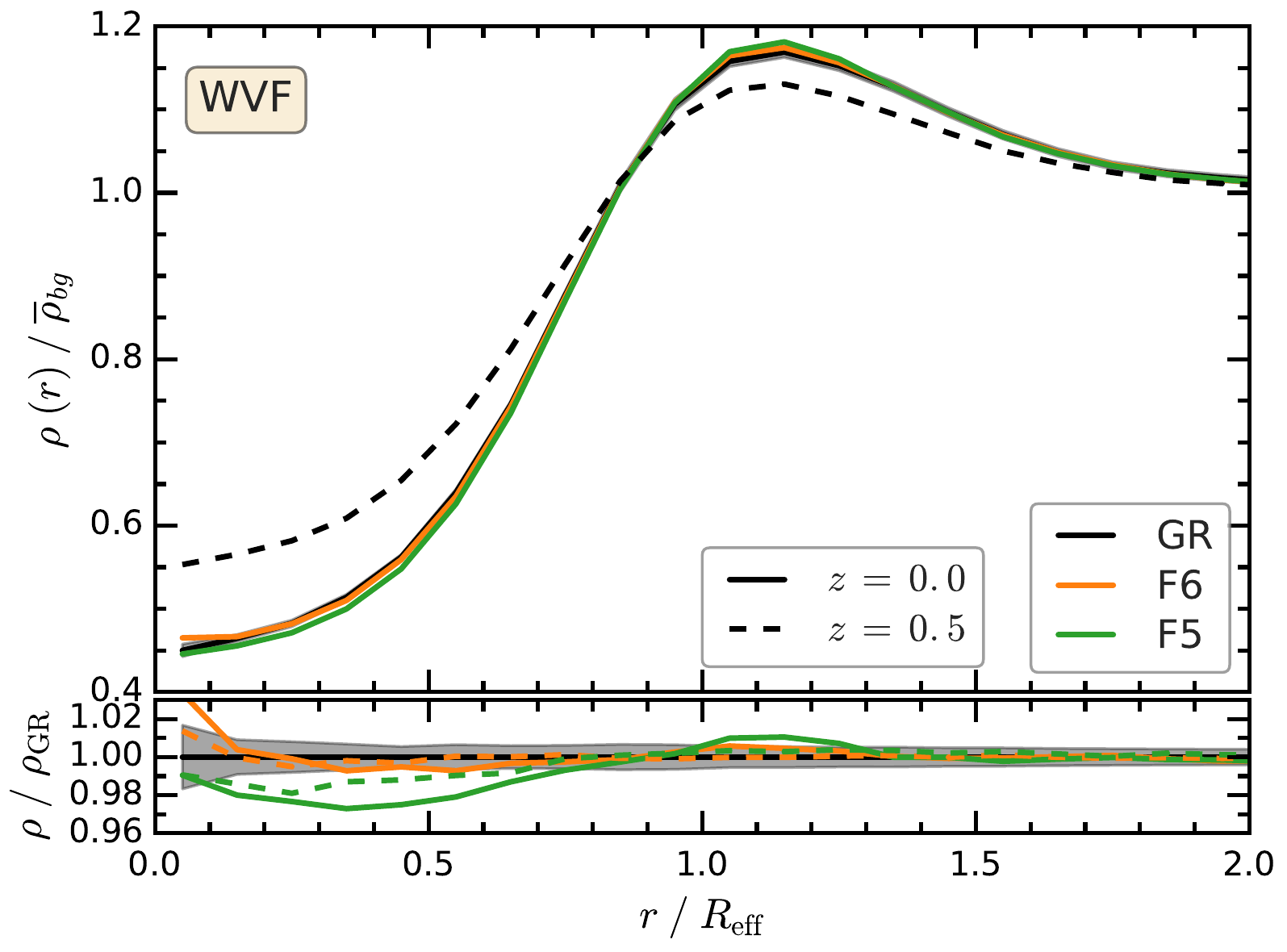} &
        \includegraphics[width=\figwidth,angle=0]{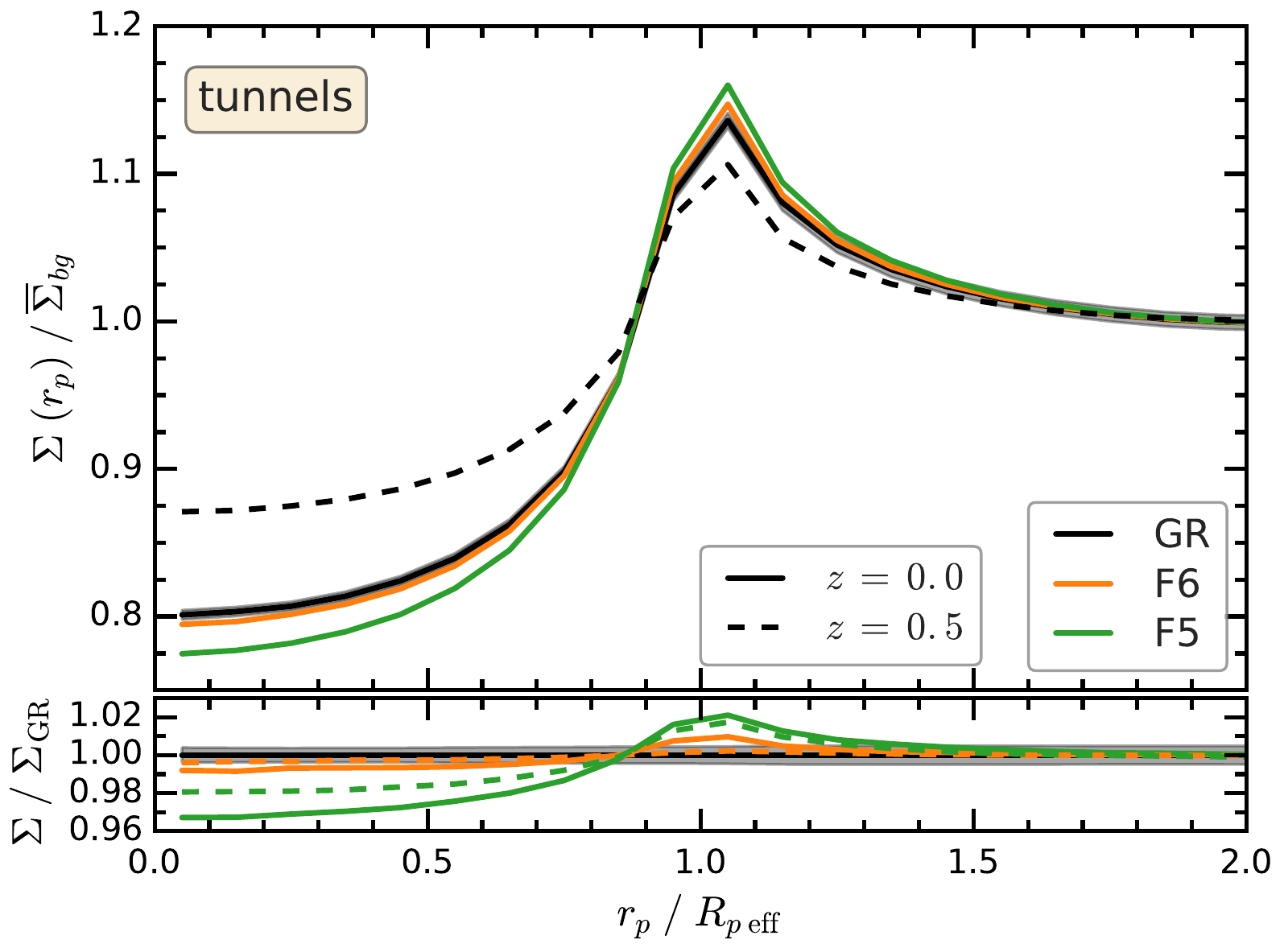} \verticaloffset
        \whitespace & \whitespace \verticaloffset
        \includegraphics[width=\figwidth,angle=0]{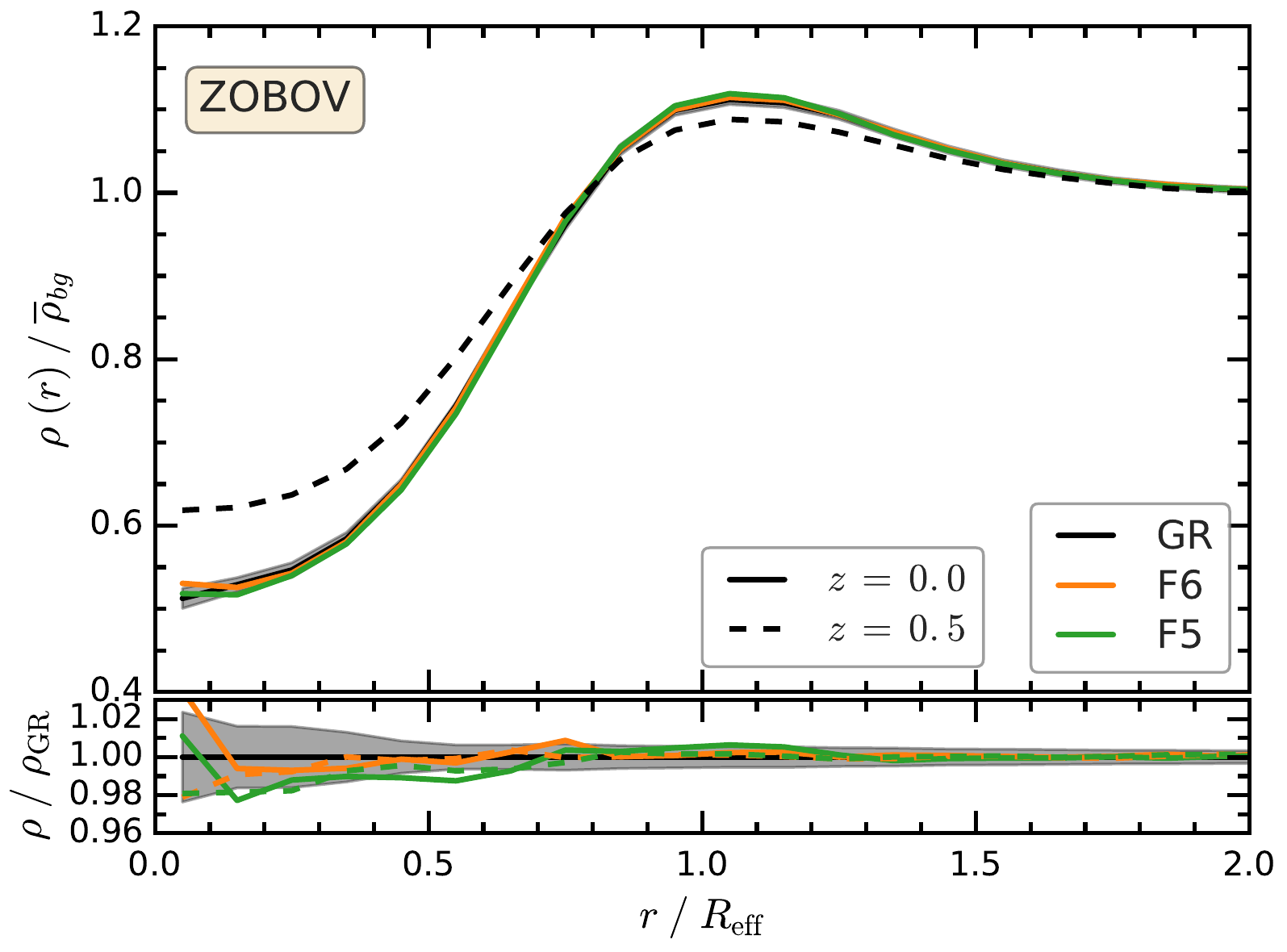} &
        \includegraphics[width=\figwidth,angle=0]{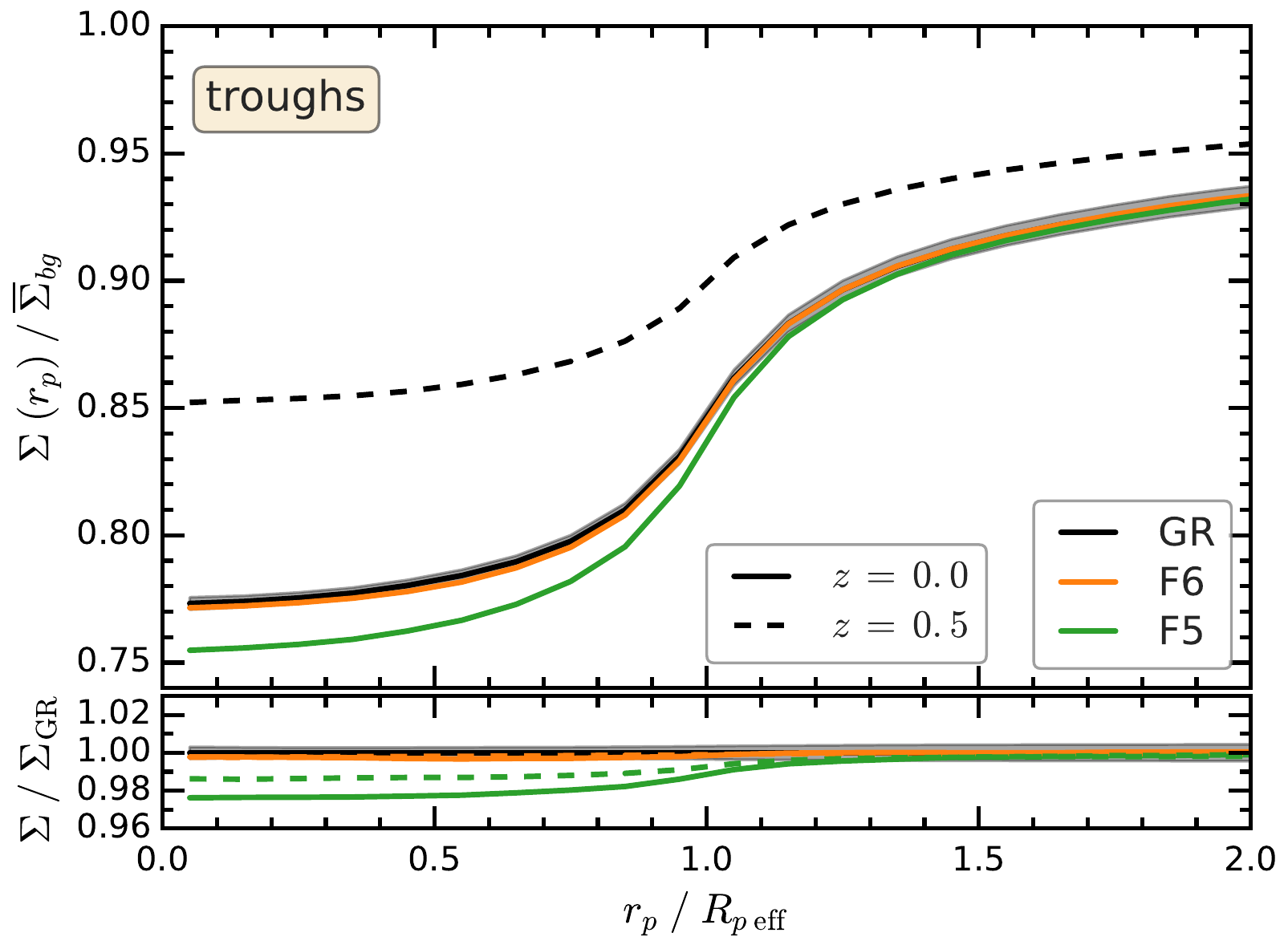}
     \end{tabular}
     \reduceVerticalOffset{}
     \caption{ Comparison of the void matter density profile between \GR{} and \FR{}. The left column shows the 3D matter density profile, $\rho/\overline{\rho}_{bg}$, of 3D voids. The right column shows the projected matter surface density, $\Sigma/\overline{\Sigma}_{bg}$, of 2D line-of-sight underdensities. The symbols and curves are the same as in \reffig{fig:abundance_fixed_n_xi}. }
     \label{fig:DM_den_fR_models}
     \vspace{-.4cm}
\end{figure*}

\subsection{Void galaxy number density profiles}
We calculate the spherically averaged number density of galaxies for each void as a function of the distance from the void centre. \changed{The galaxy number density profile, $n_g(r)$, contains the same information as the 3D void-galaxy correlation function, $\xi_{vg;~\rm{3D}}(r)$, and the two are related via
\begin{equation}
	n_g(r) = \overline{n_g}(1+\xi_{vg;~\rm{3D}}(r) )
	\label{eq:galaxy_density_profile} \;,
\end{equation}
where $\overline{n_g}$ denotes the mean number density of galaxies.}
Since the 2D voids have centres defined only in two dimensions (i.e. in the plane-of-the-sky), we compute the galaxy number density projected on the simulated plane-of-the-sky as a function of projected distance from the centre of the 2D void. \changed{\eq{eq:galaxy_density_profile} holds true for the projected number density of galaxies, $\Sigma_g(r)$, but with the 3D void-galaxy correlation function replaced by the 2D projected one, which we compute by projecting the entire simulation box in the distant observer approximation.}
After expressing the profiles in terms of the scaled radial distance, $r/\Reff{}$, we stack all the voids to obtain average profiles.

\reffig{fig:gal_den_fR_models} presents the radial galaxy number density profiles of voids. All the methods find that, on average, the void interiors are devoid of galaxies, but the average galaxy number varies between methods. For \SVF{}, $n_{g}$ shows a large increase at the void radius since \SVF{} voids are identified as the largest sphere that has mean enclosed density $\leq 0.2\overline{n_{g}}$, with $\overline{n_{g}}$ denoting the mean number density of galaxies. The watershed void finders, \WVF{} and \zobov{} both have smoothly increasing $n_{g}(r)$ profiles due to their non-spherical shape, with their overdense ridges smeared by the spherical stacking procedure \citep{Cautun2016a}. The profiles of 2D voids are expressed in terms of the projected mean galaxy number, $\Sigma_{\rm g}$, and its mean background value, $\overline{\Sigma_{\rm g}}$. Similar to the 3D \SVF{} voids, the \SVFtwoD{} voids are very underdense inside the void and show a prominent galaxy density enhancement at their edge. By definition, the tunnels are the largest circles devoid of galaxies; this explains why $\Sigma_{g}=0$ for $r<\Rpeff$ is followed by a sharp peak at $r=\Rpeff$. The troughs have $\Sigma_{g}<\overline{\Sigma_{g}}$ not only inside $\Rpeff$, but also outside their radius. This is because troughs with our selection criteria are likely to reside in large underdense regions\footnote{{Troughs are selected to correspond to the ${\sim}20\%$ most underdense regions. Moreover, unlike other voids studied in this paper, whose edge is defined by an increase in galaxy number density, the sizes of troughs are a user defined parameter, which is $2h^{-1}$Mpc in our case. These differences can explain why there is no overdense ridge in our troughs.}} (see Fig.~\ref{fig:images_3D_voids}).
With the exception of troughs, the $n_{g}(r)$ and $\Sigma_{g}$ profiles show very little dependence on redshift.

\reffig{fig:gal_den_fR_models} shows that there is no statistical difference in the galaxy number density profiles between \GR{} and \FR{} models when the latter catalogues are matched to have the same number density of tracers and the same galaxy correlation function. The only exception is inside \SVFtwoD{} voids and troughs, where the differences are likely a limitation of our mocks rather than a tell-tale signature of modified gravity. This is supported by the non-monotonic behaviour with redshift and with the parameter that determines the strength of the modifications to gravity: for example, the \Ffive{} model, which has stronger deviations from \GR{} than \Fsix{}, shows a large systematic difference at $z=0.5$ but no difference at $z=0$, while in theory we would expect bigger differences at later times. For \Fsix{}, it is especially suspicious that the $z=0$ difference in {the profile} with respect to \GR{} is bigger than the difference of the \Ffive{} model. The observed differences in the galaxy density profiles within the trough radius, which is $2\Mpch$, could be due to the fact that our \HOD{} catalogues were tuned to match the galaxy correlation functions among the different models {\it only above} separations of $2\Mpch$. In particular, we note the same qualitative (and even quantitative) behaviour of the model differences in the case of 2D spherical voids, suggesting that this is not specific to troughs. The error bars in the plot appear smaller in the case of troughs, but these are simply taken as the square root of the diagonal elements of the covariance matrix for indication, and the actual errors in different bins are indeed very strongly correlated. Therefore, we refrain from interpreting the differences shown in the galaxy density profiles of troughs and \SVFtwoD{} voids as a signature of deviation from GR, and will leave a more detailed investigation to a future work, hopefully using higher resolution simulations and mock catalogues.

\subsection{Void matter density profiles}
\label{subsec:matter_profile}

We compute the void matter profiles similarly to the galaxy number density profiles, except that now we use the full distribution of DM particles in the simulation volume. In the case of 2D underdensities, we project the particles \changed{in the full simulation box} on the simulated plane-of-the-sky as we did in the case of the \HOD{} galaxies. \changed{The 3D matter density profile, $\rho(r)$, and the projected 2D one, $\Sigma(r)$ are given by:
\begin{eqnarray}
	\rho(r) & = & \overline{\rho}_{bg} ( 1 + \xi_{vm;~\rm{3D}} ) \\
    \Sigma(r) & = & \overline{\Sigma}_{bg} ( 1 + \xi_{vm;~\rm{2D}} )
    \label{eq:matter_density_profile} \;,
\end{eqnarray}
where $\overline{\rho}_{bg}$ and $\overline{\Sigma}_{bg}$ denote the mean background 3D and the 2D projected density, respectively. The quantities $\xi_{vm;~\rm{3D}}$ and $\xi_{vm;~\rm{2D}}$ denote the void-mass cross-correlation function in 3D and in 2D projection, respectively.}

The resulting matter density profiles of voids are shown in \reffig{fig:DM_den_fR_models}. These are obtained by stacking all the voids and thus are average profiles. All methods identify regions that are underdense in the inner parts, i.e. $r\lesssim \Reff$; and that show a modest overdensity close to their edge, i.e. $r\simeq \Reff$. The only exception are troughs which are underdense even beyond $r> \Reff$ because troughs represent the most underdense regions and moreover, because their boundaries are not defined as a galaxy number overdensity. Compared to the galaxy number density profiles, the void matter profiles are both less underdense in their interiors and less overdense at their boundaries. Moreover, the matter profiles show a considerable dependence with redshift, with voids identified at higher redshift being less overdense.

We find that the interiors of both 2D and 3D voids are more underdense in \FR{} gravity than in \GR{}. The effect is the strongest for \Ffive{} and also decreases at higher redshift. This is in line with theoretical expectations, since the modifications to gravity become stronger at later times and are larger in \Ffive{} than in \Fsix{}. Due to mass conservation, since \FR{} voids have emptier interiors they should also be more overdense around the void edge. This effect is present for all void types, except \troughs{}, and it is especially prominent for \SVF{} voids, both 2D and 3D ones, and \tunnels{}. The troughs are different since they are likely embedded in underdense regions much larger than their size, and, even by going to $2\Reff$, we still have not reached the overdense ridge surrounding these underdense regions.

\changed{We studied the void density profiles of several modified gravity \HOD{} catalogues (see \refsec{subsec:HOD}): \CMASSdef{}, which uses the same \HOD{} parameter values as \GR{}, \CMASSn{}, which changes the three \HOD{} mass parameters proportionally to match the galaxy number density of \GR{}, and \CMASSnxi{} (the one shown in \reffig{fig:DM_den_fR_models}), which matches both the number density and the two-point clustering of \GR{}. For all these catalogues, the void density profiles show the same difference between \FR{} and \GR{}, suggesting that the difference between the models is robust to small changes in how galaxies populate dark matter haloes.}

\subsection{Void tangential shear profiles}
\label{subsec:void_lensing}
Weak gravitational lensing represents the most promising way of measuring the matter distribution in and around voids, and could potentially be used as a probe of modifications to gravity \citep{Cai2014,Barreira2015}. Current surveys have already measured the weak lensing imprint of voids \citep[e.g.][]{Gruen2015,Sanchez2016} and future surveys, thanks to increases in both sky coverage and image quality, would improve greatly the precision of such measurements. This motivates us to compare the weak lensing signal of the various void finders and predict which method shows the largest potential for discriminating between \GR{} and \FR{} models.

Voids have a weak, yet measurable, gravitational lensing imprint. This is most conspicuous as distortions in the shapes of background galaxies, whose image is distorted by the intervening mass distribution between source and observer. Such distortions are encapsulated in the tangential shear profile of voids, which is given by:
\begin{equation}
	\gamma_t(r) \; = \; \frac{\Delta\Sigma(r)}{\Sigma_c} \; = \; \frac{\overline{\Sigma}(<r)-\Sigma(r)}{\Sigma_c}
    \label{eq:tangential_shear} \;.
\end{equation}
The numerator is the \textit{differential surface mass density}, $\Delta\Sigma(r)$, which is the difference between the mean enclosed surface density within $r$, $\overline{\Sigma}(<r)$, and the surface density at $r$, $\Sigma(r)$, with: 
\begin{equation}
	\overline{\Sigma}(<r) = \frac{1}{\pi r^2} \int_0^r \Sigma(r') \; 2\pi r' \diff r'
    \label{eq:mean_enclosed_surface_density} \;.
\end{equation}
The denominator in \eq{eq:tangential_shear}, $\Sigma_c$, is the critical projected mass density and depends on the geometry of the lensing event, with:
\begin{equation}
	\Sigma_c \; = \; \frac{c^2}{4\pi G}\frac{D_s}{D_l D_{ls}} \; = \; \frac{c^2}{4\pi G}\frac{\chi_s}{a_l \; \chi_l \; \chi_{ls}}
     \label{eq:sigma_critical} \;,
\end{equation}
where $c$ and $G$ are the speed of light and the Newton gravitational constant, respectively. The symbols $D_l$, $D_s$ and $D_{ls}$ are angular diameter distances between the observer and the lens (the void in our case), the observer and the source galaxies, and the lens and the source galaxies. The angular diameter distances can be expressed in terms of the comoving distance, $\chi$, and the scale factor, $a$, as: $D_l=a_l\chi_l$, $D_s=a_s\chi_s$, and $D_{ls}=a_s\chi_{ls}$, which, upon insertion, results in the right hand side of \eq{eq:sigma_critical}.

\begin{figure}
     \centering
     \begin{tabular}{cc}
        \includegraphics[width=\linewidth,angle=0]{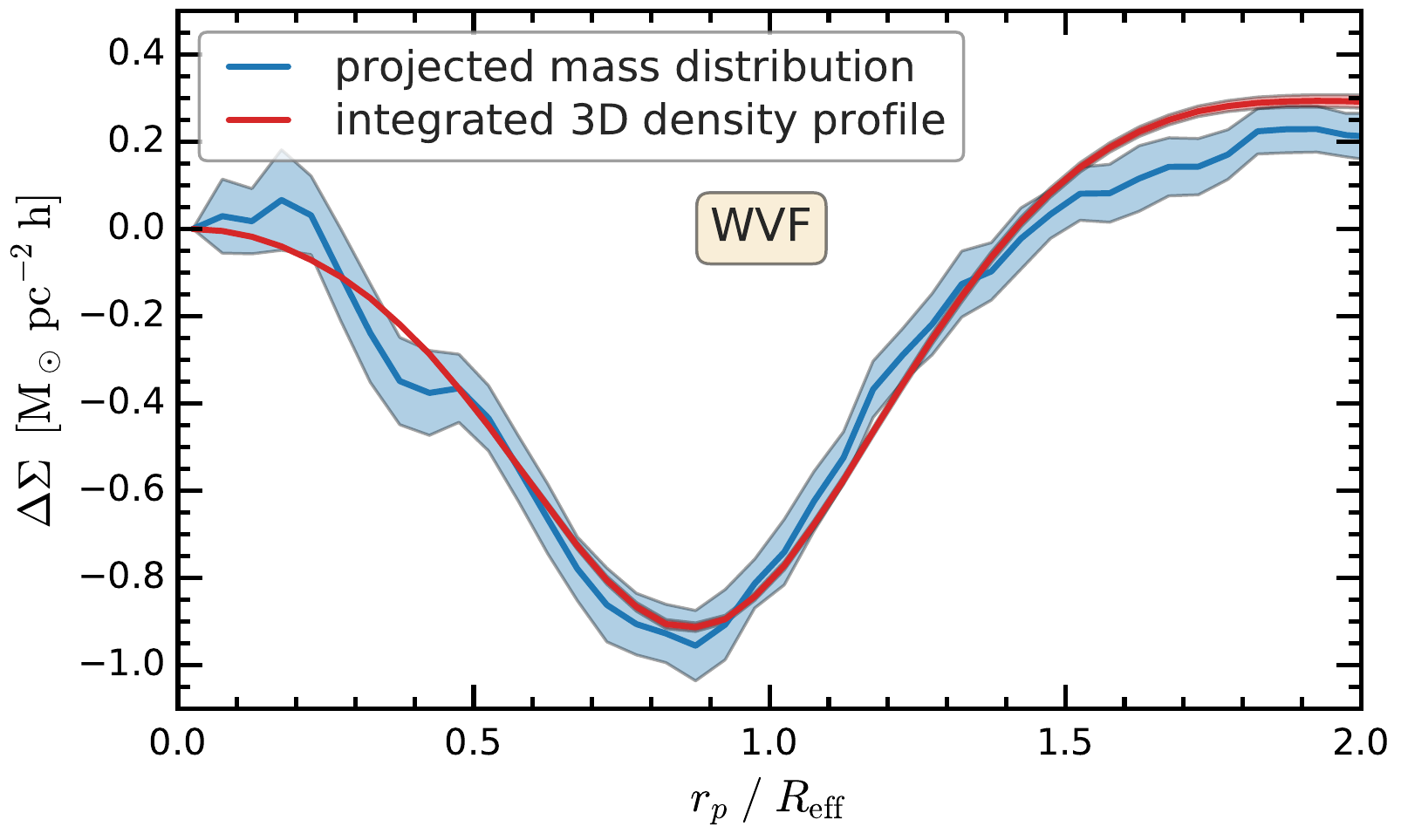}
     \end{tabular}
     \reduceVerticalOffset{}
     \caption{ Comparison of the differential surface mass density, $\Delta\Sigma$, of \WVF{} voids measured directly from the projected particle distribution and the one calculated from the 3D density profile using \eq{eq:sigma_r_integral}. Both methods were applied to the same physical volume. The uncertainty associated to the red curve is roughly the width of the curve itself and thus is barely visible. Calculating $\Delta\Sigma$ using the void 3D density profile gives a better estimate of the mean signal, but it underestimates the errors by a factor of ${\sim}7$ since this method does not account for a major source of error: the variation in the projected mass distribution of individual voids with the line-of-sight along which the void is observed. The blue shaded region accounts for this effect, and is the observationally relevant uncertainty.}
     \label{fig:lensing_compare_two_methods}
     \reduceVerticalOffset{}
\end{figure}

According to \eq{eq:tangential_shear}, up to a normalisation constant, the tangential shear is determined by the differential surface mass density, $\Delta\Sigma(r)$, which, in turn, is determined by the projected surface mass density at $r$, $\Sigma(r)$, \changed{whose calculation is described in \refsec{subsec:matter_profile}. In the case of 2D voids, $\Sigma(r)$ is shown in the right column of \reffig{fig:DM_den_fR_models}. Similarly, we compute $\Sigma(r)$ for the case of 3D voids, by projecting the voids on the simulated plane-of-the-sky using the distant observer approximation.}

\changed{In the case of 3D voids, the small amplitude of the tangential shear signal introduces an additional challenge. This is illustrated in \reffig{fig:lensing_compare_two_methods}, where we show $\Delta\Sigma(r)$ for \WVF{} voids in the \GR{} model. The differential surface density was calculated as described above, by projecting the void and matter distribution along the line-of-sight. The solid curve shows one mock observation (i.e. one simulation box and one line-of-sight choice), while the uncertainty region corresponds to the $1\sigma$ sample variance calculated using multiple realizations and line-of-sights. The $1\sigma$ uncertainty is around 10 per cent of the signal strength at $r\sim\Reff$ and is much larger than the typical difference in the void density profile between \FR{} and \GR{} models, which is only a few per cent (see \reffig{fig:DM_den_fR_models}). This means that to systematically characterize the lensing differences between \FR{} and \GR{}, we would need a large number of mocks.}

Alternatively, $\Sigma(r)$ can be computed by integrating the 3D density profile of voids, $\rho_{3D}$, via the expression \citep{Barreira2015}:
\begin{equation}
	\Sigma(r) \; = \; \int_{-L}^{L} \rho_{3D}(\sqrt{r^2+l^2}) \; \diff l 
     \label{eq:sigma_r_integral} \;,
\end{equation}
where the integral is along the line-of-sight, $l$, up to some line-of-sight distance from the void, $L$, large enough to account for large scale correlations in the mass distribution. In practice, we take $L$ to be 3 times the size of each void since the matter density profiles are well converged to unity beyond those distances. We perform the calculation outlined in \eq{eq:sigma_r_integral} using directly the density profiles of 3D voids shown in the left column of \reffig{fig:DM_den_fR_models}. In contrast to other studies \citep[e.g.,][{which focus on different models than here}]{Barreira2015,Falck2017}, we prefer not to fit a functional form since the differences between \GR{} and \FR{} profiles are small. Using a fitting function incurs the danger of artificially increasing or decreasing the differences between \GR{} and \FR{}, thereby biasing our predictions.

\begin{figure*}
     \centering
     \begin{tabular}{cc}
        \includegraphics[width=\figwidth,angle=0]{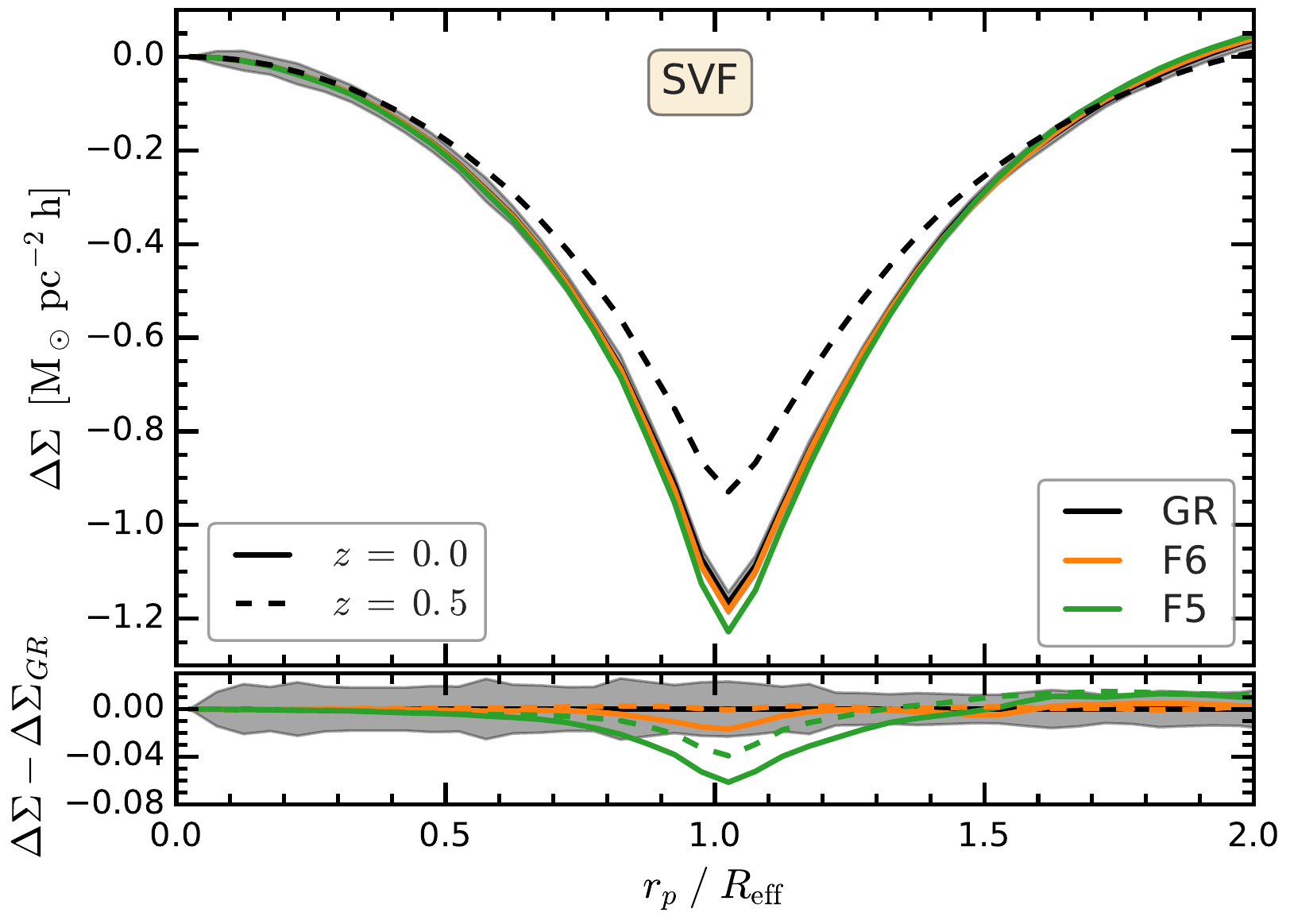} &
        \includegraphics[width=\figwidth,angle=0]{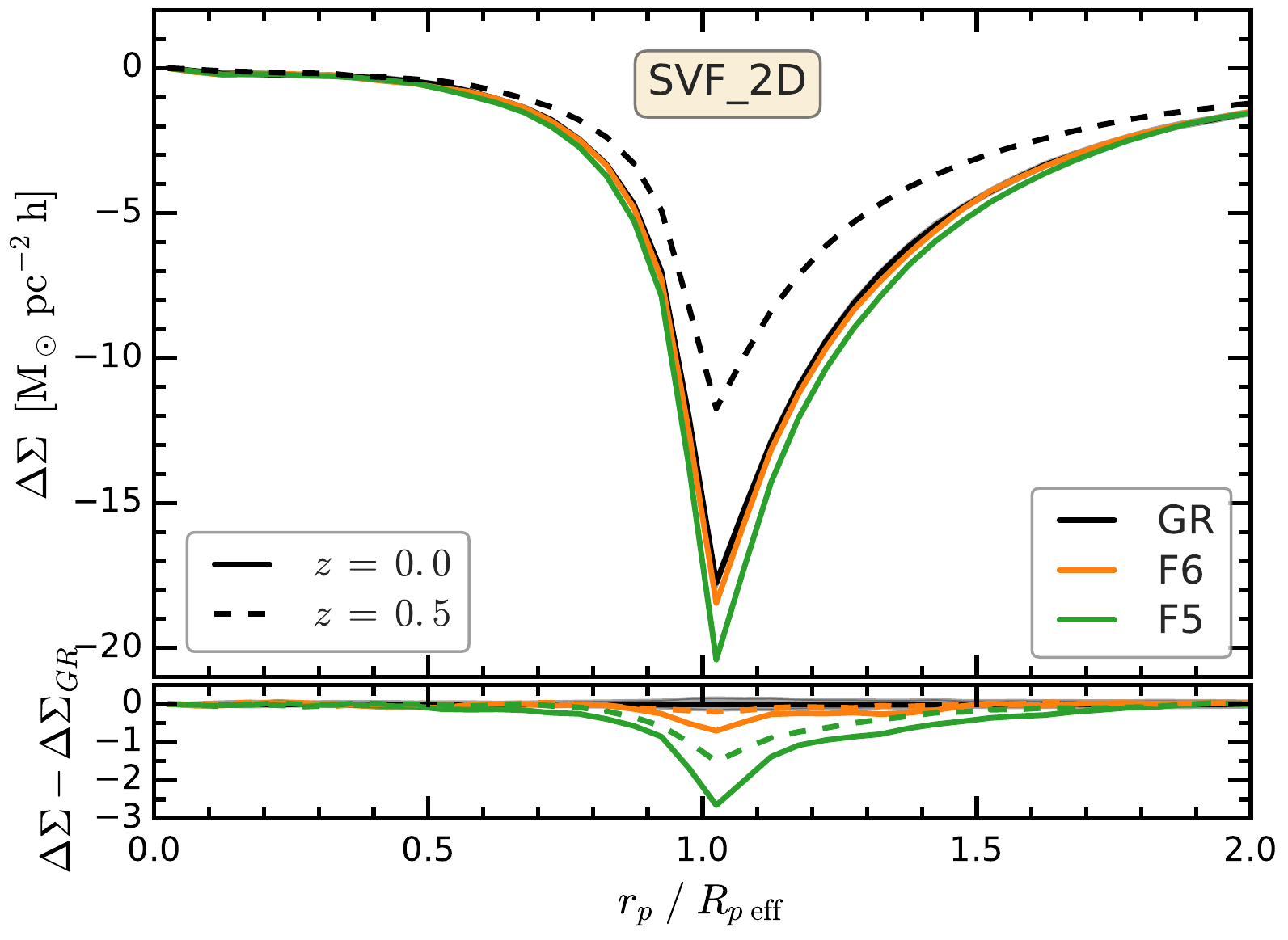} \verticaloffset
        \whitespace & \whitespace \verticaloffset
        \includegraphics[width=\figwidth,angle=0]{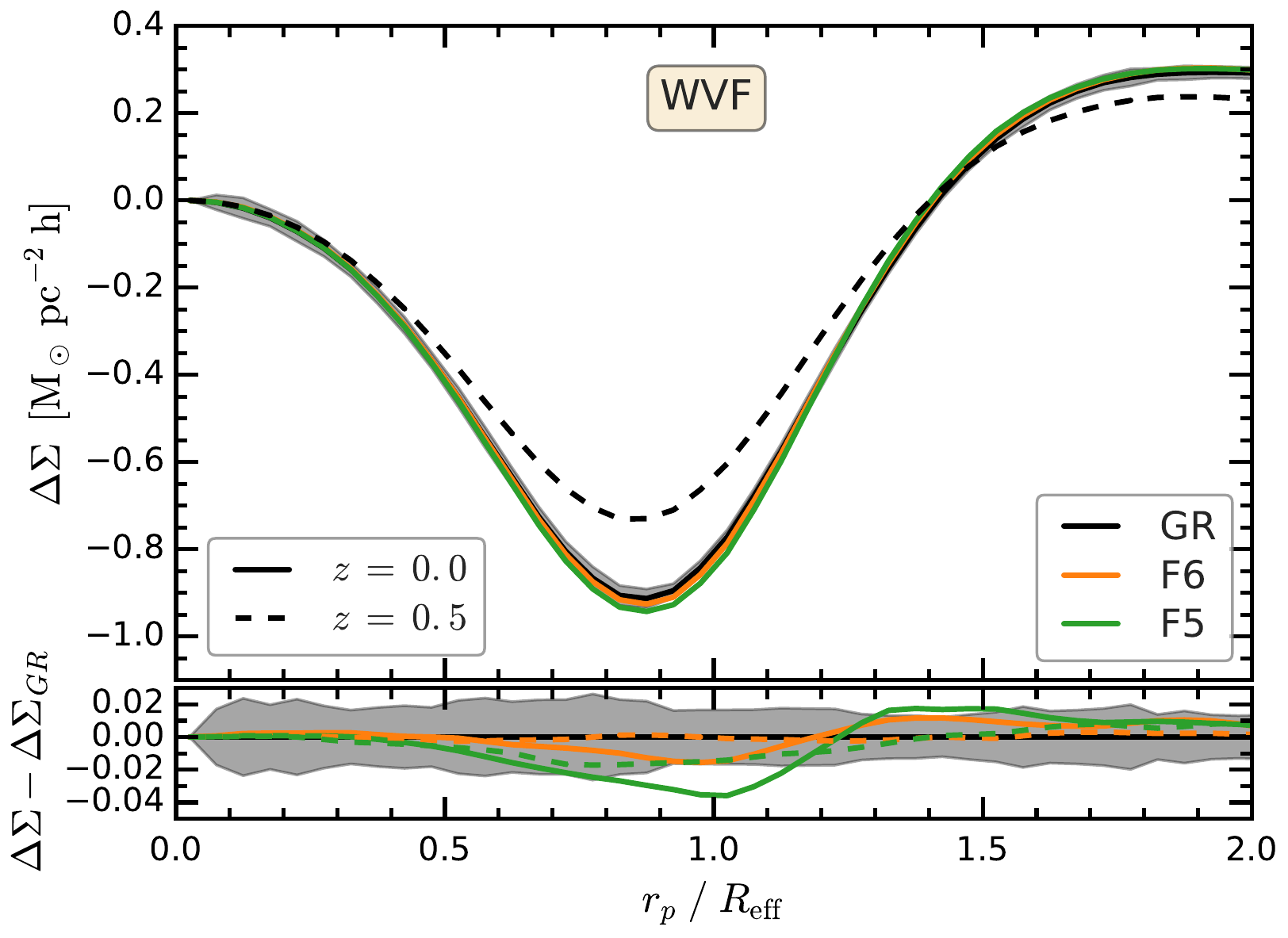} &
        \includegraphics[width=\figwidth,angle=0]{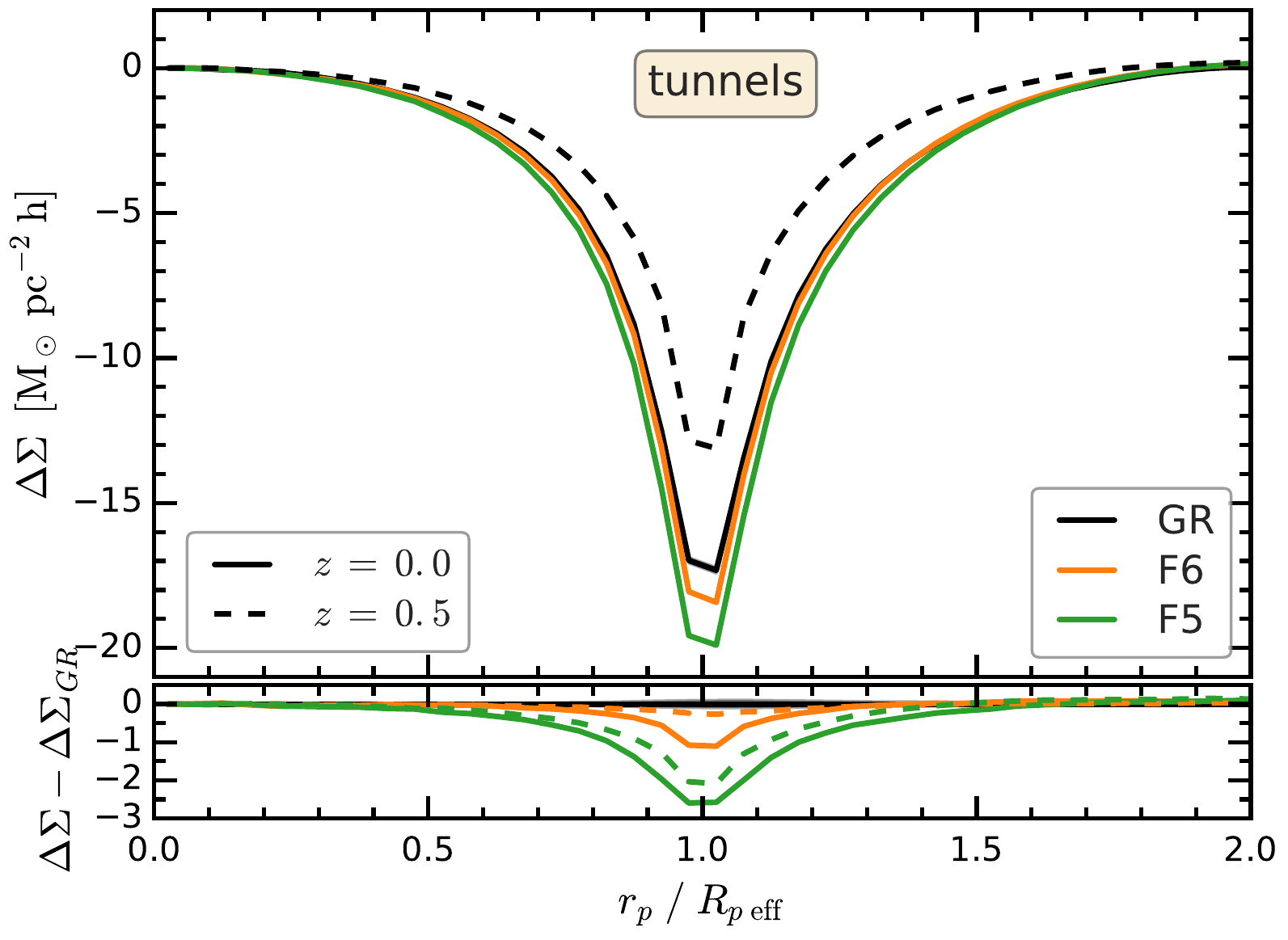} \verticaloffset
        \whitespace & \whitespace \verticaloffset
        \includegraphics[width=\figwidth,angle=0]{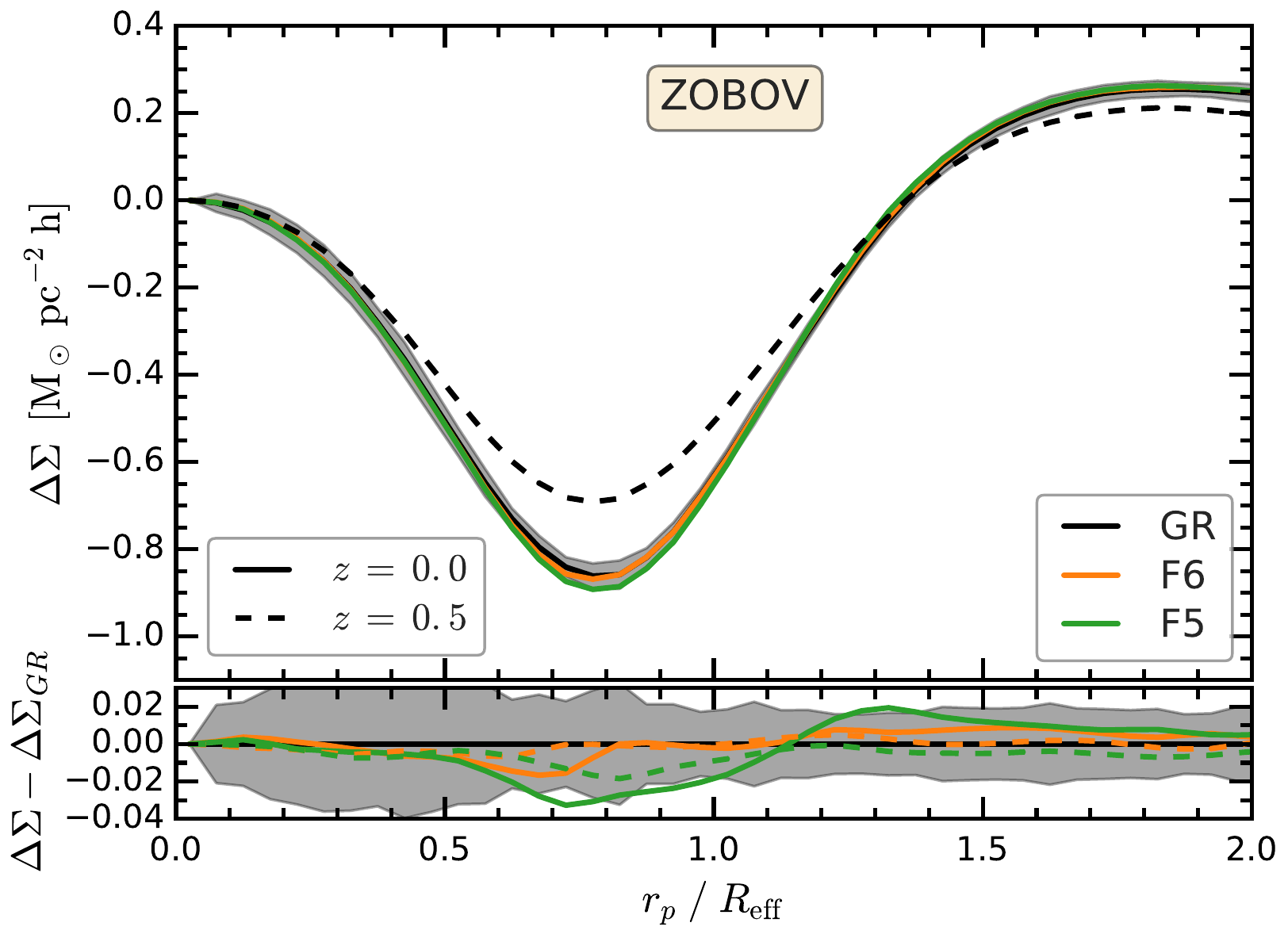} &
        \includegraphics[width=\figwidth,angle=0]{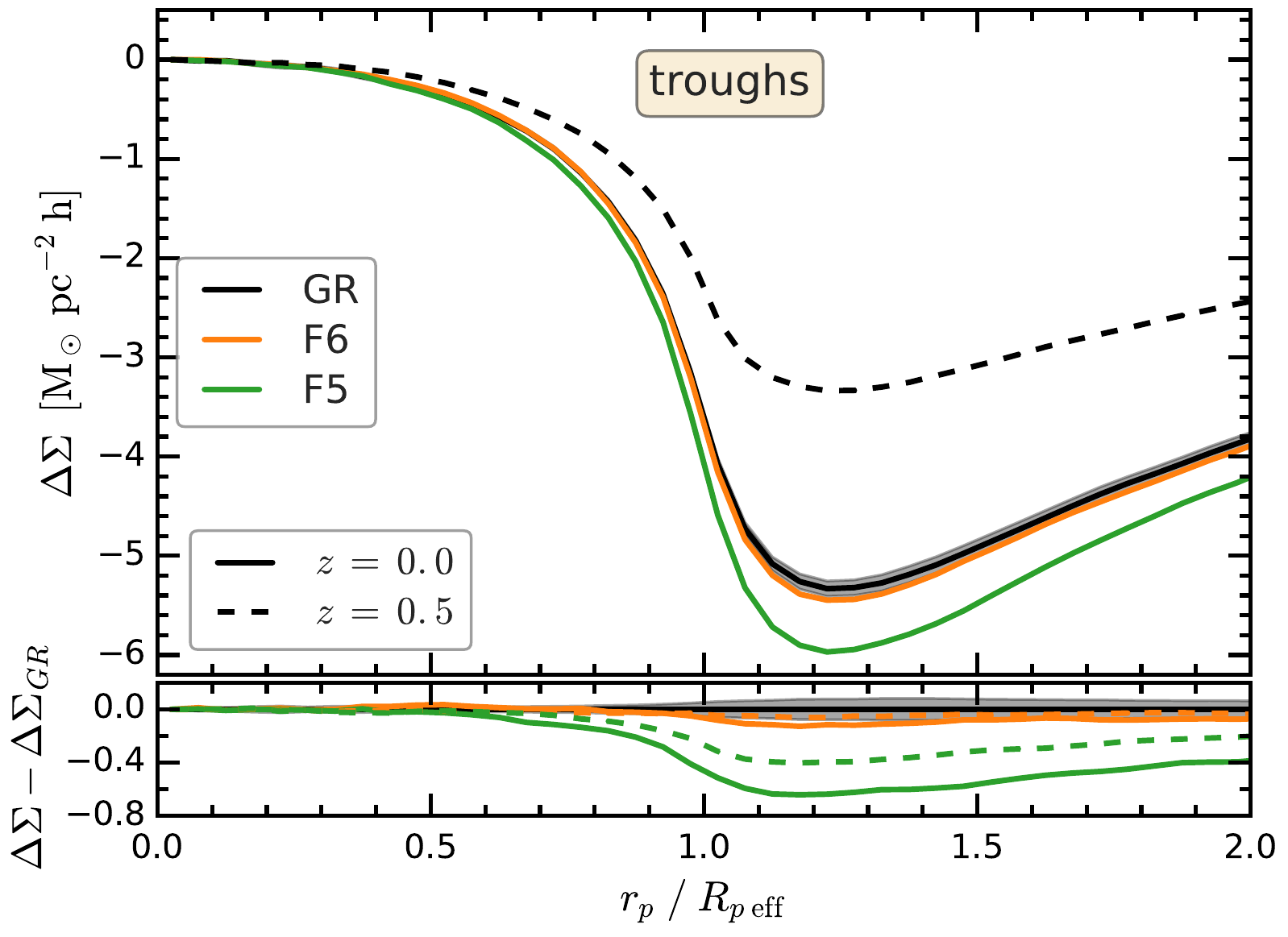}
     \end{tabular}
     \reduceVerticalOffset{}
     \caption{ Comparison of the differential surface mass density, $\Delta\Sigma$, profile of voids in \GR{} and \FR{} models. \changed{These were calculated by projecting the mass distribution along one of the axes of the simulation box.} The grey shaded regions show the $1\sigma$ uncertainties corresponding to the sample variance of the \GR{} signal. For 2D voids, the errors correspond to the sample variance of a $(1.024\Gpch)^3$ volume, which is the volume of each of our five simulation boxes. For 3D voids, the errors are very large so we show the sample variance for an \Euclid{}-like survey with a $10\GpchVolume$ volume, which is $9.3$ times larger than the volume of each of our simulation boxes. The symbols and curves are the same as in \reffig{fig:abundance_fixed_n_xi}. }
     \label{fig:lensing_fR_models}
     \reduceVerticalOffset{}
\end{figure*}

A second curve in \reffig{fig:lensing_compare_two_methods} shows the outcome of calculating the differential surface mass density of \WVF{} voids using \eq{eq:sigma_r_integral}. As expected, we find good agreement between the mean values of the two calculations. The small discrepancy at $r>1.5\Reff{}$ is a combination of correlated errors and a slight overestimation of $\Delta\Sigma(r)$ due to the fact that we limit the integration in \eq{eq:sigma_r_integral} to 3 times the void radius. \changed{The sample variance $1\sigma$ uncertainty region, which was found using multiple realizations of the \GR{} box combined with bootstrap sampling, is much smaller than the uncertainty resulting from projecting the voids and matter distribution along the line-of-sight. This suggests that integrating the 3D matter density profile is a more computationally efficient method of calculating the mean $\Delta\Sigma(r)$ value of 3D voids. However, the same calculation underestimates the size of the sample variance error by a factor of ${\sim}7$.} This discrepancy is due to \eq{eq:sigma_r_integral} neglecting one major source of scatter. The $\Delta\Sigma(r)$ uncertainty is a combination of two effects. First, it is affected by the void-to-void variation in their radial mass distribution. Secondly, since voids are highly non-spherical, the projected mass distribution around each void shows plenty of variation depending on the viewing direction. The 3D density profile of each void corresponds to an average over all possible line-of-sights and thus does not include this latter source of scatter. Therefore, the observationally relevant uncertainty is the one computed using the projected particle distribution. 

To summarize, for 3D voids we follow a hybrid approach for calculating $\Delta\Sigma$. The mean $\Delta\Sigma$ for both \GR{} and \FR{} models was computed using \eq{eq:sigma_r_integral}, that is by integrating the 3D density profiles along the line-of-sight. This means that we can measure to a high accuracy systematic differences between \GR{} and \FR{} models. To compute the observationally relevant \GR{} sample variance, we used the projected particle distribution around each void. Similarly to previous error estimates, we compute the covariance matrix using 500 bootstrap samples constructed from 5 simulation boxes (for details see the fourth paragraph in \refsec{sect:results}).

\reffig{fig:lensing_fR_models} shows the differential surface mass density, $\Delta\Sigma$,  for the six voids studied here. In all cases, we find that $\Delta\Sigma$ is negative at least up to, $r\lesssim1.5\Reff$, indicating that voids, due to their underdense interiors, produce divergent lensing, which is similar to a concave lens. In contrast, high density regions (e.g. clusters) give rise to convergent lensing, which is similar to a convex lens. For most void types the diverging weak lensing signal peaks at the void radius, $r=\Reff$, with troughs being the exception for which the signal peaks at $r\simeq1.2\Reff$. Of the 3D underdensities, \SVF{} voids produce the strongest tangential shear, which is ${\sim}30\%$ larger than the signal of the other two 3D voids. The lensing signature of \WVF{} voids, and probably that of \zobov{} ones, can be increased by a factor of ${\sim}2$ by stacking with respect to the boundary of these non-spherical objects \citep{Cautun2016a}, which would result in a stronger lensing signal than the \SVF{} one. The 2D underdensities have an even stronger weak lensing imprint than the 3D ones, with troughs producing a ${\sim}5$ times larger signal than \SVF{} voids. The \SVFtwoD{} voids and tunnels have an even larger tangential shear, roughly $15$ times larger than that of \SVF{} voids. This is due to the fact that 2D voids are much smaller than 3D ones and therefore they are measuring the matter fluctuations at much smaller scales.

\reffig{fig:lensing_fR_models} also compares the $\Delta\Sigma(r)$ profiles in \GR{} and \FR{} gravity. Voids in modified gravity models show a stronger lensing signal than in \GR{}, with the enhancement being largest for the \Ffive{} model at late times. To assess the significance to which these differences can be measured, \reffig{fig:lensing_fR_models} shows as a grey shaded region the $1\sigma$ sample variance for \GR{}. For 3D voids, the uncertainty corresponding to the volume of the simulation box is very large, so we present a rescaled error for an \Euclid{}-like survey with a $10\GpchVolume$ volume (9.3 times larger than our simulation box), which practically means reducing the uncertainty by a factor of $\sqrt{9.3}$. Among the 3D voids, \SVF{} voids show the largest difference compared to the fiducial \GR{} model, both in absolute terms as well as when compared to the error bars, but note that the $\Delta\Sigma_{\FR{}}-\Delta\Sigma_{\GR}$ difference is very similar to the cosmic variance. In contrast, for 2D voids, the signature of modified gravity models is significantly larger than the associated uncertainties in the lensing signal, which makes these objects ideal for testing \FR{} models.

\subsection{Predictions for \Euclid{} and \LSST{}}
\label{subsec:euclid_lsst_predictions}
Here we investigate the extent to which void lensing in future surveys can be used to constrain modified gravity theories. In particular, we focus our attention to the \Euclid{} \citep{euclid} and the \LSST{} \citep{lsst} lensing surveys which cover $20,000$ and $18,000$ square degrees, respectively. For both surveys we assume that their sky coverage area overlaps with spectroscopic surveys that have a galaxy number density in the redshift range $z=0.3$ to $0.7$ at least as high as that of the \textsc{SDSS} \CMASS{} sample. In many cases there will be overlapping spectroscopic surveys with  higher galaxy number densities, in which case our analysis quantifies the modified gravity constraints that can be inferred when using only the brightest galaxies. Using the full spectroscopic survey would probably result in even tighter constraints, but our simulations lack the resolution to make predictions for such observations.

Let us consider a survey with lenses in the redshift range $[z_{l;\; min},z_{l;\; max}]$ with a number density of lenses given by $W_l(z_{l})$. The lensing signal of these objects is measured using sources distributed according to $W_s(z_{s})$ in the redshift range $[z_{s;\; min},z_{s;\; max}]$. Then, the mean tangential shear is given by:
\begin{equation}
	\overline{\gamma}_t \; = \; \int_{z_{l;\; min}}^{z_{l;\; max}} \diff z_l \int_{z_{s;\; min}}^{z_{s;\; max}} \diff z_s  \; \frac{\Delta\Sigma(z_l)}{\Sigma_c(z_l,z_s)} \; W_l(z_{l}) W_s(z_{s})
     \label{eq:gamma_t_integral_general} \;,
\end{equation}
where $\Delta\Sigma(z_l)$ is the mean differential surface density of lenses at redshift $z_l$ and $\Sigma_c(z_l,z_s)$ is the corresponding critical surface density for lens redshift $z_l$ and source redshift $z_s$. 

We are interested in obtaining an approximate estimate for the tangential shear, so we consider a simplified set-up. First, we take all the source galaxies to have a single redshift which is the median redshift of the distribution, $z_{s\; med}$. Secondly, we take the lenses to have a uniform comoving number density, in which case $W_l(z_{l}) \propto \chi^2(z_l)$, with $\chi(z_l)$ the comoving distance to redshift $z_l$. Thirdly, we take the differential surface density, $\Delta\Sigma(z_l)$, to be independent of redshift.  The goal is to make predictions for a lens distribution in the redshift range $[0.3,0.7]$, so we take the value of $\Delta\Sigma(z_l=0.5)$ calculated in \reffig{fig:lensing_fR_models}. After accounting for all these simplifications, \eq{eq:gamma_t_integral_general} becomes:
\begin{align}
    \overline{\gamma}_t \; & = \; \Delta\Sigma(z_l=0.5) \; \frac{4\pi G}{c^2\chi_s} \;\; \frac{ \displaystyle \int_{\chi_{l;\; min}}^{\chi_{l;\; max}} \diff \chi_l \; (1+z_l)^{-1} \chi^3_l \left( \chi_l - \chi_s \right) }{ \displaystyle \int_{\chi_{l;\; min}}^{\chi_{l;\; max}} \diff \chi_l \; \chi^2_l } 
    \label{eq:gamma_t_integral_here_1} \\
      \; & \equiv \; \frac{ \Delta\Sigma(z_l=0.5) } { \Sigma_{c;\; eff} }
     \label{eq:gamma_t_integral_here_2} \;,
\end{align}
where $\chi_s=\chi(z_{s;\;med})$, $\chi_l=\chi(z_l)$, $\chi_{l;\;min}=\chi(z_{l;\;min})$ and $\chi_{l;\;max}=\chi(z_{l;\;max})$. All the terms of \eq{eq:gamma_t_integral_here_1} to the right of $\Delta\Sigma(z_l=0.5)$ can be grouped together into the inverse of an \textit{effective critical surface density for the survey}, $\Sigma_{c;\; eff}$. Using this notation, \eq{eq:gamma_t_integral_here_1} can be rewritten as \eq{eq:gamma_t_integral_here_2}, which is similar in form to \eq{eq:tangential_shear}. 

Here we adopt $z_{s;\;med}=0.8$ and $1.2$, which corresponds to the median source redshift for the \Euclid{} and the \LSST{} surveys, respectively, to obtain $\Sigma_{c;\; eff} = 6770$ and $3960\shearUnit$. For the \Euclid{} survey we predict tangential shear values at the position of the dip of $\gamma_t \simeq -1\times10^{-4}$ for 3D voids, $\gamma_t = -5\times10^{-4}$ for troughs and  $\gamma_t = -2\times10^{-3}$ for \SVFtwoD{} voids and tunnels. For the \LSST{} survey we predict a lensing signal that is a factor of $1.7$ times larger than for \Euclid{}. Thus, depending on the method used to identify underdense regions, the weak lensing signal can vary by a factor of $20$, being lowest for 3D underdensities and highest for 2D underdensities, with \SVFtwoD{} voids and tunnels having the highest lensing imprint.

\changed{The tangential shear measurements are affected by three important sources of uncertainty: void sample variance, the covariance of uncorrelated large-scale structure along the path of the light rays and shape noise \citep[e.g. see][]{Krause2013}.} 
\changedNew{The first two error sources can be obtained by calculating the void tangential shear profile due to the mass distribution between the source plane and the observer. For this, we construct a mock light cone for each GR realization. First, the mass distribution in the redshift range $z=0.3$ to $z=0.7$ is given by the $z=0.5$ snapshot of the respective GR realization. To account for uncorrelated large-scale structure, the mass distribution for $z<0.3$ and for $z>0.7$ is taken from the $z=0.5$ snapshot of the other GR realizations.} 
For simplicity, our light cone mocks use the mass distribution at $z=0.5$, which neglects the time evolution of the clustering, and, secondly, we use a cylindrical geometry while in practice observations have a conical geometry. 
\changedNew{We calculate the tangential shear of each individual void by applying \eq{eq:tangential_shear} to thin slices along the line-of-sight and summing the contribution of all these slices. We do so for all the 5 \GR{} realizations.}

\changed{The remaining source of uncertainty, shape noise, comes from the intrinsic ellipticity distribution, characterized by its variance, $\sigma_\epsilon$, of the source galaxies used to measure $\gamma_t$. 
We measure shape noise by generating a random distribution of source galaxies in our simulated plane-of-the-sky (i.e. projected simulation box), with each source having a randomly assigned and randomly oriented ellipticity, with the ellipticity variance being given by $\sigma_\epsilon$ (this is similar to the procedure described in \citealt{Sanchez2016}, but applied to mock catalogues and not to the data). 
For each void in the catalogue, we calculate the mean source galaxy ellipticity for the same radial bins used to estimate the weak lensing signal.} 
For this calculation, we adopt and intrinsic source ellipticity, $\sigma_\epsilon=0.22$, and a number density of sources, $n_{sources}=30$ and $40~\rm{arcmin}^{-2}$ for \Euclid{} and \LSST{}, respectively. 
\changedNew{Then, we add the shape noise to the tangential shear signal for each of the voids found in the 5 \GR{} realizations. To compute the covariance matrix, we split each \GR{} void catalogue into 64 subregions and generate 100 bootstrap realizations; the resulting $N=500$ estimates are used to calculate the total covariance matrix. The resulting covariance matrix, which corresponds to a survey with the same volume as each of our simulation boxes, is rescaled  by multiplying with the factor $V_{\rm sim}/V_{\rm survey}$, where $V_{\rm sim}$ and $V_{\rm survey}$ are the comoving volumes of the simulation box and the survey, respectively.}

\changed{Since the inverse of a noisy estimate of the covariance matrix is biased high, we correct for this effect by multiplying the inverse covariance by the Anderson-Hartlap factor \citep{Anderson2003,Hartlap2007}, 
\begin{equation}
	\alpha=\frac{N-N_{bin}-2}{N-1}
    \label{eq:Anderson-Hartlap_factor}
\end{equation} 
where $N=500$ is the number of realizations used to estimate the covariance matrix and $N_{bin}$ is the number of bins.}

\begin{figure}
     \centering
     \begin{tabular}{c}
        \includegraphics[width=\linewidth,angle=0]{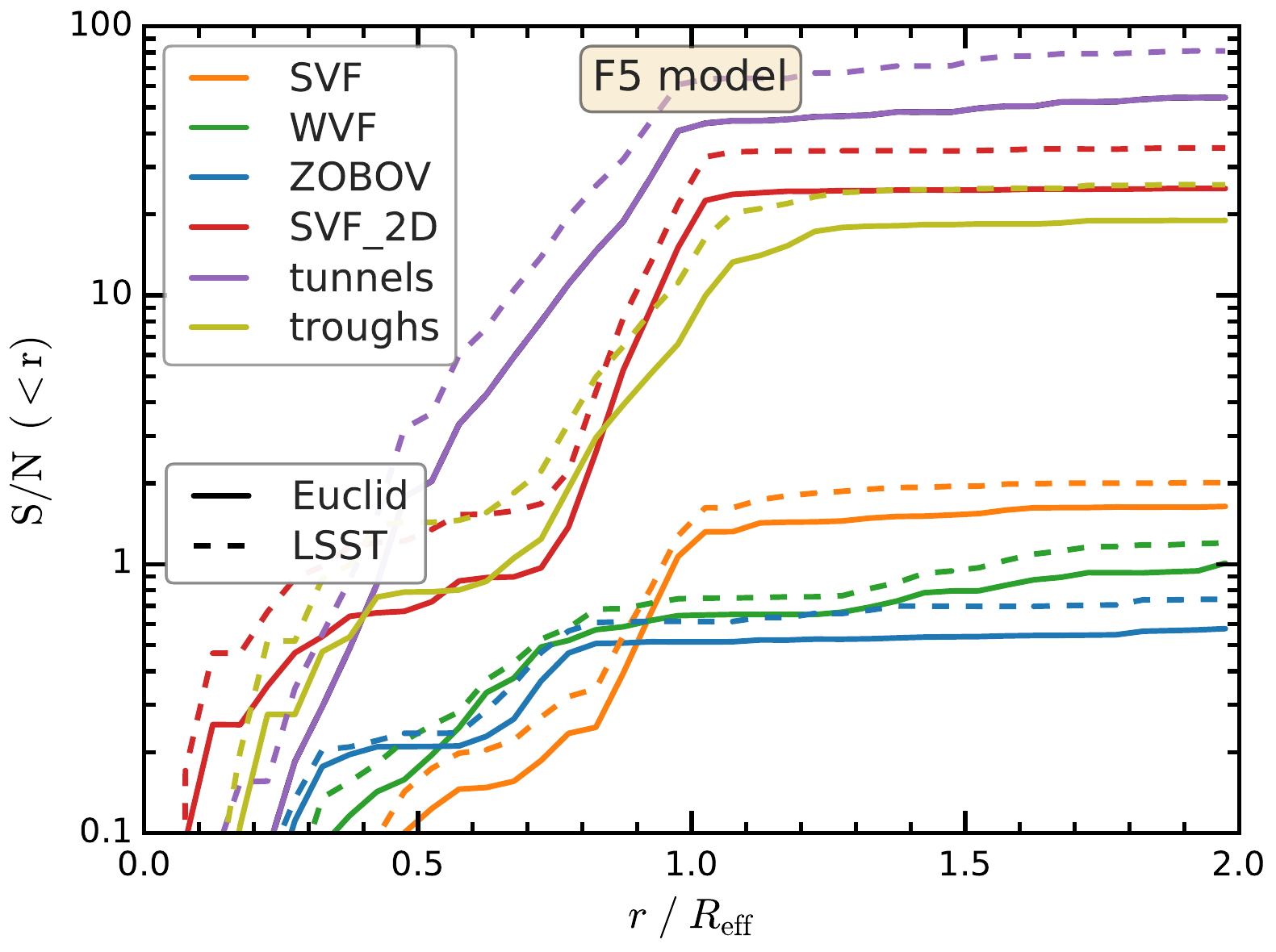} \verticaloffset
        \includegraphics[width=\linewidth,angle=0]{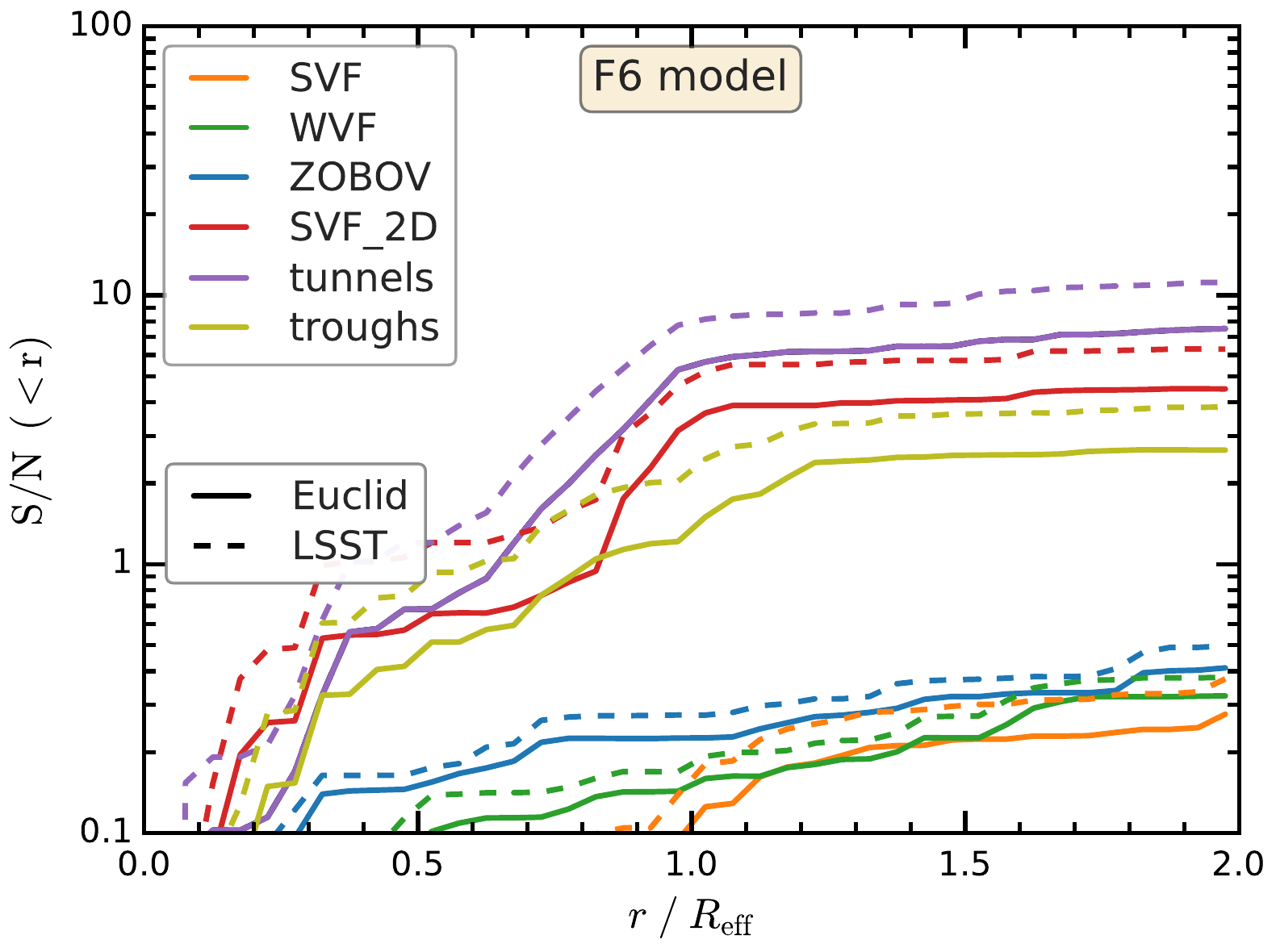}
     \end{tabular}
     \reduceVerticalOffset{}
     \caption{ The cumulative (from small to large radius) signal to noise, S/N, of the differences in tangential shear between \FR{} and \GR{}. The top and bottom panels show the S/N of the \Ffive{} and \Fsix{} models, respectively. Each colour corresponds to one of the six void types studied in this paper. The predictions for an Euclid- and a LSST-like lensing surveys are shown as solid and dashed lines, respectively. }
     \label{fig:lensing_FR_SN}
     \reduceVerticalOffset{}
\end{figure}

A useful way to quantify the degree to which \FR{} models can be distinguished from the fiducial case is to compute the signal-to-noise ratio (S/N) for the tangential shear signal of the various void catalogues. In this respect, we define the cumulative S/N up to a radial bin $k$ as:
\begin{equation}
     \left(\rm{S/N}\right)^2(<k) = \displaystyle \sum_{i\leq k;\; j\leq k} \delta\gamma_t(i) \;\; \rm{cov}^{-1}(i,j) \;\; \delta\gamma_t(j)
     \label{eq:signal-to-noise} \;,
\end{equation}
where $\delta\gamma_t=\gamma_{t\; \FR{}}-\gamma_{t\; \GR{}}$ is the excess tangential shear signal in \FR{} gravity compared to \GR{} and $\rm{cov}^{-1}$ is the inverse of the tangential shear covariance matrix. The sum is over radial bins $i$ and $j$ which take values from $1$ to a maximum of $k$. 
\changed{For each value of $k$, we calculate the $\rm{cov}^{-1}$ matrix by inverting the first $k\times k$ entries of the covariance matrix, with the number of bins in the Anderson-Hartlap correction factor (see Eq. \ref{eq:Anderson-Hartlap_factor}) being given by $N_{bin}=k$. The radial range $r/\Reff\leq 2$ is split into 40 equal width bins.}
The S/N values correspond to the number of sigma that the \FR{} models can be distinguished from the standard \GR{} one.

We compute the cumulative S/N ratios for all the six void catalogues for both the \Ffive{} and \Fsix{} models. The results are shown in \reffig{fig:lensing_FR_SN}. For all methods, the cumulative S/N increases up to the void radius, after which it stays relatively constant, which suggests that most of the power for distinguishing gravity models comes from the region $r\lesssim\Reff{}$. The 2D voids show the largest S/N for both \FR{} models, with \tunnels{} being the most promising method. The S/N is larger for \LSST{} than for \Euclid{} since the source galaxies of the former survey are at higher redshift and there are 25 percent more of them. In the case of \Euclid{}, tunnels have a maximum S/N of 50 and 7 for respectively the \Ffive{} and \Fsix{} models. For \LSST{}, the tunnels' S/N of the same two models increases to 80 and 11, respectively. 

For the \Ffive{} case, the lensing signal of 3D voids peaks at a S/N${\sim}1-2$. This value is lower than previous studies, with \citet{Cai2014} predicting that \SVF{} voids in a $1\GpchVolume{}$ volume can distinguish \Ffive{} with S/N${\sim}7$. Extrapolating this result to the \Euclid{} volume would result in a $\sqrt{10}$ larger value, i.e., S/N${\sim}22$, while in this work we only find a S/N${\sim}2$. The discrepancy is due to several differences between our analysis and the \citeauthor{Cai2014} one. We estimate the S/N using the $z=0.5$ snapshots, while \citeauthor{Cai2014} used the $z=0$ matter distribution; as seen in \reffig{fig:lensing_fR_models}, the difference in void tangential shear between \FR{} and \GR{} are smaller at higher redshift. Also, we include two additional error sources, shape noise and the contribution of uncorrelated line-of-sight large scale structure, which are similar in magnitude to the sample covariance of voids.

\begin{figure}
     \centering
     \begin{tabular}{c}
        \includegraphics[width=\linewidth,angle=0]{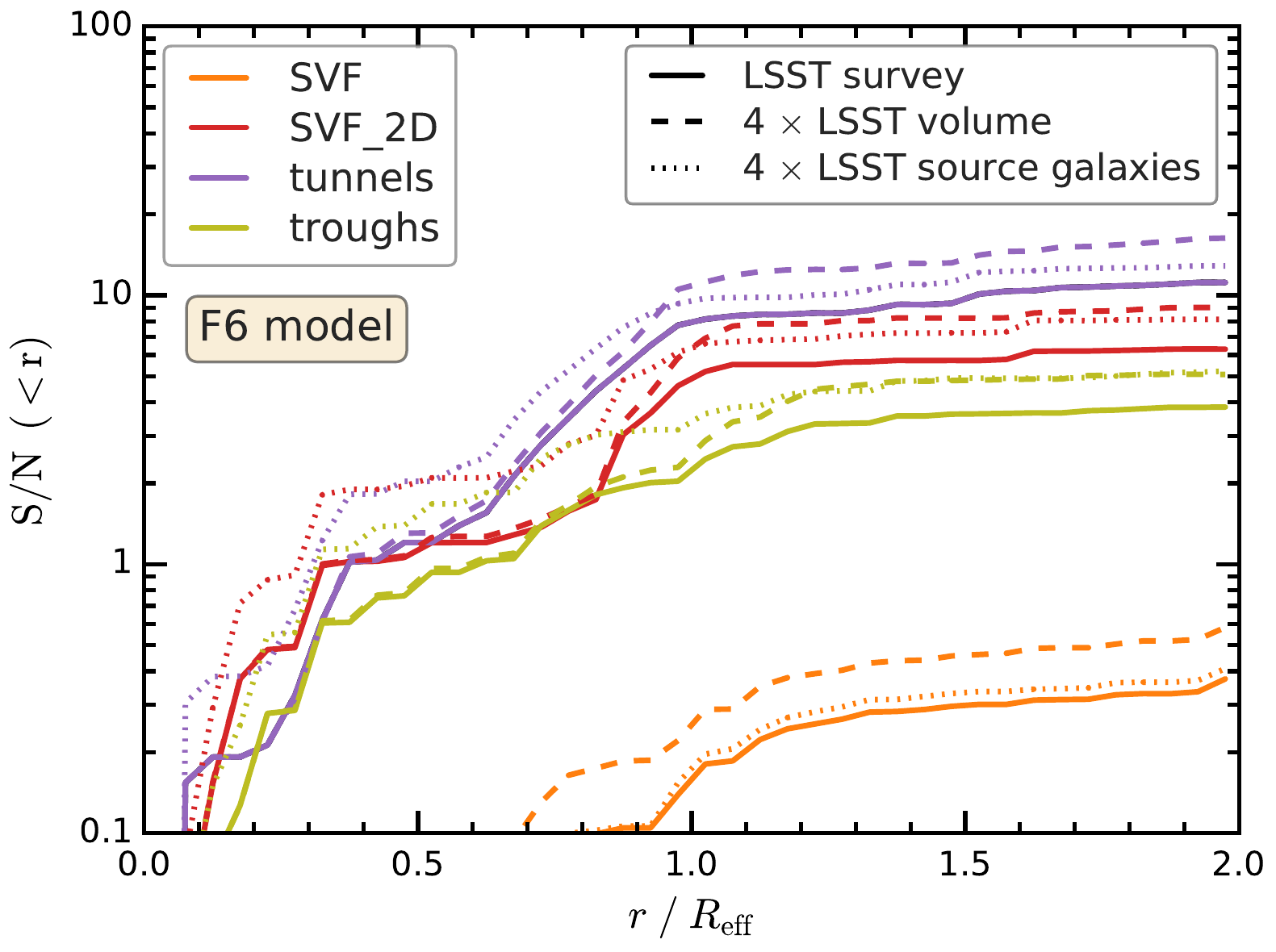}
     \end{tabular}
     \reduceVerticalOffset{}
     \caption{ Comparison of which sources of error dominate the S/N of the difference in tangential shear between \Fsix{} and \GR{}. Each colour corresponds to a different void finder; for clarity we only show the \SVF{} results for 3D voids. The solid curves show the S/N for the \LSST{} survey; the dashed lines show the change in S/N if instead the survey would have the same number of lensing sources but cover a 4 times larger volume; the dotted lines show the S/N if instead the survey would have 4 times more lensing source galaxies. }
     \label{fig:lensing_F6_SN_no_shape_noise}
     \reduceVerticalOffset{}
\end{figure}

In \reffig{fig:lensing_F6_SN_no_shape_noise} we study which of two possible observational strategies, that is surveying a larger volume versus having a higher density of source galaxies, optimizes the gain of using void lensing to distinguish between the \FR{} and \GR{} models. We analyse the case of \Fsix{} and we limit the analysis to the \LSST{} survey, which in \reffig{fig:lensing_FR_SN} gives the largest S/N. For each void finder, \reffig{fig:lensing_F6_SN_no_shape_noise} shows three sets of S/N curves: (i) \LSST{} (solid lines), (ii) 4 times the \LSST{} volume, but equal number of lensing source galaxies as \LSST{} (dotted lines), and (iii) \LSST{} volume with 4 times the number of \LSST{} lensing source galaxies. The case (ii) and (iii) corresponds to decreasing separately the covariance matrix contribution from respectively cosmic variance and shape noise. In the case of 3D underdensities, for clarity, we only show \SVF{} voids, but \WVF{} and \zobov{} voids show the same trends. We find that the S/N is boosted more when increasing the survey volume compared to when increasing the density of source galaxies by the same factor, although for troughs and \SVFtwoD{} both strategies lead to similar S/N gains. In the case of 3D voids, shape noise represents a small fraction of the error budget, so having larger survey volumes brings the largest gain. For 2D underdensities, the shape noise error contribution is larger than in the case of their 3D counterparts, but it is still smaller than uncertainties due to sample variance and uncorrelated line-of-sight structures.

\section{Discussion and conclusions}
\label{sect:discussion}

{We have carried out a detailed comparison of the extent to which voids can test models of modified gravity. Underdense regions are typically unscreened in modified gravity theories, which suggests that voids can be potentially good discriminators of such models. There are multiple ways of defining voids, with different algorithms designed to identify structures at different scales and with different geometries. Moreover, different void finders may have differing systematic and statistical error budgets. This motivated the need for a comprehensive comparison of different void finding algorithms in order to precisely quantify which voids are best suited to test modified gravity models in light of the coming data from big galaxy surveys. In particular, we would like to know what level of constraints future surveys such as \Euclid{} and \LSST{} can lead to when using void statistics. In this analysis, we compare six existing and new methods to identify cosmic voids, with three algorithms finding underdensities in the 3D galaxy distribution and three in the line-of-sight projected 2D galaxy distribution.}

We expect that the constraining power of different void finders to be model dependent, so this work focuses on a class of very popular theoretical models: \FR{} gravity proposed by \citet{Hu2007}, which is a representative example of the general models known as chameleons \citep{Khoury2004}. We studied two \FR{} models, \Ffive{} and \Fsix{}, with \Ffive{} corresponding to stronger deviations from \GR{} than \Fsix{}. The underlying philosophy is to use this very example to quantitatively understand the future constraints given by void statistics, and the conclusions from this study will be indicative for the more general models.

We improved upon previous studies, which used either dark matter particles or dark matter haloes as tracers, by identifying the voids using tracer galaxies, and therefore special care has been taken to create realistic mock galaxy catalogues. We have used a five-parameter HOD model to populate dark matter haloes with galaxies. The \GR{} \HOD{} uses the \cite{Manera2013} parameters, which reproduce the \textsc{sdss} \CMASS{}-sample galaxy clustering, while the \FR{} \HOD{} parameters were tuned to result in the same galaxy number density and projected two-point correlation functions as the GR ones. The resulting number densities of galaxies are $n_g=3$ and $5\times10^{-4}$ (Mpc$/h$)$^{-3}$ for $z=0.5$ and $z=0$, respectively, and we leave a study of the effect of varying $n_g$ (which requires higher-resolution simulations than used here) to future work.

The main conclusions of this work are:

\vspace{-.3cm}
\subsubsection*{1. Void abundance}

The abundance of voids is sensitive to the tracers used to identify them. For example, \cite{Cai2014} found that the different $f(R)$ models predict very different abundances for voids found using the dark matter field directly. For sparse tracers, such as dark matter haloes and galaxies whose number densities are in the region of ${\sim}5\times10^{-4}$($h^{-1}$Mpc)$^{-3}$, we find the same void abundance across all models  when matching the number densities, and the agreement becomes even better when we further match the galaxy correlation functions (see \reffig{fig:abundance_fixed_n_xi}). Thus, void abundances do not have additional discrimination power of modified gravity models once the number density and the correlation function is the same across all models. 

The above conclusion holds when comparing voids identified using the same method for all models, and there can still be a lot of variation if we compare the abundances from different void finders. 

\vspace{-.3cm}
\subsubsection*{2. Void profiles}

Similar to void abundances, the galaxy number density profiles of voids are almost identical across all models when we have matched their two-point correlation functions (see \reffig{fig:gal_den_fR_models}), and it suggests that the observed void galaxy number density profiles cannot be used to distinguish the models studied here.

However, the modification of gravity does affect the distribution of the underlying dark matter field. The agreement of the galaxy correlation functions in the different models has been achieved by tuning the HOD parameters -- which are empirical parameters describing how galaxies populate dark matter haloes from a simulation -- in the models, and this means that these models must have different galaxy bias in order to match the same observational data. When studying the void matter density profiles, we find that voids in $f(R)$ gravity are more underdense (see \reffig{fig:DM_den_fR_models}) due to the presence of a fifth force that evacuates underdense regions more efficiently.

Gravitational lensing is a way to directly probe the total matter distribution between the source and the observer, and so we compared the stacked lensing signal (tangential shear) by voids between $f(R)$ and GR gravity to find that  the former generally predicts stronger void lensing (see \reffig{fig:lensing_fR_models}). The tangential shear profiles of voids depend on the way in which the voids are identified. In particular, we confirm that among the 3D void finders, the \SVF{} gives rise to the strongest model difference in the tangential shear profile, since, by construction, spherical voids show a stronger density variation with distance at the void edge. We also find that 2D voids, in particular tunnels and 2D \SVF{}, show both stronger lensing signals and stronger model differences compared to 3D voids.

\vspace{-.3cm}
\subsubsection*{3. Constraining power of void lensing in future surveys}

In order to quantify the significance of the lensing signals for future galaxy surveys such as \Euclid{} and \LSST{}, and to assess more accurately the potential of using voids to test chameleon-type models, we have performed a comprehensive calculation of the signal-to-noise of void tangential shear to distinguish \FR{} and \GR{} models. Our analysis has taken into account the major sources of uncertainties: the number densities of voids in the survey volume, the number density of source galaxies, line-of-sight projection effects, and the shape noise due to the intrinsic ellipticity of source galaxies.

We find that 2D voids are the most promising underdensities for probing \FR{} gravity. Of these, the tangential shear by tunnels has the largest constraining power, with a \Euclid{}-like survey being able to distinguish the \FR{} gravity \Ffive{} and \Fsix{} models at a $50$ and $7\sigma$ confidence level, respectively (see \reffig{fig:lensing_FR_SN}). The \LSST{} data will have an even higher constraining power corresponding to a confidence level of $80$ and $11\sigma$ for the \Ffive{} and \Fsix{} models. The 2D \SVF{} voids have somewhat less distinguishing power than tunnels, with troughs being the least constraining of the 2D voids.

We also find that the weak lensing signal of 3D voids has a poor power to distinguishing \FR{} models. The most promising of them, the \SVF{}, applied to the \LSST{} survey can constrain the \Ffive{} and \Fsix{} models to a confidence level of $2$ and $0.3\sigma$. This distinguishing power is significantly lower than found in previous literature. The discrepancy with previous studies is due to (1) the use of the $z=0.5$ snapshots, which show smaller differences between \FR{} and \GR{} than present at $z=0$, and (2) our inclusion of additional error sources, such as shape noise and the contribution of uncorrelated line-of-sight large scale structure.

We investigated the dominant void lensing uncertainty sources for the \LSST{} survey to find that while source galaxy shape noise is less important than errors due to sample variance and line-of-sight large scale structures, it still has a significant contribution (see \reffig{fig:lensing_F6_SN_no_shape_noise} ). Thus, the prospects of using void lensing to distinguish \FR{} models from \GR{} can be best boosted by increasing the survey volume. The best way of increasing the survey volume is by extending the sky coverage of the \LSST{} survey; going to higher redshift will cover more volume, but may not help as much because at higher redshift the difference between models also decreases.

\vspace{.3cm}
This paper uses a conservative sample of tracer galaxies with a number density corresponding to that of the \CMASS{} galaxy sample. \Euclid{} and \LSST{} are expected to have a larger number density of galaxies that will allow for the identification of more voids, but with smaller sizes. It remains to be studied how the larger density of tracers can affect the constraining power of void weak lensing. We leave this for future work, since it needs higher-volume simulations than used here with much better mass resolution, the latter needed to resolve the lower mass dark matter haloes that host galaxies fainter than in \CMASS{}. Another important question, which is not addressed in this paper, is the potential degeneracy between the effects of modified gravity and other cosmological parameters such as $\Omega_m$ and $\sigma_8$ \citep{Cai2014}. 
\changed{In \FR{} gravity, the convergence power spectra show a scale-dependent enhancement compared to \GR{} predictions \citep[e.g.,][]{Tessore2015,Li2017}, which suggests that the lensing signature of \FR{} models is not degenerate with cosmological parameters, as explicitly checked by \cite{Shirasaki2017}. Extending this conclusion to void lensing is non-trivial because the convergence power spectrum represents an average over the entire volume while voids sample mostly underdense regions, and, furthermore, there is an environmental dependence of the fifth force in \FR{} models. Understanding the degeneracy between cosmological parameters and void lensing} will involve new simulations to be carried out by varying these cosmological parameters, and will be left for future work.

\vspace{-.3cm}
\section*{Acknowledgements}
The authors would like to thank the anonymous referee for detailed comments that have helped us improve the paper and to Mark Neyrinck for insightful discussions. MC was supported by Science and Technology Facilities Council (STFC) [ST/L00075X/1, ST/P000451/1]; and by an ERC Advanced Investigator grant COSMIWAY [GA 267291]. EP is supported by CONICYT-PCHA/Doctorado Nacional (2017-21170093). SB is indebted to the generous grant of an ITC Fellowship from Harvard University and, formerly, the STFC through grant ST/K501979/1. YC was supported by the European Research Council under grant numbers 670193. NP received support from Fondecyt Regular 1150300, BASAL CATA PFB-06, Anillo ACT-1417. BL is supported by an European Research Council Starting Grant (ERC-StG-716532-PUNCA) and STFC (ST/L00075X/1, ST/P000541/1). 
This project has received funding from the European Union's Horizon 2020 Research and Innovation Programme under the Marie Sk\l odowska-Curie grant agreement No 734374.     
This work used the DiRAC Data Centric system at Durham University, operated by the Institute for Computational Cosmology on behalf of the STFC DiRAC HPC Facility (www.dirac.ac.uk). This equipment was funded by BIS National E-infrastructure capital grant ST/K00042X/1, STFC capital grants ST/H008519/1 and ST/K00087X/1, STFC DiRAC Operations grant ST/K003267/1 and Durham University. DiRAC is part of the National E-Infrastructure.
Part of the analysis was done on the Geryon cluster at the Centre for Astro-Engineering UC. The Anillo ACT-86, FONDEQUIP AIC-57, and QUIMAL 130008 provided funding for several improvements to the Geryon cluster.


\vspace{-.3cm}
\renewcommand{\jcap}{JCAP}
\bibliographystyle{mnras}



\appendix

\section{Selection of tunnels and troughs }
\label{appendix:tunnels_troughs}
\begin{figure}
     \centering
     \begin{tabular}{c}
        \includegraphics[width=.93\linewidth,angle=0]{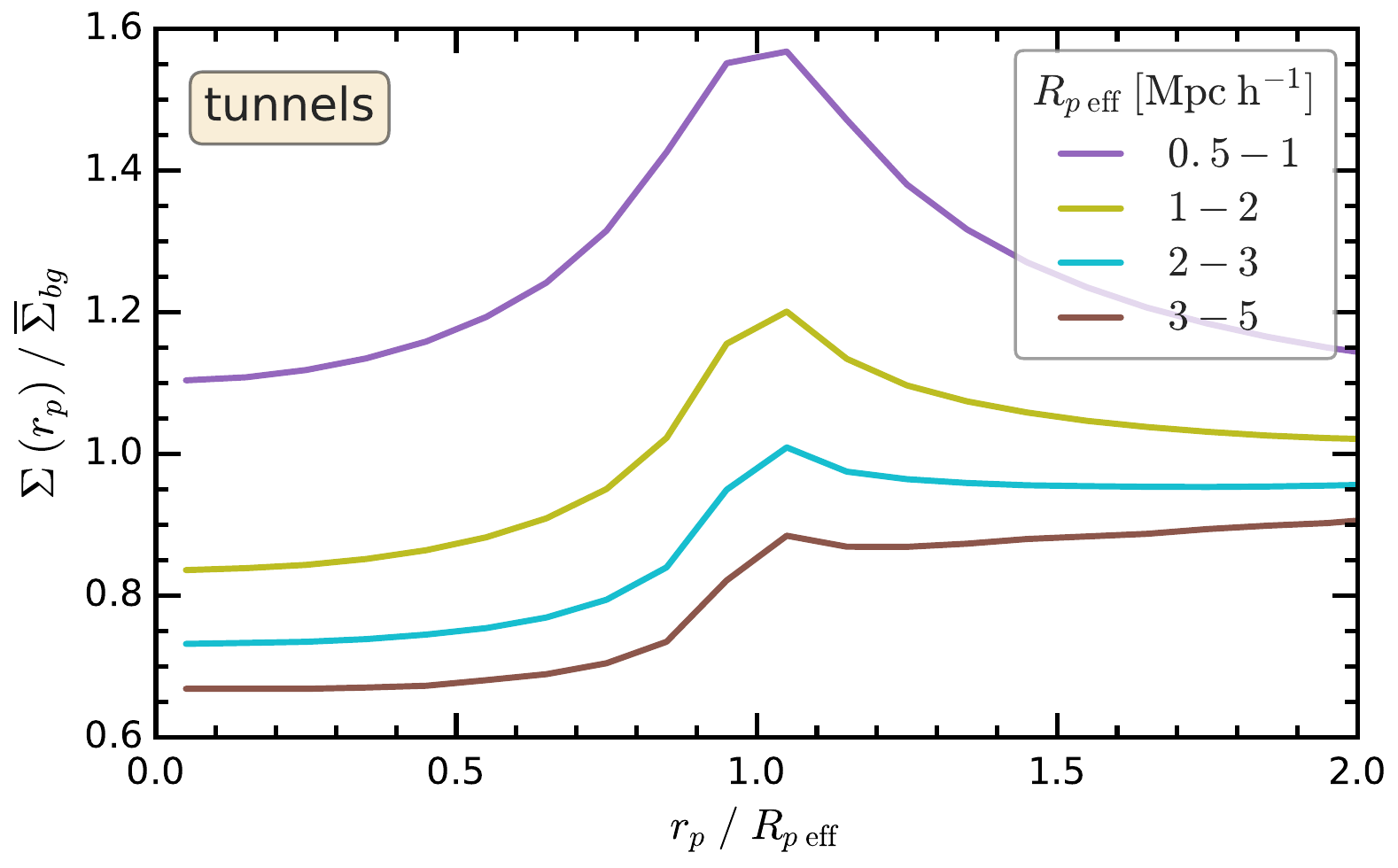} \verticaloffset
        \whitespaceFull \verticaloffset
        \includegraphics[width=.93\linewidth,angle=0]{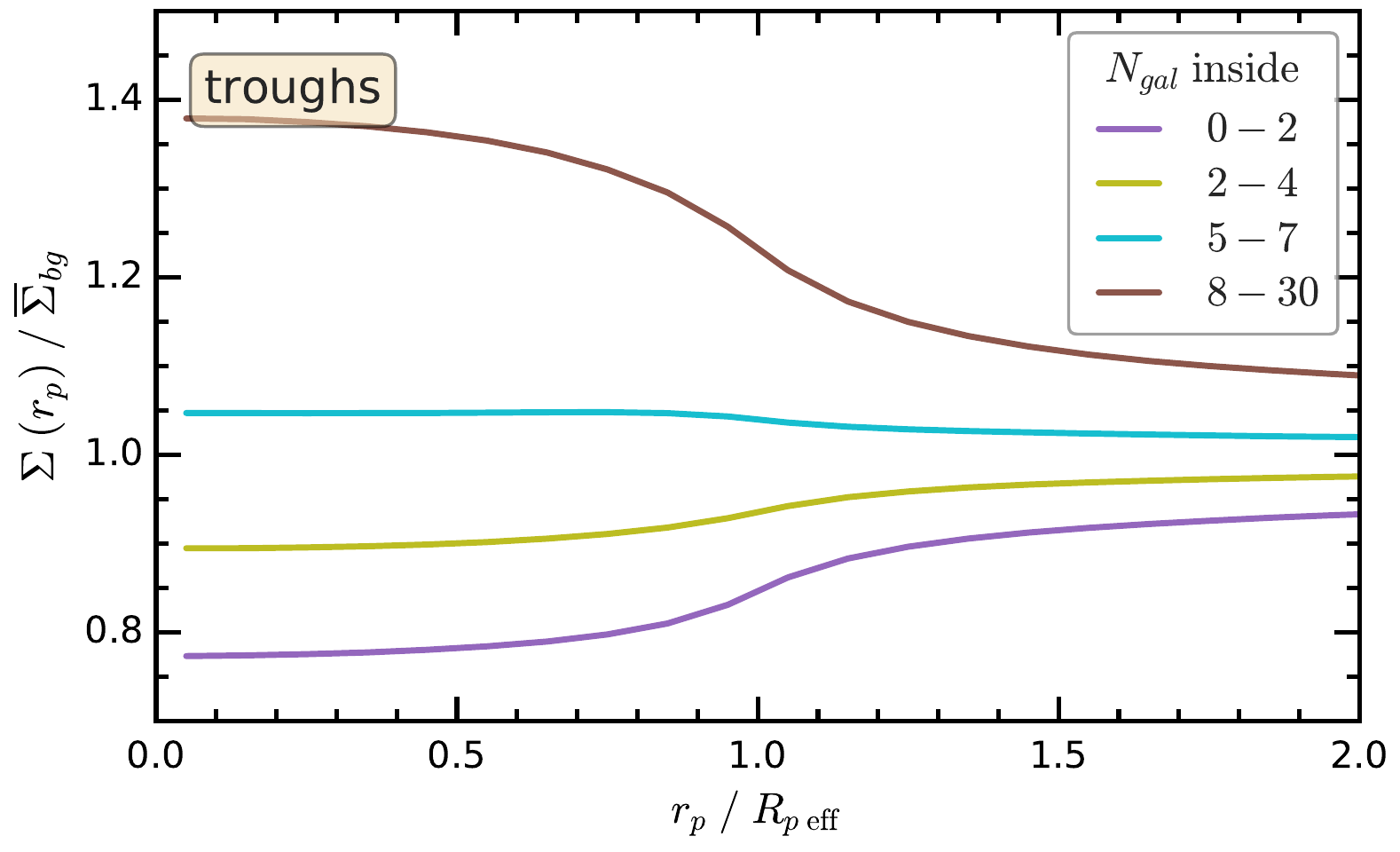}
     \end{tabular}
     \reduceVerticalOffset{}
     \caption{ The projected matter density, $\Sigma$, profile of \tunnels{} (top) and \troughs{} (bottom) at $z=0$. The \tunnels{} are split according to their effective projected radius, $\Rpeff{}$. We use only \tunnels{} with $\Rpeff \ge 1\Mpch$ that correspond to underdense regions. The \troughs{} are split according to the number of tracer galaxies they contain. We use only \troughs{} with $N_{\rm{gal}} \leq 2$ that correspond to a similar projected underdensity as the \tunnels{}. }
     \label{fig:DM_den_different_2D_radii}
     \reduceVerticalOffset{}
\end{figure}

\renewcommand\thefigure{B\arabic{figure}} 
\setcounter{figure}{0} 

\begin{figure*}
     \centering
     \begin{tabular}{cc}
        \includegraphics[width=\figwidthTwoTwo,angle=0]{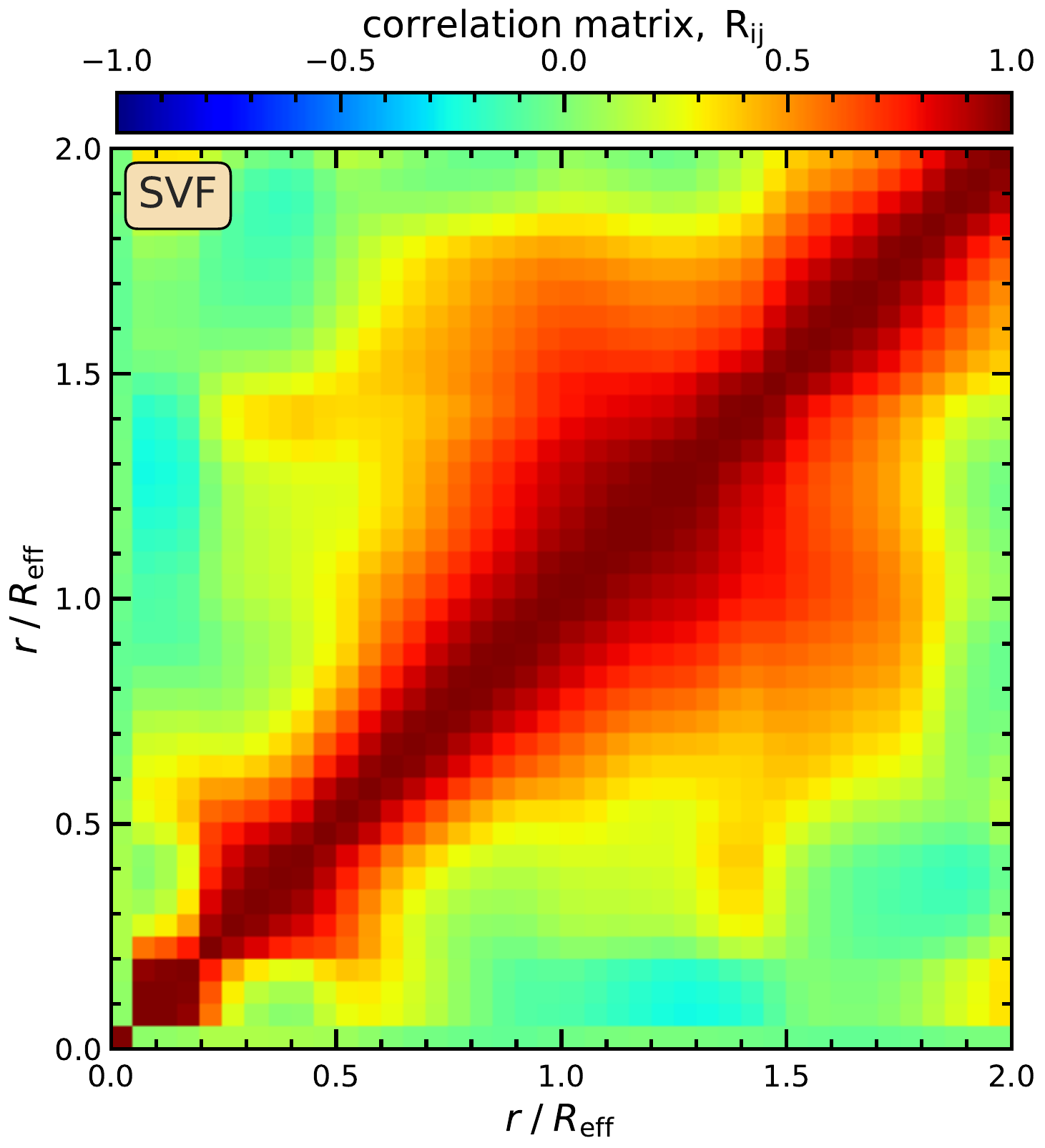} &
        \includegraphics[width=\figwidthTwoTwo,angle=0]{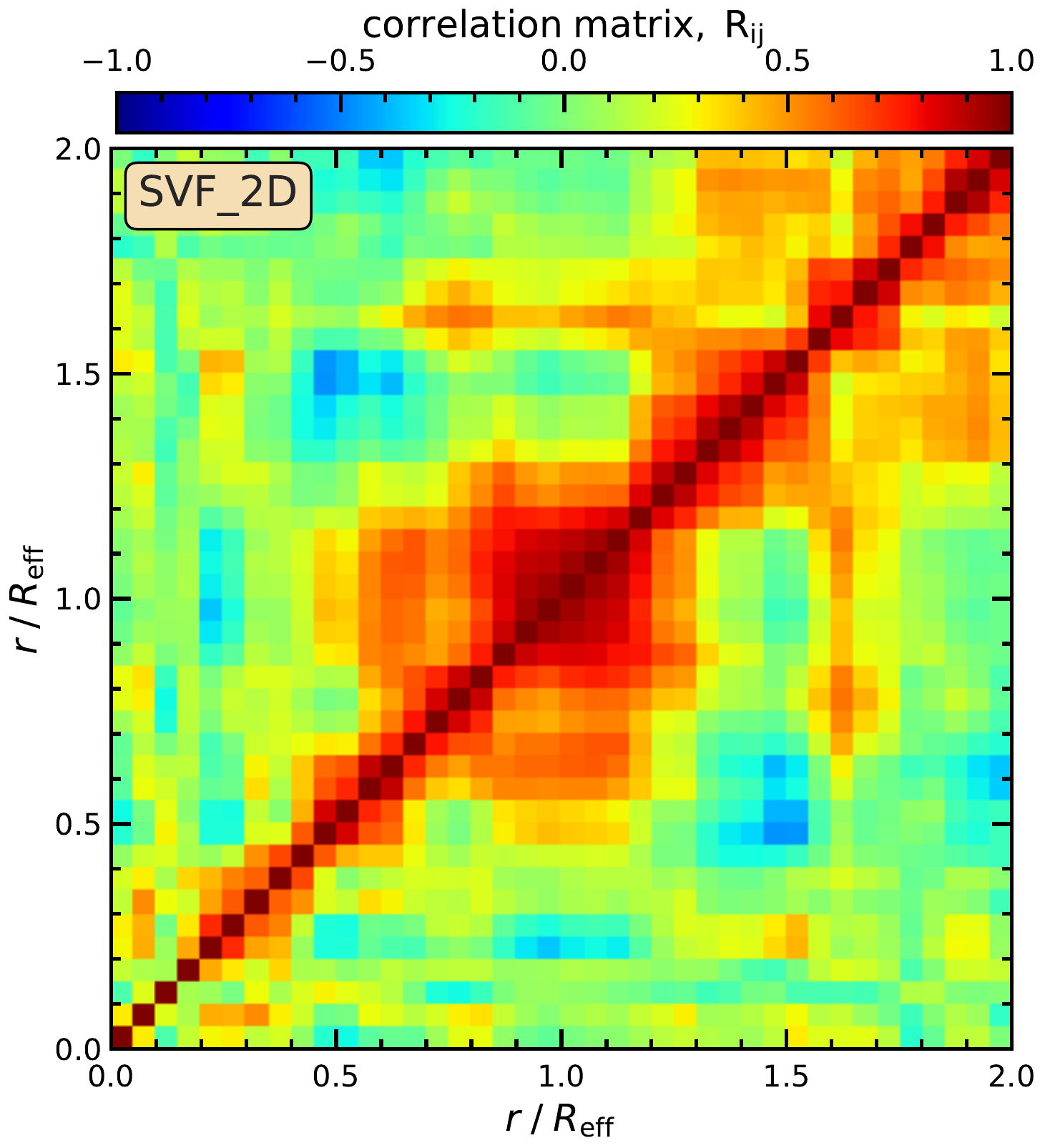} \verticaloffset
        \whitespaceTwoTwo & \whitespaceTwoTwo \verticaloffsetTwoTwo
        \includegraphics[width=\figwidthTwoTwo,angle=0]{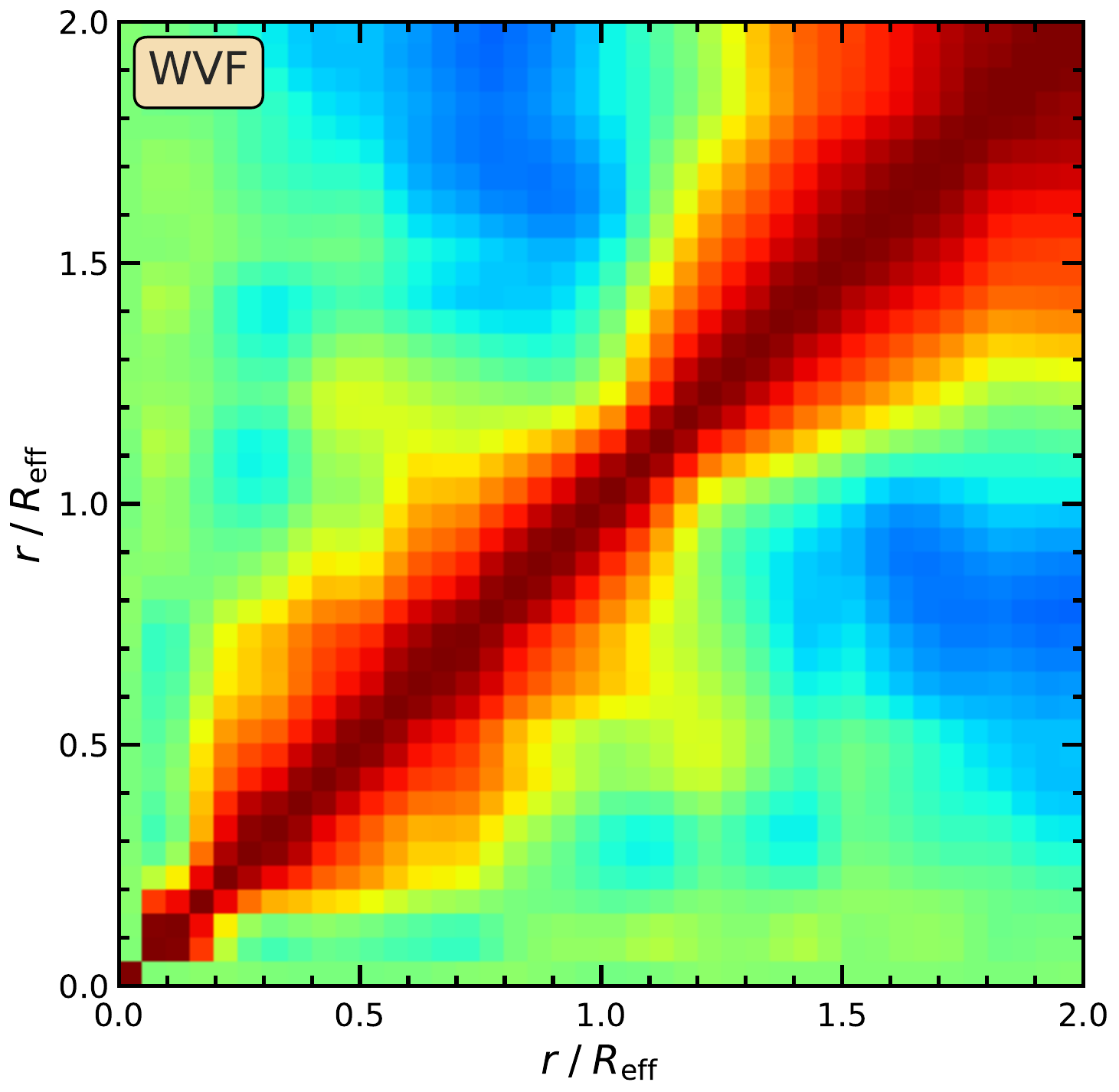} &
        \includegraphics[width=\figwidthTwoTwo,angle=0]{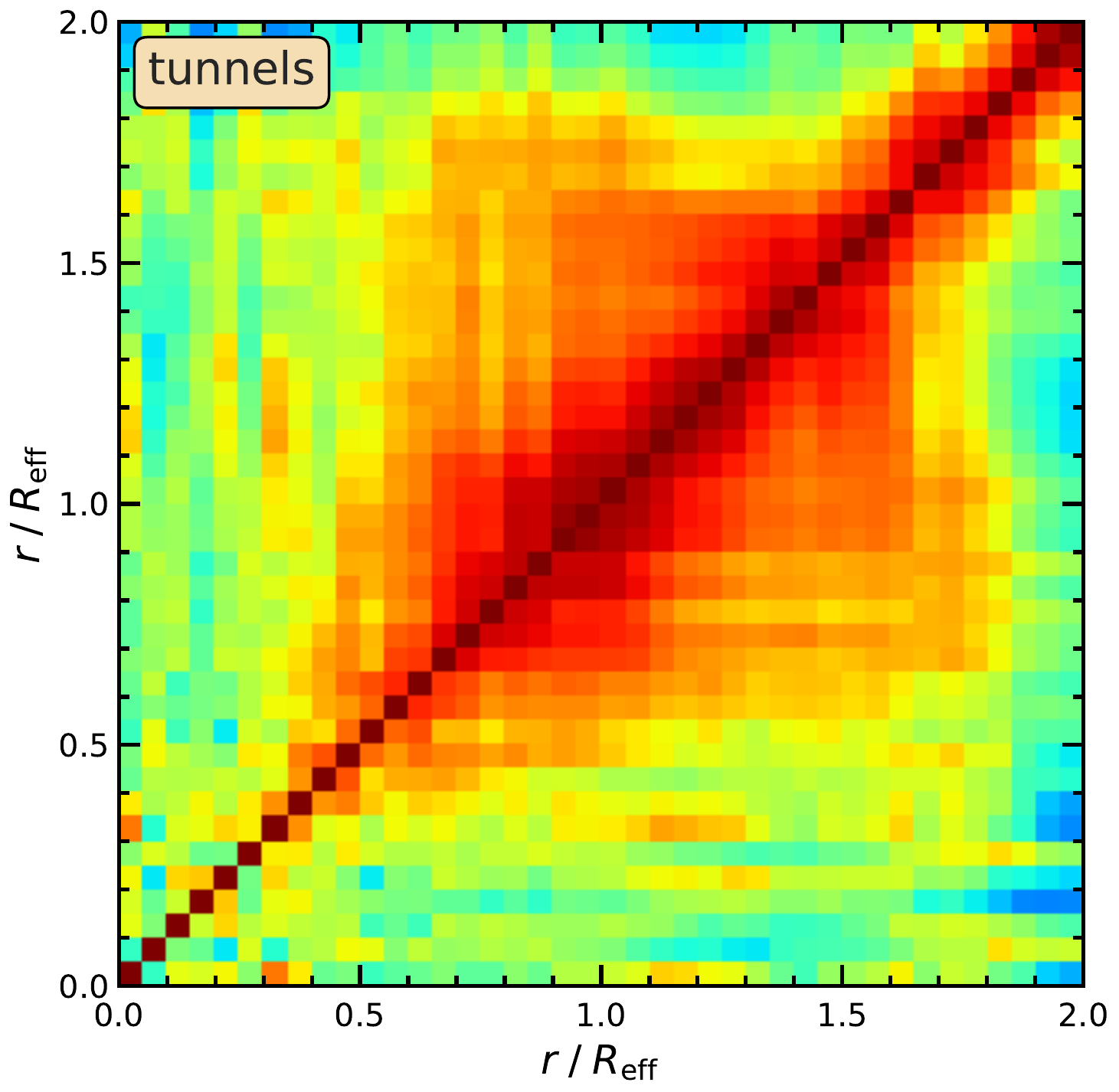} \verticaloffset
        \whitespaceTwoTwo & \whitespaceTwoTwo \verticaloffsetTwoTwo
        \includegraphics[width=\figwidthTwoTwo,angle=0]{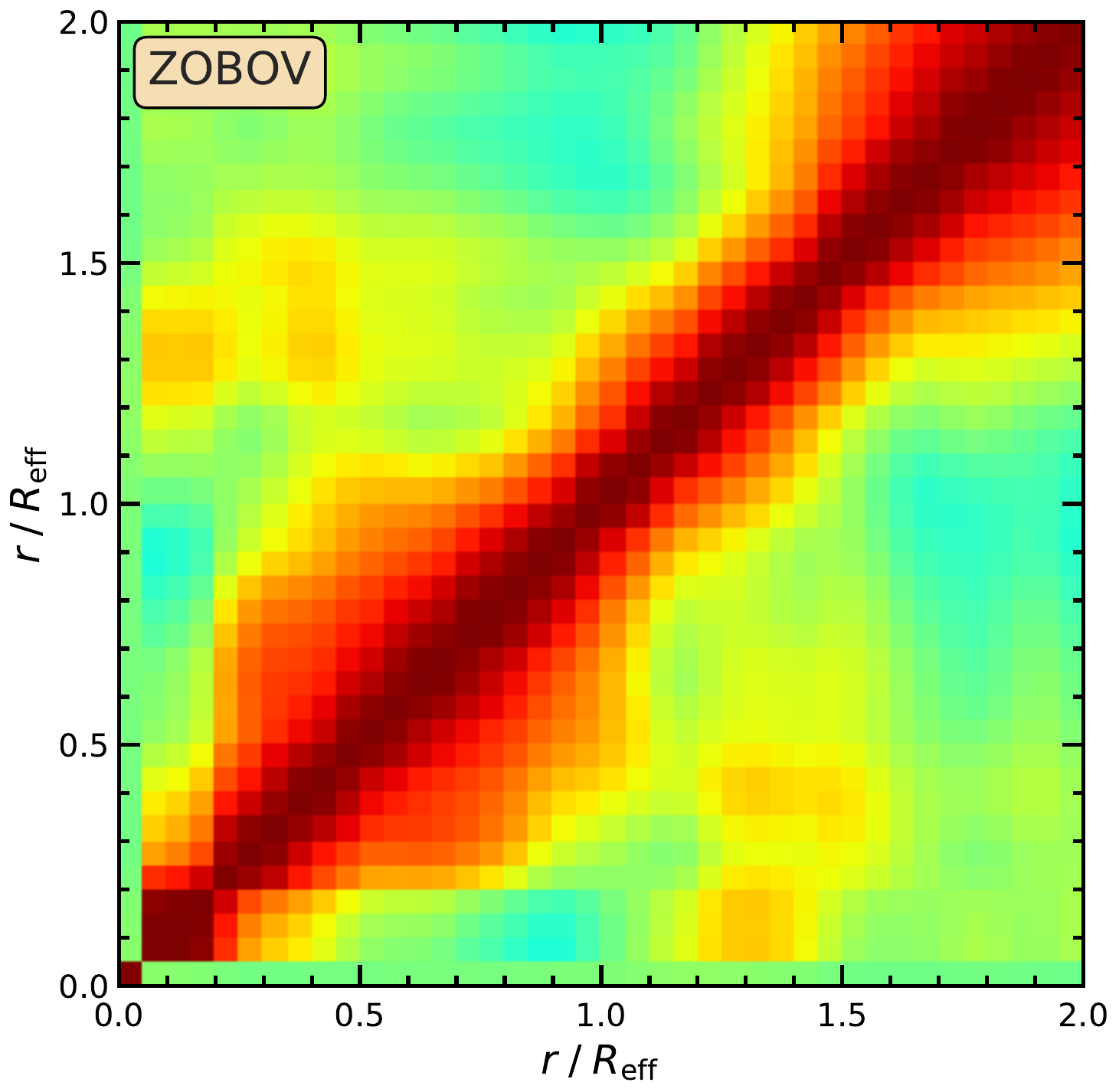} &
        \includegraphics[width=\figwidthTwoTwo,angle=0]{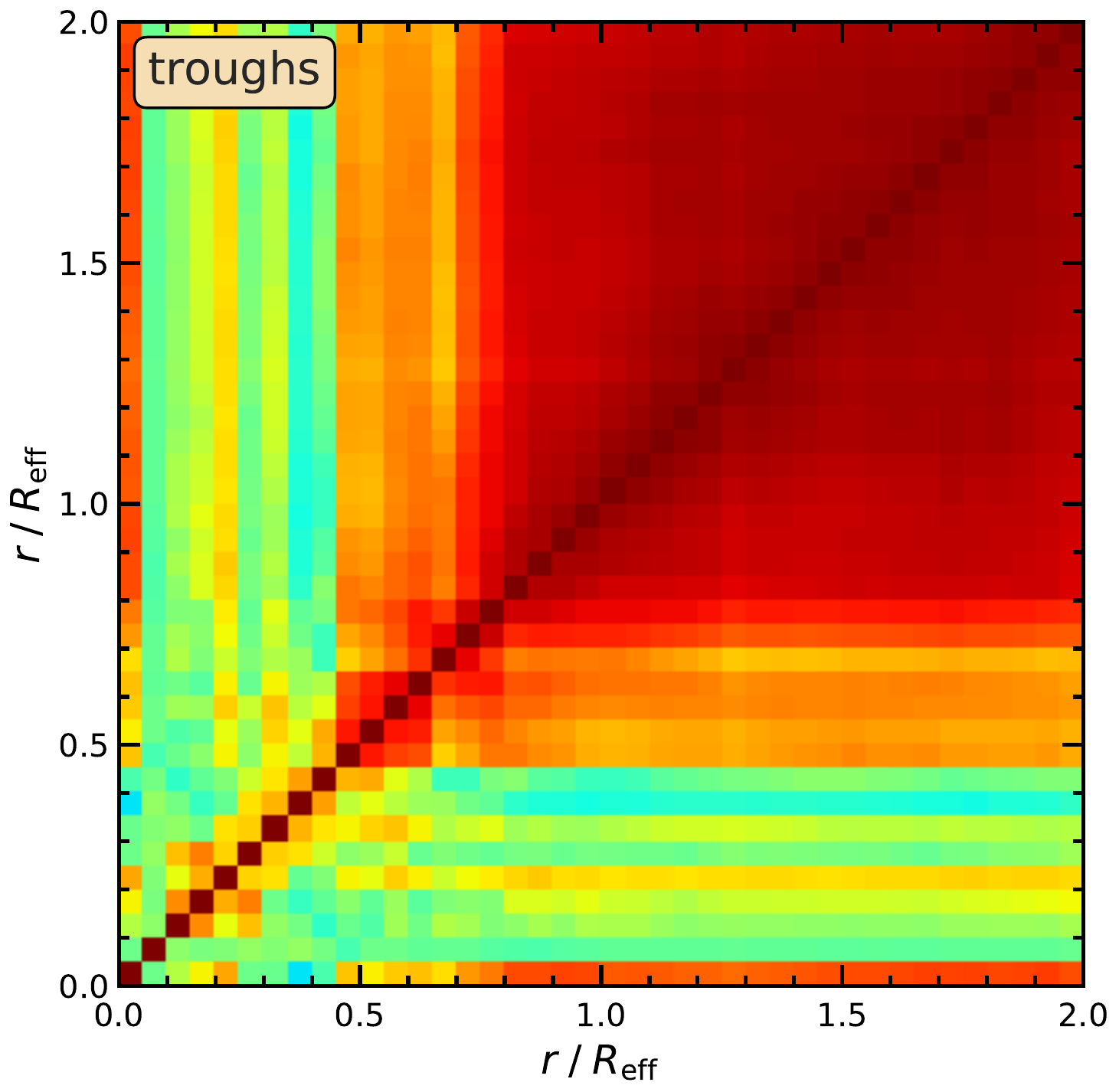}
     \end{tabular}
     \reduceVerticalOffset{}
     \caption{ The \changed{sample variance} correlation matrices, $R_{ij}$, of the differential surface mass density, $\Delta\Sigma$, for the six void finders used in this work. }
     \label{fig:covariance_matrix}
     \reduceVerticalOffset{}
\end{figure*}

\begin{figure}
     \centering
     \begin{tabular}{c}
        \includegraphics[width=.75\linewidth,angle=0]{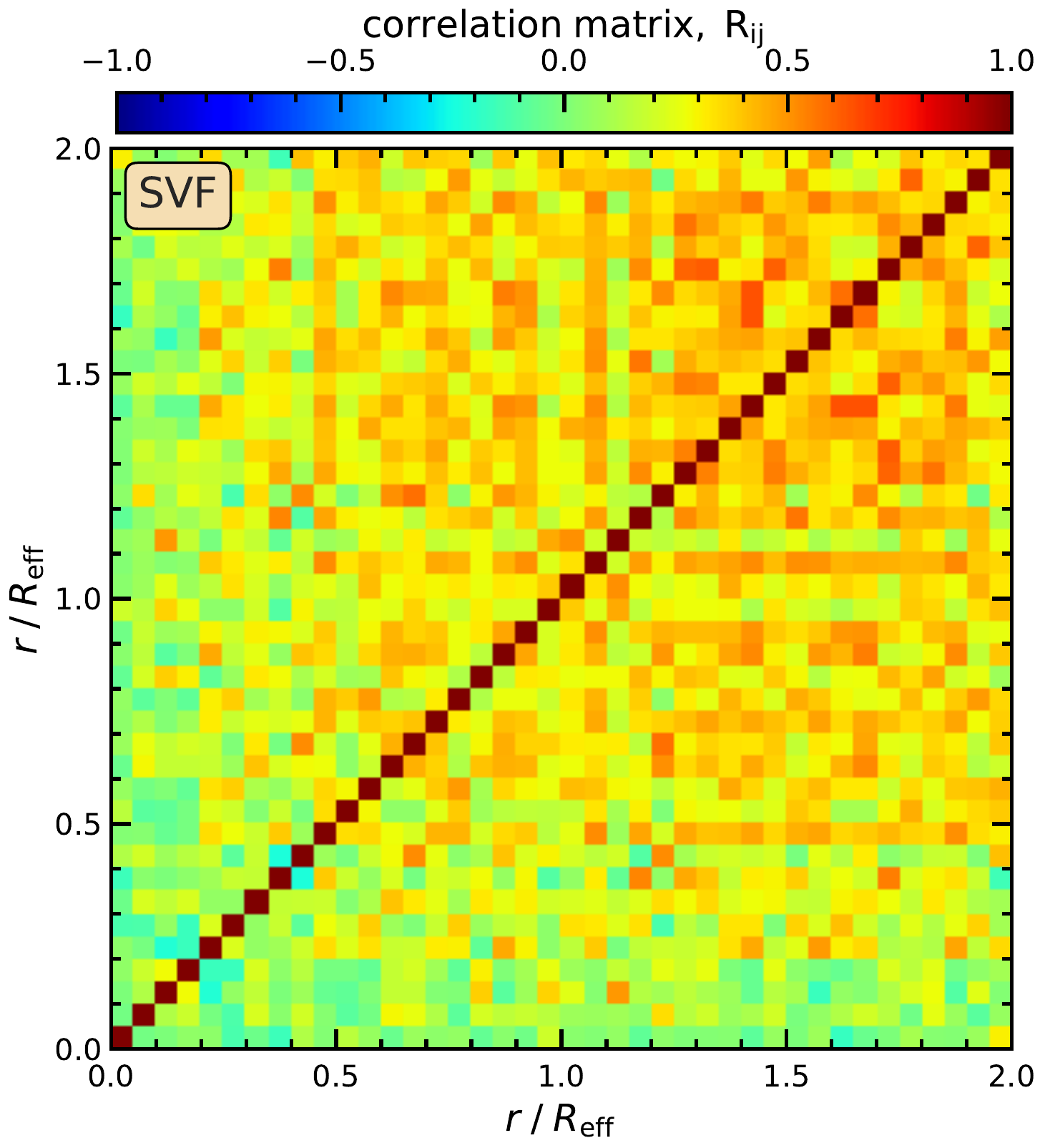} \verticaloffset
        \\[-.18cm]
        \includegraphics[width=.75\linewidth,angle=0]{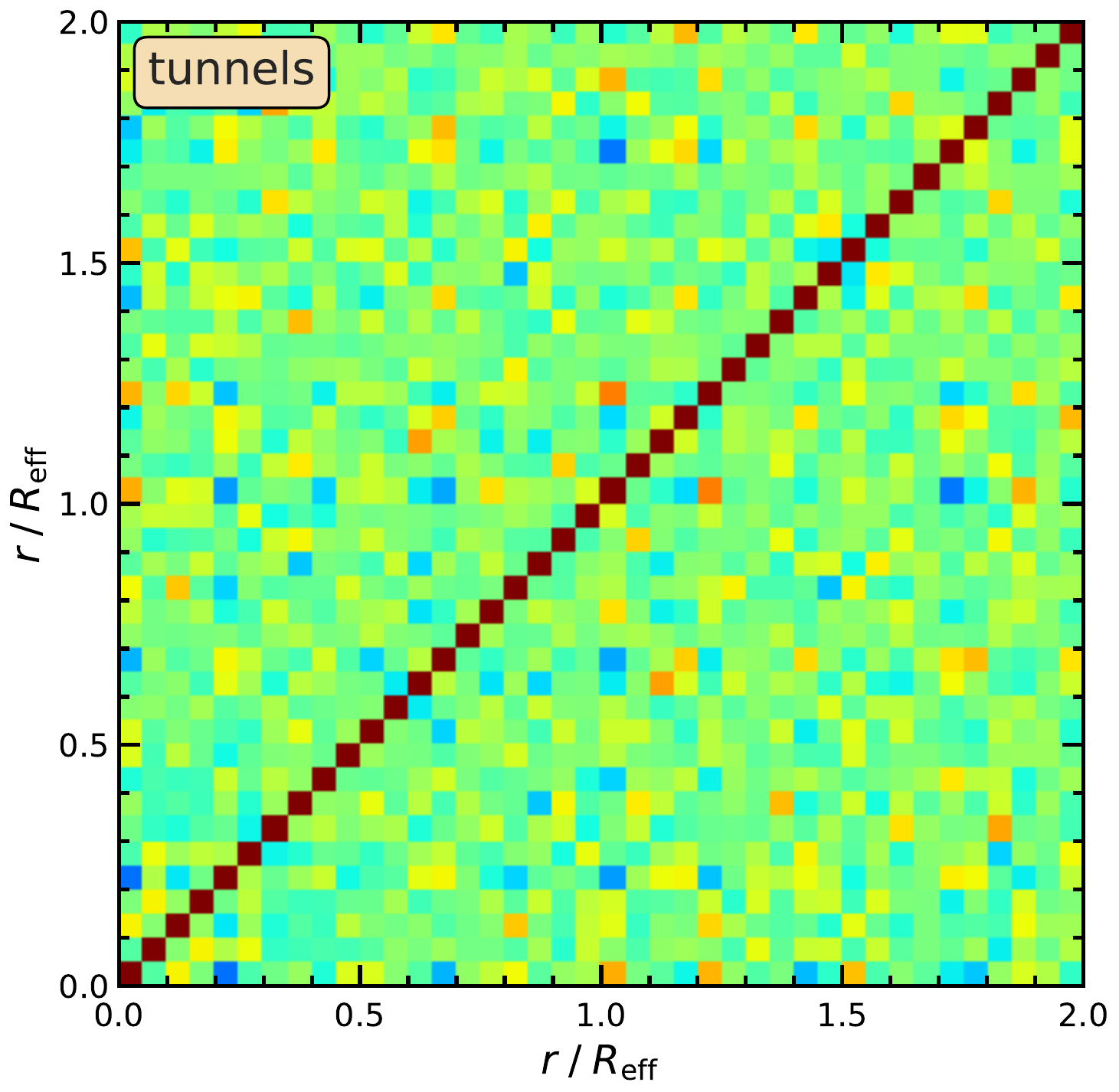} \verticaloffset
        \\[-.18cm]
        \includegraphics[width=.75\linewidth,angle=0]{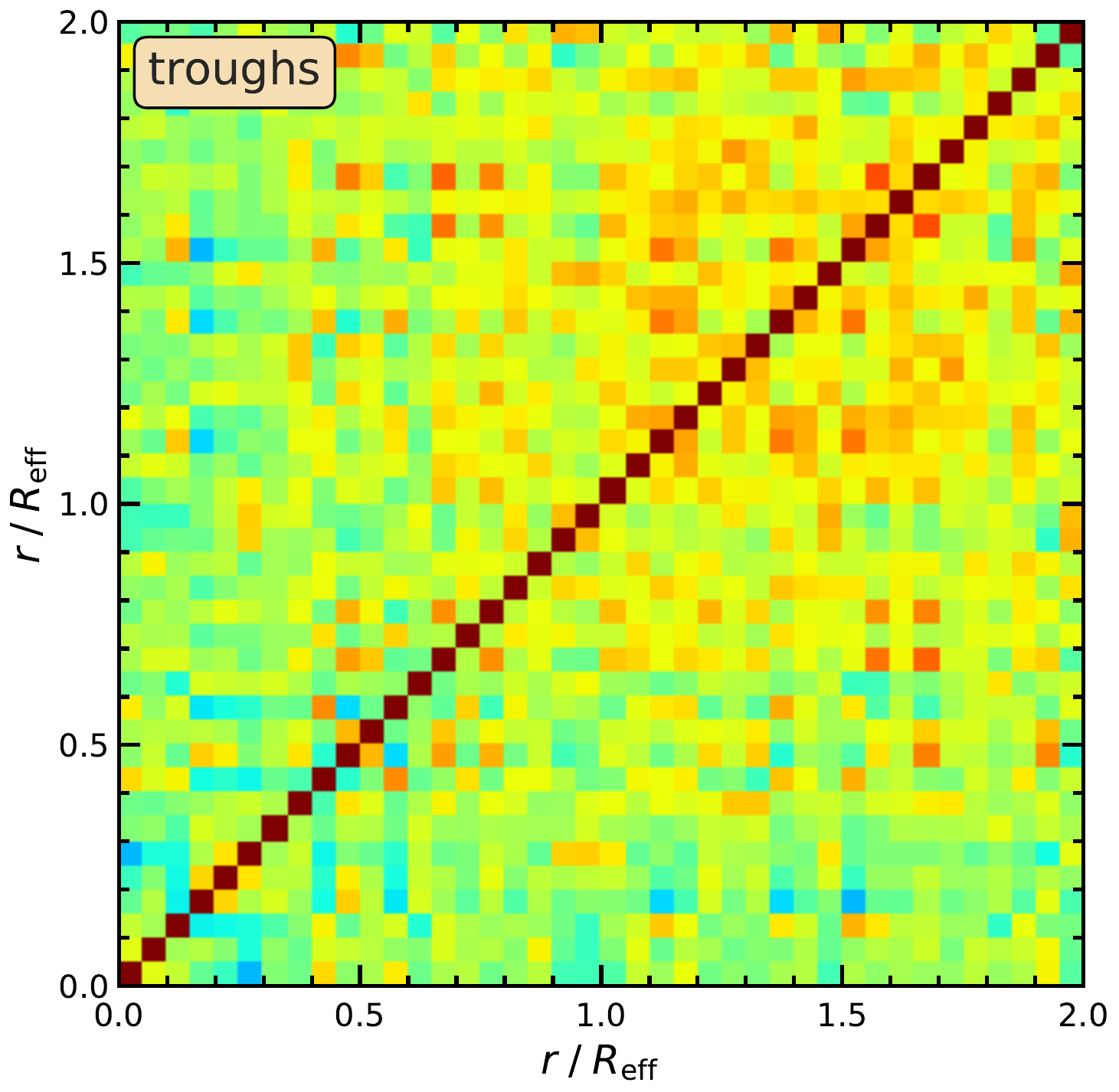}
     \end{tabular}
     \reduceVerticalOffset{}
     \caption{ \changed{The shape noise correlation matrices for three of the void finders used in this work. The correlation matrices shown here are: top panel for \SVF{} voids (we checked that all 3D voids have a similar matrix), central panel for tunnel (we checked that the \SVFtwoD{} result looks almost identical), and bottom panel for troughs.} }
     \label{fig:covariance_matrix_shape_noise}
     \reduceVerticalOffset{}
\end{figure}

We are interested in studying the signature of \FR{} modified gravity in underdense regions, but not all \tunnels{} and \troughs{} correspond to underdense regions. Here we present one simple method, which can be applied to observations, of segregating \tunnels{} and \troughs{} according to their projected density contrast, $\Sigma/\meanSigma$. For \tunnels{}, their radius is correlated with their projected density, as shown in the top panel of \reffig{fig:DM_den_different_2D_radii}. Large \tunnels{} correspond to regions sparsely populated with galaxies and thus have low $\Sigma$ values; in contrast small tunnels are found in regions crowded with galaxies, typically corresponding to high mass haloes. A simple cut in radius, $\Rpeff \ge 1\Mpch$, represents a good compromise between including as many objects as possible while still considering mainly underdense \tunnels{}.

All \troughs{} are defined to have the same radius, $2\Mpch$, so to discriminate between under- and overdense ones we use the number of galaxies inside each trough. Troughs that contain few galaxies are typically found in regions of low projected matter densities, as shown in the bottom panel of \reffig{fig:DM_den_different_2D_radii}. The four bins in trough galaxy count were selected to contain roughly $20$, $30$, $30$ and $20$ percent of the population, respectively. Our analysis uses only troughs that contain at most 2 galaxies inside them, which corresponds to $22$ and $30$ percent of objects at $z=0$ and $z=0.5$, respectively. The $z=0.5$ \HOD{} has a lower galaxy number density and thus the same cut in trough galaxy count corresponds to a larger fraction of the population. The threshold of 2 galaxy counts in troughs is motivated by finding a compromise between large differences in the $\Sigma$ profiles between \FR{} and \GR{}, which are maximal for the most underdense troughs, and including a sufficient fraction of the population.

\section{Lensing covariance matrix }
\label{appendix:cov_matrix}
\changed{Here we describe and present the tangential shear covariance matrix for each of the six voids used in our analysis. The void sample variance is estimated by first measuring the differential surface mass density, $\Delta\Sigma$. This is calculated by correlating the void distribution of each of the six void finders with the 2D matter distribution obtained from projecting the entire simulation box along a principal axis. In the case of the 2D voids, the principal axis of the projection is the same as the principal axis of the projection used to identify the voids in the first place. We estimate $\Delta\Sigma$ for $N=500$ realisations, which are obtained by using a mix of 5 simulation boxes with 100 bootstrap realizations for each box (see Sect.~\ref{sect:results} for more details). Then, the $(i,j)$ entry of the sample variance is given by
\begin{equation}
	{\rm cov}_{\Delta\Sigma}(i,j) = \frac{1}{N-1}\sum_{k=1}^{N} \left[ \Delta\Sigma_k(i)-\overline{\Delta\Sigma}(i) \right]   \left[ \Delta\Sigma_k(j)-\overline{\Delta\Sigma}(j) \right] 
    \label{eq:lensing_sample_variance} \;,
\end{equation}
where $\Delta\Sigma_k$ is the differential surface mass density for realization $k$ and $\overline{\Delta\Sigma}=\tfrac{1}{N}\sum\Delta\Sigma_k$ is the mean value of $\Delta\Sigma$ across all realizations.
}

\reffig{fig:covariance_matrix} shows the correlation matrix, $R_{ij}$, of $\Delta\Sigma$ for the voids studied here. The correlation matrix is given by 
\begin{equation}
	R_{ij} = \frac{ {\rm cov}_{\Delta\Sigma}(i,j) }{\sigma_i\sigma_j} \;,
\end{equation}
where ${\rm cov}_{\Delta\Sigma}(i,j)$ is the covariance matrix of $\Delta\Sigma$ and $\sigma^2_i=\rm{cov}_{\Delta\Sigma}(i,i)$, which are the diagonal entries of the covariance matrix. For most void finders, the correlation matrix is mainly diagonal suggesting that while there is a strong correlation between neighbouring bins, bins which are farther apart are only very weakly correlated. The only exception is for troughs, where we find that all bins with $r\gtrsim0.5\Reff$ show a large degree of correlation, independently of how far apart they are. This is probably a consequence of the high degree of overlap between different troughs (see \reffig{fig:images_3D_voids}).


\changed{The shape noise covariance was calculated by generating a spatially random catalogue of source galaxies whose number density and intrinsic ellipticity distribution matches that of the target lensing survey. Furthermore, the source galaxies are oriented randomly, which corresponds to a null tangential shear signal. We calculate the stacked mean galaxy ellipticity for each void catalogue by correlating the void distribution with the source galaxies. We obtain 500 bootstrap realizations which are then used to compute the shape noise covariance, which is shown in \reffig{fig:covariance_matrix_shape_noise}. For all void finders, we find that the diagonal terms are dominant. This is especially the case for tunnels and \SVFtwoD{} voids, which show the smallest degree of overlap (see \reffig{fig:images_3D_voids}). The remaining void catalogues have various degrees of overlap that results in many off-diagonal elements being non-zero.}


\renewcommand{\jcap}{JCAP}
\bibliographystyle{mnras}
\bibliography{references}

\begin{thebibliography}{}
\makeatletter
\relax
\def\mn@urlcharsother{\let\do\@makeother \do\$\do\&\do\#\do\^\do\_\do\%\do\~}
\def\mn@doi{\begingroup\mn@urlcharsother \@ifnextchar [ {\mn@doi@}
  {\mn@doi@[]}}
\def\mn@doi@[#1]#2{\def\@tempa{#1}\ifx\@tempa\@empty \href
  {http://dx.doi.org/#2} {doi:#2}\else \href {http://dx.doi.org/#2} {#1}\fi
  \endgroup}
\def\mn@eprint#1#2{\mn@eprint@#1:#2::\@nil}
\def\mn@eprint@arXiv#1{\href {http://arxiv.org/abs/#1} {{\tt arXiv:#1}}}
\def\mn@eprint@dblp#1{\href {http://dblp.uni-trier.de/rec/bibtex/#1.xml}
  {dblp:#1}}
\def\mn@eprint@#1:#2:#3:#4\@nil{\def\@tempa {#1}\def\@tempb {#2}\def\@tempc
  {#3}\ifx \@tempc \@empty \let \@tempc \@tempb \let \@tempb \@tempa \fi \ifx
  \@tempb \@empty \def\@tempb {arXiv}\fi \@ifundefined
  {mn@eprint@\@tempb}{\@tempb:\@tempc}{\expandafter \expandafter \csname
  mn@eprint@\@tempb\endcsname \expandafter{\@tempc}}}

\bibitem[\protect\citeauthoryear{{Achitouv}}{{Achitouv}}{2016}]{Achitouv2016}
{Achitouv} I.,  2016, \mn@doi [\prd] {10.1103/PhysRevD.94.103524}, \href
  {http://adsabs.harvard.edu/abs/2016PhRvD..94j3524A} {94, 103524}

\bibitem[\protect\citeauthoryear{{Anderson}}{{Anderson}}{2003}]{Anderson2003}
{Anderson} T.~W.,  2003, Wiley-Interscience

\bibitem[\protect\citeauthoryear{{Anderson} et~al.,}{{Anderson}
  et~al.}{2012}]{Anderson2012}
{Anderson} L.,  et~al., 2012, \mn@doi [\mnras]
  {10.1111/j.1365-2966.2012.22066.x}, \href
  {http://adsabs.harvard.edu/abs/2012MNRAS.427.3435A} {427, 3435}

\bibitem[\protect\citeauthoryear{{Banerjee} \& {Dalal}}{{Banerjee} \&
  {Dalal}}{2016}]{Banerjee2016}
{Banerjee} A.,  {Dalal} N.,  2016, \mn@doi [\jcap]
  {10.1088/1475-7516/2016/11/015}, \href
  {http://adsabs.harvard.edu/abs/2016JCAP...11..015B} {11, 015}

\bibitem[\protect\citeauthoryear{Barreira, Cautun, Li, Baugh  \&
  Pascoli}{Barreira et~al.}{2015}]{Barreira2015}
Barreira A.,  Cautun M.,  Li B.,  Baugh C.,   Pascoli S.,  2015, \mn@doi
  [\jcap] {10.1088/1475-7516/2015/08/028}, 1508, 028

\bibitem[\protect\citeauthoryear{{Barreira}, {Bose}, {Li}  \&
  {Llinares}}{{Barreira} et~al.}{2017}]{Barreira2017a}
{Barreira} A.,  {Bose} S.,  {Li} B.,   {Llinares} C.,  2017, \mn@doi [\jcap]
  {10.1088/1475-7516/2017/02/031}, \href
  {http://adsabs.harvard.edu/abs/2017JCAP...02..031B} {2, 031}

\bibitem[\protect\citeauthoryear{{Behroozi}, {Wechsler}  \& {Wu}}{{Behroozi}
  et~al.}{2013}]{Behroozi2013}
{Behroozi} P.~S.,  {Wechsler} R.~H.,   {Wu} H.-Y.,  2013, \mn@doi [\apj]
  {10.1088/0004-637X/762/2/109}, \href
  {http://adsabs.harvard.edu/abs/2013ApJ...762..109B} {762, 109}

\bibitem[\protect\citeauthoryear{{Benson}, {Cole}, {Frenk}, {Baugh}  \&
  {Lacey}}{{Benson} et~al.}{2000}]{Benson2000}
{Benson} A.~J.,  {Cole} S.,  {Frenk} C.~S.,  {Baugh} C.~M.,   {Lacey} C.~G.,
  2000, \mn@doi [\mnras] {10.1046/j.1365-8711.2000.03101.x}, \href
  {http://adsabs.harvard.edu/abs/2000MNRAS.311..793B} {311, 793}

\bibitem[\protect\citeauthoryear{{Berlind} \& {Weinberg}}{{Berlind} \&
  {Weinberg}}{2002}]{Berlind2002}
{Berlind} A.~A.,  {Weinberg} D.~H.,  2002, \mn@doi [\apj] {10.1086/341469},
  \href {http://adsabs.harvard.edu/abs/2002ApJ...575..587B} {575, 587}

\bibitem[\protect\citeauthoryear{{Blumenthal}, {da Costa}, {Goldwirth}, {Lecar}
   \& {Piran}}{{Blumenthal} et~al.}{1992}]{Blumenthal1992}
{Blumenthal} G.~R.,  {da Costa} L.~N.,  {Goldwirth} D.~S.,  {Lecar} M.,
  {Piran} T.,  1992, \mn@doi [\apj] {10.1086/171147}, \href
  {http://adsabs.harvard.edu/abs/1992ApJ...388..234B} {388, 234}

\bibitem[\protect\citeauthoryear{{Bond}, {Kofman}  \& {Pogosyan}}{{Bond}
  et~al.}{1996}]{Bond1996}
{Bond} J.~R.,  {Kofman} L.,   {Pogosyan} D.,  1996, \mn@doi [\nat]
  {10.1038/380603a0}, \href {http://adsabs.harvard.edu/abs/1996Natur.380..603B}
  {380, 603}

\bibitem[\protect\citeauthoryear{{Bos}, {van de Weygaert}, {Dolag}  \&
  {Pettorino}}{{Bos} et~al.}{2012}]{Bos2012}
{Bos} E.~G.~P.,  {van de Weygaert} R.,  {Dolag} K.,   {Pettorino} V.,  2012,
  \mn@doi [\mnras] {10.1111/j.1365-2966.2012.21478.x}, \href
  {http://adsabs.harvard.edu/abs/2012MNRAS.426..440B} {426, 440}

\bibitem[\protect\citeauthoryear{{Bose}, {Hellwing}  \& {Li}}{{Bose}
  et~al.}{2015}]{Bose2015}
{Bose} S.,  {Hellwing} W.~A.,   {Li} B.,  2015, \mn@doi [\jcap]
  {10.1088/1475-7516/2015/02/034}, \href
  {http://adsabs.harvard.edu/abs/2015JCAP...02..034B} {2, 034}

\bibitem[\protect\citeauthoryear{{Bose}, {Li}, {Barreira}, {He}, {Hellwing},
  {Koyama}, {Llinares}  \& {Zhao}}{{Bose} et~al.}{2017}]{Bose:2016wms}
{Bose} S.,  {Li} B.,  {Barreira} A.,  {He} J.-h.,  {Hellwing} W.~A.,  {Koyama}
  K.,  {Llinares} C.,   {Zhao} G.-B.,  2017, \mn@doi [\jcap]
  {10.1088/1475-7516/2017/02/050}, \href
  {http://adsabs.harvard.edu/abs/2017JCAP...02..050B} {2, 050}

\bibitem[\protect\citeauthoryear{{Cai}, {Padilla}  \& {Li}}{{Cai}
  et~al.}{2015}]{Cai2014}
{Cai} Y.-C.,  {Padilla} N.,   {Li} B.,  2015, \mn@doi [\mnras]
  {10.1093/mnras/stv777}, \href
  {http://adsabs.harvard.edu/abs/2015MNRAS.451.1036C} {451, 1036}

\bibitem[\protect\citeauthoryear{Cai, Taylor, Peacock  \& Padilla}{Cai
  et~al.}{2016}]{Cai2016}
Cai Y.-C.,  Taylor A.,  Peacock J.~A.,   Padilla N.,  2016, \mn@doi [\mnras]
  {10.1093/mnras/stw1809}, 462, 2465

\bibitem[\protect\citeauthoryear{{Cai}, {Neyrinck}, {Mao}, {Peacock}, {Szapudi}
   \& {Berlind}}{{Cai} et~al.}{2017}]{Cai2017}
{Cai} Y.-C.,  {Neyrinck} M.,  {Mao} Q.,  {Peacock} J.~A.,  {Szapudi} I.,
  {Berlind} A.~A.,  2017, \mn@doi [\mnras] {10.1093/mnras/stw3299}, \href
  {http://adsabs.harvard.edu/abs/2017MNRAS.466.3364C} {466, 3364}

\bibitem[\protect\citeauthoryear{{Cautun} \& {van de Weygaert}}{{Cautun} \&
  {van de Weygaert}}{2011}]{Cautun2011}
{Cautun} M.~C.,  {van de Weygaert} R.,  2011, preprint, \href
  {http://adsabs.harvard.edu/abs/2011arXiv1105.0370C} {} (\mn@eprint {arXiv}
  {1105.0370})

\bibitem[\protect\citeauthoryear{{Cautun}, {van de Weygaert}  \&
  {Jones}}{{Cautun} et~al.}{2013}]{Cautun2013}
{Cautun} M.,  {van de Weygaert} R.,   {Jones} B.~J.~T.,  2013, \mn@doi [MNRAS]
  {10.1093/mnras/sts416}, \href
  {http://adsabs.harvard.edu/abs/2013MNRAS.429.1286C} {429, 1286}

\bibitem[\protect\citeauthoryear{{Cautun}, {van de Weygaert}, {Jones}  \&
  {Frenk}}{{Cautun} et~al.}{2014}]{Cautun2014a}
{Cautun} M.,  {van de Weygaert} R.,  {Jones} B.~J.~T.,   {Frenk} C.~S.,  2014,
  \mn@doi [MNRAS] {10.1093/mnras/stu768}, \href
  {http://adsabs.harvard.edu/abs/2014MNRAS.441.2923C} {441, 2923}

\bibitem[\protect\citeauthoryear{{Cautun}, {Cai}  \& {Frenk}}{{Cautun}
  et~al.}{2016}]{Cautun2016a}
{Cautun} M.,  {Cai} Y.-C.,   {Frenk} C.~S.,  2016, \mn@doi [\mnras]
  {10.1093/mnras/stw154}, \href
  {http://adsabs.harvard.edu/abs/2016MNRAS.457.2540C} {457, 2540}

\bibitem[\protect\citeauthoryear{{Clampitt} \& {Jain}}{{Clampitt} \&
  {Jain}}{2015}]{Clampitt2014}
{Clampitt} J.,  {Jain} B.,  2015, \mn@doi [\mnras] {10.1093/mnras/stv2215},
  \href {http://adsabs.harvard.edu/abs/2015MNRAS.454.3357C} {454, 3357}

\bibitem[\protect\citeauthoryear{Clampitt, Cai  \& Li}{Clampitt
  et~al.}{2013}]{Clampitt2013}
Clampitt J.,  Cai Y.-C.,   Li B.,  2013, \mn@doi [\mnras]
  {10.1093/mnras/stt219}, 431, 749

\bibitem[\protect\citeauthoryear{{Colberg} et~al.,}{{Colberg}
  et~al.}{2008}]{Colberg2008}
{Colberg} J.~M.,  et~al., 2008, \mn@doi [\mnras]
  {10.1111/j.1365-2966.2008.13307.x}, \href
  {http://adsabs.harvard.edu/abs/2008MNRAS.387..933C} {387, 933}

\bibitem[\protect\citeauthoryear{{Demchenko}, {Cai}, {Heymans}  \&
  {Peacock}}{{Demchenko} et~al.}{2016}]{Demchenko2016}
{Demchenko} V.,  {Cai} Y.-C.,  {Heymans} C.,   {Peacock} J.~A.,  2016, \mn@doi
  [\mnras] {10.1093/mnras/stw2030}, \href
  {http://adsabs.harvard.edu/abs/2016MNRAS.463..512D} {463, 512}

\bibitem[\protect\citeauthoryear{Erickcek, Barnaby, Burrage  \& Huang}{Erickcek
  et~al.}{2013}]{Erickcek:2013oma}
Erickcek A.~L.,  Barnaby N.,  Burrage C.,   Huang Z.,  2013, \mn@doi [Phys.
  Rev. Lett.] {10.1103/PhysRevLett.110.171101}, 110, 171101

\bibitem[\protect\citeauthoryear{{Falck}, {Koyama}, {Zhao}  \&
  {Cautun}}{{Falck} et~al.}{2017}]{Falck2017}
{Falck} B.,  {Koyama} K.,  {Zhao} G.,   {Cautun} M.,  2017, preprint, \href
  {http://adsabs.harvard.edu/abs/2017arXiv170408942F} {} (\mn@eprint {arXiv}
  {1704.08942})

\bibitem[\protect\citeauthoryear{{Gruen} et~al.,}{{Gruen}
  et~al.}{2016}]{Gruen2015}
{Gruen} D.,  et~al., 2016, \mn@doi [\mnras] {10.1093/mnras/stv2506}, \href
  {http://adsabs.harvard.edu/abs/2016MNRAS.455.3367G} {455, 3367}

\bibitem[\protect\citeauthoryear{Hamaus, Sutter, Lavaux  \& Wandelt}{Hamaus
  et~al.}{2015}]{Hamaus2015}
Hamaus N.,  Sutter P.~M.,  Lavaux G.,   Wandelt B.~D.,  2015, \mn@doi [JCAP]
  {10.1088/1475-7516/2015/11/036}, 1511, 036

\bibitem[\protect\citeauthoryear{Hamaus, Pisani, Sutter, Lavaux, Escoffier,
  Wandelt  \& Weller}{Hamaus et~al.}{2016}]{Hamaus2016}
Hamaus N.,  Pisani A.,  Sutter P.~M.,  Lavaux G.,  Escoffier S.,  Wandelt
  B.~D.,   Weller J.,  2016, \mn@doi [Phys. Rev. Lett.]
  {10.1103/PhysRevLett.117.091302}, 117, 091302

\bibitem[\protect\citeauthoryear{{Hartlap}, {Simon}  \& {Schneider}}{{Hartlap}
  et~al.}{2007}]{Hartlap2007}
{Hartlap} J.,  {Simon} P.,   {Schneider} P.,  2007, \mn@doi [\aap]
  {10.1051/0004-6361:20066170}, \href
  {http://adsabs.harvard.edu/abs/2007A%26A...464..399H} {464, 399}

\bibitem[\protect\citeauthoryear{{Hinshaw} et~al.,}{{Hinshaw}
  et~al.}{2013}]{WMAP9}
{Hinshaw} G.,  et~al., 2013, \mn@doi [\apjs] {10.1088/0067-0049/208/2/19},
  \href {http://adsabs.harvard.edu/abs/2013ApJS..208...19H} {208, 19}

\bibitem[\protect\citeauthoryear{{Hu} \& {Sawicki}}{{Hu} \&
  {Sawicki}}{2007}]{Hu2007}
{Hu} W.,  {Sawicki} I.,  2007, \mn@doi [\prd] {10.1103/PhysRevD.76.064004},
  \href {http://adsabs.harvard.edu/abs/2007PhRvD..76f4004H} {76, 064004}

\bibitem[\protect\citeauthoryear{{Jennings}, {Li}  \& {Hu}}{{Jennings}
  et~al.}{2013}]{Jennings2013}
{Jennings} E.,  {Li} Y.,   {Hu} W.,  2013, \mn@doi [\mnras]
  {10.1093/mnras/stt1169}, \href
  {http://adsabs.harvard.edu/abs/2013MNRAS.434.2167J} {434, 2167}

\bibitem[\protect\citeauthoryear{Joyce, Jain, Khoury  \& Trodden}{Joyce
  et~al.}{2015}]{Joyce2014}
Joyce A.,  Jain B.,  Khoury J.,   Trodden M.,  2015, \mn@doi [Phys. Rept.]
  {10.1016/j.physrep.2014.12.002}, 568, 1

\bibitem[\protect\citeauthoryear{{Khoury} \& {Weltman}}{{Khoury} \&
  {Weltman}}{2004}]{Khoury2004}
{Khoury} J.,  {Weltman} A.,  2004, \mn@doi [\prd] {10.1103/PhysRevD.69.044026},
  \href {http://adsabs.harvard.edu/abs/2004PhRvD..69d4026K} {69, 044026}

\bibitem[\protect\citeauthoryear{{Koyama}}{{Koyama}}{2016}]{Koyama2016}
{Koyama} K.,  2016, \mn@doi [Rep. Prog. Phys.] {10.1088/0034-4885/79/4/046902},
  \href {http://adsabs.harvard.edu/abs/2016RPPh...79d6902K} {79, 046902}

\bibitem[\protect\citeauthoryear{{Krause}, {Chang}, {Dor{\'e}}  \&
  {Umetsu}}{{Krause} et~al.}{2013}]{Krause2013}
{Krause} E.,  {Chang} T.-C.,  {Dor{\'e}} O.,   {Umetsu} K.,  2013, \mn@doi
  [\apjl] {10.1088/2041-8205/762/2/L20}, \href
  {http://adsabs.harvard.edu/abs/2013ApJ...762L..20K} {762, L20}

\bibitem[\protect\citeauthoryear{{Kravtsov}, {Berlind}, {Wechsler}, {Klypin},
  {Gottl{\"o}ber}, {Allgood}  \& {Primack}}{{Kravtsov}
  et~al.}{2004}]{Kravtsov2004}
{Kravtsov} A.~V.,  {Berlind} A.~A.,  {Wechsler} R.~H.,  {Klypin} A.~A.,
  {Gottl{\"o}ber} S.,  {Allgood} B.,   {Primack} J.~R.,  2004, \mn@doi [\apj]
  {10.1086/420959}, \href {http://adsabs.harvard.edu/abs/2004ApJ...609...35K}
  {609, 35}

\bibitem[\protect\citeauthoryear{{LSST Science Collaboration} et~al.,}{{LSST
  Science Collaboration} et~al.}{2009}]{lsst}
{LSST Science Collaboration} et~al., 2009, preprint, \href
  {http://adsabs.harvard.edu/abs/2009arXiv0912.0201L} {} (\mn@eprint {arXiv}
  {0912.0201})

\bibitem[\protect\citeauthoryear{Lam, Clampitt, Cai  \& Li}{Lam
  et~al.}{2015}]{Lam2014}
Lam T.~Y.,  Clampitt J.,  Cai Y.-C.,   Li B.,  2015, \mn@doi [\mnras]
  {10.1093/mnras/stv797}, 450, 3319

\bibitem[\protect\citeauthoryear{Laureijs, Amiaux, Arduini, Auguères,
  Brinchmann, Cole  et~al.}{Laureijs et~al.}{2011}]{euclid}
Laureijs R.,  Amiaux J.,  Arduini S.,  Auguères J.~L.,  Brinchmann J.,  Cole
  R.,   et~al., 2011, preprint (\mn@eprint {arXiv} {1110.3193})

\bibitem[\protect\citeauthoryear{{Li}}{{Li}}{2011}]{Li2011}
{Li} B.,  2011, \mn@doi [\mnras] {10.1111/j.1365-2966.2010.17867.x}, \href
  {http://adsabs.harvard.edu/cgi-bin/bib_query?arXiv:1009.1406} {411, 2615}

\bibitem[\protect\citeauthoryear{{Li} \& {Shirasaki}}{{Li} \&
  {Shirasaki}}{2017}]{Li2017}
{Li} B.,  {Shirasaki} M.,  2017, preprint, \href
  {http://adsabs.harvard.edu/abs/2017arXiv171007291L} {} (\mn@eprint {arXiv}
  {1710.07291})

\bibitem[\protect\citeauthoryear{{Li} \& {Zhao}}{{Li} \& {Zhao}}{2010}]{Li2010}
{Li} B.,  {Zhao} H.,  2010, \mn@doi [\prd] {10.1103/PhysRevD.81.104047}, \href
  {http://adsabs.harvard.edu/abs/2010PhRvD..81j4047L} {81, 104047}

\bibitem[\protect\citeauthoryear{{Li}, {Zhao}, {Teyssier}  \& {Koyama}}{{Li}
  et~al.}{2012a}]{ecosmog}
{Li} B.,  {Zhao} G.,  {Teyssier} R.,   {Koyama} K.,  2012a, \mn@doi [\jcap]
  {10.1088/1475-7516/2012/01/051}, \href
  {http://adsabs.harvard.edu/abs/2012JCAP...01..051L} {1, 051}

\bibitem[\protect\citeauthoryear{Li, Zhao  \& Koyama}{Li
  et~al.}{2012b}]{Li2012}
Li B.,  Zhao G.,   Koyama K.,  2012b, \mn@doi [\mnras]
  {10.1111/j.1365-2966.2012.20573.x}, \href
  {http://adsabs.harvard.edu/cgi-bin/bib_query?arXiv:1111.2602} {421, 3481}

\bibitem[\protect\citeauthoryear{{Lombriser}}{{Lombriser}}{2014}]{Lombriser2014}
{Lombriser} L.,  2014, \mn@doi [Annalen der Physik] {10.1002/andp.201400058},
  \href {http://adsabs.harvard.edu/abs/2014AnP...526..259L} {526, 259}

\bibitem[\protect\citeauthoryear{{Manera} et~al.,}{{Manera}
  et~al.}{2013}]{Manera2013}
{Manera} M.,  et~al., 2013, \mn@doi [\mnras] {10.1093/mnras/sts084}, \href
  {http://adsabs.harvard.edu/abs/2013MNRAS.428.1036M} {428, 1036}

\bibitem[\protect\citeauthoryear{{Massara}, {Villaescusa-Navarro}, {Viel}  \&
  {Sutter}}{{Massara} et~al.}{2015}]{Massara2015}
{Massara} E.,  {Villaescusa-Navarro} F.,  {Viel} M.,   {Sutter} P.~M.,  2015,
  \mn@doi [\jcap] {10.1088/1475-7516/2015/11/018}, \href
  {http://adsabs.harvard.edu/abs/2015JCAP...11..018M} {11, 018}

\bibitem[\protect\citeauthoryear{{Neyrinck}}{{Neyrinck}}{2008}]{Neyrinck2008}
{Neyrinck} M.~C.,  2008, \mn@doi [\mnras] {10.1111/j.1365-2966.2008.13180.x},
  \href {http://adsabs.harvard.edu/abs/2008MNRAS.386.2101N} {386, 2101}

\bibitem[\protect\citeauthoryear{{Padilla}, {Ceccarelli}  \&
  {Lambas}}{{Padilla} et~al.}{2005}]{Padilla2005}
{Padilla} N.~D.,  {Ceccarelli} L.,   {Lambas} D.~G.,  2005, \mn@doi [\mnras]
  {10.1111/j.1365-2966.2005.09500.x}, \href
  {http://adsabs.harvard.edu/abs/2005MNRAS.363..977P} {363, 977}

\bibitem[\protect\citeauthoryear{{Paillas}, {Lagos}, {Padilla}, {Tissera},
  {Helly}  \& {Schaller}}{{Paillas} et~al.}{2017}]{Paillas2017}
{Paillas} E.,  {Lagos} C.~D.~P.,  {Padilla} N.,  {Tissera} P.,  {Helly} J.,
  {Schaller} M.,  2017, \mn@doi [\mnras] {10.1093/mnras/stx1514}, \href
  {http://adsabs.harvard.edu/abs/2017MNRAS.470.4434P} {470, 4434}

\bibitem[\protect\citeauthoryear{Paranjape, Lam  \& Sheth}{Paranjape
  et~al.}{2012}]{Paranjape2012}
Paranjape A.,  Lam T.~Y.,   Sheth R.~K.,  2012, \mn@doi [\mnras]
  {10.1111/j.1365-2966.2011.20154.x}, 420, 1648

\bibitem[\protect\citeauthoryear{{Peacock} \& {Smith}}{{Peacock} \&
  {Smith}}{2000}]{Peacock2000}
{Peacock} J.~A.,  {Smith} R.~E.,  2000, \mn@doi [\mnras]
  {10.1046/j.1365-8711.2000.03779.x}, \href
  {http://adsabs.harvard.edu/abs/2000MNRAS.318.1144P} {318, 1144}

\bibitem[\protect\citeauthoryear{{Perlmutter} et~al.,}{{Perlmutter}
  et~al.}{1999}]{Perlmutter1999}
{Perlmutter} S.,  et~al., 1999, \mn@doi [\apj] {10.1086/307221}, \href
  {http://adsabs.harvard.edu/abs/1999ApJ...517..565P} {517, 565}

\bibitem[\protect\citeauthoryear{{Pisani}, {Sutter}, {Hamaus}, {Alizadeh},
  {Biswas}, {Wandelt}  \& {Hirata}}{{Pisani} et~al.}{2015}]{Pisani2015}
{Pisani} A.,  {Sutter} P.~M.,  {Hamaus} N.,  {Alizadeh} E.,  {Biswas} R.,
  {Wandelt} B.~D.,   {Hirata} C.~M.,  2015, \mn@doi [\prd]
  {10.1103/PhysRevD.92.083531}, \href
  {http://adsabs.harvard.edu/abs/2015PhRvD..92h3531P} {92, 083531}

\bibitem[\protect\citeauthoryear{{Platen}, {van de Weygaert}  \&
  {Jones}}{{Platen} et~al.}{2007}]{Platen2007}
{Platen} E.,  {van de Weygaert} R.,   {Jones} B.~J.~T.,  2007, \mn@doi [\mnras]
  {10.1111/j.1365-2966.2007.12125.x}, \href
  {http://adsabs.harvard.edu/abs/2007MNRAS.380..551P} {380, 551}

\bibitem[\protect\citeauthoryear{{Pollina}, {Baldi}, {Marulli}  \&
  {Moscardini}}{{Pollina} et~al.}{2016}]{Pollina2016}
{Pollina} G.,  {Baldi} M.,  {Marulli} F.,   {Moscardini} L.,  2016, \mn@doi
  [\mnras] {10.1093/mnras/stv2503}, \href
  {http://adsabs.harvard.edu/abs/2016MNRAS.455.3075P} {455, 3075}

\bibitem[\protect\citeauthoryear{{Prunet}, {Pichon}, {Aubert}, {Pogosyan},
  {Teyssier}  \& {Gottloeber}}{{Prunet} et~al.}{2008}]{mpgrafic}
{Prunet} S.,  {Pichon} C.,  {Aubert} D.,  {Pogosyan} D.,  {Teyssier} R.,
  {Gottloeber} S.,  2008, \mn@doi [\apjs] {10.1086/590370}, \href
  {http://adsabs.harvard.edu/abs/2008ApJS..178..179P} {178, 179}

\bibitem[\protect\citeauthoryear{{Riess} et~al.,}{{Riess}
  et~al.}{1998}]{Riess1998}
{Riess} A.~G.,  et~al., 1998, \mn@doi [\apj] {10.1086/300499}, \href
  {http://adsabs.harvard.edu/abs/1998AJ....116.1009R} {116, 1009}

\bibitem[\protect\citeauthoryear{Sanchez et~al.}{Sanchez
  et~al.}{2017}]{Sanchez2016}
Sanchez C.,  et~al., 2017, \mn@doi [\mnras] {10.1093/mnras/stw2745}, 465, 746

\bibitem[\protect\citeauthoryear{{Schaap} \& {van de Weygaert}}{{Schaap} \&
  {van de Weygaert}}{2000}]{Schaap2000}
{Schaap} W.~E.,  {van de Weygaert} R.,  2000, \aap, \href
  {http://adsabs.harvard.edu/abs/2000A%26A...363L..29S} {363, L29}

\bibitem[\protect\citeauthoryear{{Scoccimarro}, {Sheth}, {Hui}  \&
  {Jain}}{{Scoccimarro} et~al.}{2001}]{Scoccimarro2001}
{Scoccimarro} R.,  {Sheth} R.~K.,  {Hui} L.,   {Jain} B.,  2001, \mn@doi [\apj]
  {10.1086/318261}, \href {http://adsabs.harvard.edu/abs/2001ApJ...546...20S}
  {546, 20}

\bibitem[\protect\citeauthoryear{Sheth \& van~de Weygaert}{Sheth \& van~de
  Weygaert}{2004}]{Sheth2003}
Sheth R.~K.,  van~de Weygaert R.,  2004, \mn@doi [\mnras]
  {10.1111/j.1365-2966.2004.07661.x}, 350, 517

\bibitem[\protect\citeauthoryear{{Shirasaki}, {Nishimichi}, {Li}  \&
  {Higuchi}}{{Shirasaki} et~al.}{2017}]{Shirasaki2017}
{Shirasaki} M.,  {Nishimichi} T.,  {Li} B.,   {Higuchi} Y.,  2017, \mn@doi
  [\mnras] {10.1093/mnras/stw3254}, \href
  {http://adsabs.harvard.edu/abs/2017MNRAS.466.2402S} {466, 2402}

\bibitem[\protect\citeauthoryear{{Tessore}, {Winther}, {Metcalf}, {Ferreira}
  \& {Giocoli}}{{Tessore} et~al.}{2015}]{Tessore2015}
{Tessore} N.,  {Winther} H.~A.,  {Metcalf} R.~B.,  {Ferreira} P.~G.,
  {Giocoli} C.,  2015, \mn@doi [\jcap] {10.1088/1475-7516/2015/10/036}, \href
  {http://adsabs.harvard.edu/abs/2015JCAP...10..036T} {10, 036}

\bibitem[\protect\citeauthoryear{{Teyssier}}{{Teyssier}}{2002}]{ramses}
{Teyssier} R.,  2002, \mn@doi [\aap] {10.1051/0004-6361:20011817}, \href
  {http://adsabs.harvard.edu/abs/2002A%26A...385..337T} {385, 337}

\bibitem[\protect\citeauthoryear{Will}{Will}{2014}]{Will2014}
Will C.~M.,  2014, \mn@doi [Living Rev. Rel.] {10.12942/lrr-2014-4}, 17, 4

\bibitem[\protect\citeauthoryear{{Zheng}, {Coil}  \& {Zehavi}}{{Zheng}
  et~al.}{2007}]{Zheng2007}
{Zheng} Z.,  {Coil} A.~L.,   {Zehavi} I.,  2007, \mn@doi [\apj]
  {10.1086/521074}, \href {http://adsabs.harvard.edu/abs/2007ApJ...667..760Z}
  {667, 760}

\bibitem[\protect\citeauthoryear{{Zivick}, {Sutter}, {Wandelt}, {Li}  \&
  {Lam}}{{Zivick} et~al.}{2015}]{Zivick2015}
{Zivick} P.,  {Sutter} P.~M.,  {Wandelt} B.~D.,  {Li} B.,   {Lam} T.~Y.,  2015,
  \mn@doi [\mnras] {10.1093/mnras/stv1209}, \href
  {http://adsabs.harvard.edu/abs/2015MNRAS.451.4215Z} {451, 4215}

\bibitem[\protect\citeauthoryear{van~de Weygaert \& Platen}{van~de Weygaert \&
  Platen}{2011}]{vandeWeygaert2009}
van~de Weygaert R.,  Platen E.,  2011, \mn@doi [Int. J. Mod. Phys.]
  {10.1142/S2010194511000092}, 1, 41

\makeatother
\end{thebibliography}

\bsp	
\label{lastpage}
\end{document}